\documentclass{jfm}

\usepackage{graphicx}
\usepackage{newtxtext}
\usepackage{newtxmath}
\usepackage{natbib}
\usepackage{hyperref}
\hypersetup{
    colorlinks = true,
    urlcolor   = blue,
    citecolor  = black,
}

\newcommand{\RomanNumeralCaps}[1]
\linenumbers

\usepackage{amsmath}
\usepackage{amssymb}
\usepackage{enumitem}

\newcommand{\bk}{\mathbf{k}}

\newcommand{\br}{\mathbf{r}}

\newcommand{\bu}{\mathbf{u}}
\newcommand{\bv}{\mathbf{v}}

\newcommand{\rd}{\mathrm{d}}

\newcommand{\R}{\mathbb{R}}

\title{Bathyspheres, dust devils, lunar plumes and Martian drones:  a generalised mathematical model for particle-fluid flow fields generated by static particle bed entrainment across macroscopic scales}

\author{S. Langdon\aff{1}\corresp{\email{stephen.langdon@brunel.ac.uk}}
  \and D. J. Needham\aff{2}}

\affiliation{\aff{1}Department of Mathematics, Brunel University of London, Uxbridge, UB8 3PH, United Kingdom.
\aff{2}School of Mathematics, University of Birmingham, Birmingham, B15 2TT, United Kingdom.
}

\begin{document}
\maketitle

\begin{abstract}
We consider a recently derived mathematical model for the mechanism by which a wind driven flow entrains particles into a fluid from an otherwise static particle bed, a process that is fundamental in initiating and driving the phenomenon of particle cloud formation. This model was previously applied to the ``helicopter brownout'' problem, in which, as a helicopter descends towards a bed of sand,  the downwash and swirling flow from the rotors entrain particles into the air, forming a high-velocity particle-laden cloud around the helicopter.  Here, we define a new measure of particle cloud formation potential, outline how the model can be incorporated into existing Computational Fluid Dynamics software, and extend the model to scenarios including: formation of particle clouds in the absence of swirl, such as those caused by a bathysphere descending to the ocean bed, or a rocket landing on the lunar surface, in which case we can precisely predict the critical height above the surface at which particle cloud formation is initiated; generation of dust clouds by less powerful downwash and swirling flows caused by drones or other smaller rotorcraft, such as NASA's Ingenuity Mars helicopter, in which case gravitational effects are included in the model;  formation of dust devils (dust clouds formed within and around an upwardly circulating columnar air flow), in which case downwash is replaced by updraft.  In each case, predictive models are derived through minor adaptations to key modelling assumptions.
\end{abstract}

{\bf MSC Codes }  76-10, 76T99


\section{Introduction}
\label{sec:intro}

In this paper, we consider a mathematical model for the generation of wind-driven particle flow fields from otherwise static particle beds, with the region above the static particle bed considered as a fluidized region with particles in suspension in the fluid, forming a two-phase flow.  A crucial aspect of developing an accurate model is including a rational mechanism to account for the process of entrainment/detrainment of particles into or from the flow to the particle bed.  This process is fundamental in initiating and driving the phenomenon of the particle cloud, and its omission is likely to lead to models that may fail to accurately simulate the two-phase flow phenomena in the main fluidized region.

In~\citet{NL24}, the authors developed a rational mathematical model for this process, consistent with the fundamental Newtonian laws of mechanics.  This was achieved by considering the physical model across three structured regions, namely: the fluidized region in which the particles lifted from the surface of the static particle bed are in suspension in the fluid;  the interfacial layer in which the interaction mechanisms between the fluid flow and the particle bed enable particle entrainment into (or detrainment from) the fluid;  the static particle bed, in which the particles are stored.  Careful consideration of the local force balance between the fluid flow and the incipiently fluidized particles
within the interfacial layer enabled the derivation of a macroscopic boundary condition that could be applied at the interface of the fully fluidized region and the particle bed surface, closing the macroscopic two-phase flow problem in the fully fluidized region, with the entrainment and detrainment mechanism fully accounted for in a rational manner.  Further details of this mathematical model are provided in~\S\ref{sec:model}.

One key application of this model, which has received significant attention in the literature in recent years (see, e.g., \citet{PhBr08,PhKiBr11,GLG,PoMiPaPeVe20,TaGaBaLiZhHu21,TGZWCW,TaYoHeYuWa22,CaWaJi22,RoBaSt23,LSXW,LXJHL, WuRe26}), is the scenario in which a helicopter (or other rotorcraft) descends towards an otherwise static bed of sand, with the downwash and swirling flow from the rotors entraining particles into the air, forming a high-velocity particle-laden cloud around the helicopter.  This can potentially lead to a significant deterioration in visibility and is often referred to as the ``brownout problem'' in the aero-engineering literature.  The specific application of the model to this problem was presented in~\citet[\S6]{NL24}, and in~\citet{LN25} it was shown that key features of the field (pilot visibility and surface deformation) could easily be extracted from the solution, enabling an analysis of how those features may depend on key model parameters (such as helicopter height, ratio of swirl velocity to downwash velocity, and physical properties of the particle bed). 

Here, we show how the model derived in~\citet{NL24} can be applied (with some modifications) to address a variety of related scenarios and introduce some new ideas for using this model as a predictive tool.  For each application considered, we revisit the model of~\citet{NL24} and consider whether any additional features that were excluded from the helicopter brownout model of~\citet[\S6]{NL24} should be included, for that particular application, and whether additional analysis could enable a clearer understanding, due to the specific nature of each scenario under consideration.

We begin, in~\S\ref{sec:model}, by summarizing the mathematical model, as presented in~\citet{NL24}.  We then introduce, in \S\ref{sec:cloud}, a new measure of the tendency of a primary fluid flow field above a static particle bed to generate a significant fluidized particle cloud above the bed. This measure can be computed solely from the boundary data of the primary fluid flow field, enabling an a priori assessment of how the particle cloud formation potential of this fluid flow field may depend on key parameters in the specific application.  Some properties of this measure are described in Appendix~\ref{sec:PC}.

In \S\ref{sec:no_swirl}, we consider the case in which there is no swirl component driving the primary fluid flow (that is, the case in which the particle entrainment is driven solely by primary fluid flow downwash).  In this scenario, we can determine a priori, solely from a consideration of the boundary data, critical values of key parameters for which the nature of the primary two-phase fluid flow changes, and in particular, for which significant particle cloud formation becomes possible. Two applications in which the swirl component of the flow can be neglected are: the problem in which a bathysphere descends towards the ocean bed (see, e.g., \citet{Beeb}); the problem in which a lander descends towards a celestial body, such as the Moon or Mars (often referred to as the ``lunar plume'' problem in the aero-engineering literature, see, e.g., \citet{JC22,JKJKBJL25}).  Here, we show how each of these applications can be modelled through an appropriate choice of the relevant parameters, as introduced in~\S\ref{sec:model}.  The surface deformation calculation in \citet{LN25}, measuring how the initially flat particle bed is deformed, may be particularly relevant to these applications.

For the case in which the primary fluid flow field has both a downwash and swirl generated by a helicopter rotor (as was considered in~\citet[\S6]{NL24} and~\citet{LN25}), the effect of gravity on the particles in the flow was neglected, due to the relative effect of the different forces in such a rapid primary fluid flow.  However, for comparable applications in which the downwash and swirling fluid flow are less powerful, such as those generated by a drone, it may be important to include the effect of gravity in the model.  In \S\ref{sec:gravity}, we show how the mathematical model derived in~\citet{NL24} can be adapted to include gravity and consider two relevant applications: the problem in which a small drone hovers over a sand bed on Earth; the problem in which a small drone, such as the NASA Ingenuity Mars helicopter, is hovering over a particle bed on Mars (see, e.g., \citet{RaLoSlWa21,CaBaDiYo25}).  Again, these separate applications can be modelled simply through an appropriate choice of parameters.

Through our numerical experiments in~\S\ref{sec:gravity}, we notice new features arising in the two-phase flow field whose presence appears to be due to small values of the parameter $\alpha$ (defined precisely in \S\ref{sec:model} below), which measures the ratio of particle collision pressure to drag force in the particle flow phase.  For sand fluidized in air, we tentatively estimate $\alpha\approx0.1$, but for smaller and lighter particles (such as Martian dust), we hypothesize that $\alpha$ might take much smaller values.  We explore further how these new features depend on $\alpha$ (both with and without inclusion of gravity in the model) in Appendix~\ref{sec:alpha}.

As a final physical application, in \S\ref{sec:DD} we consider the particle cloud generated by the naturally occurring  phenomenon often referred to as a dust devil, modelling of which has been considered recently by \citet{dust_devil_D}.  This is a dust cloud that can form within and around an upwardly circulating columnar air flow.  For this application, the key change to our modelling assumptions is the replacement of downwash with updraft, and our results show that this leads to particle cloud formation in our model broadly consistent with that seen in photographs of dust devils (see, e.g., \citet[Figure 1]{dust_devil_B}).

The mathematical model considered in this paper contains a number of simplifying assumptions to enable consideration of the fundamental processes of the key two-phase flow without requiring consideration of the complexity of the fine-scale details of the primary fluid flow field.  Although this is sufficient to enable an investigation of how the most significant features in the two-phase flow field depend on specific parameters, there are many applications for which fine-scale detail of the primary fluid flow field is required, such as modelling blade erosion  (see, e.g., \citet{TaYaXuZhWaFa26}) or air intake duct sand ingestion (see, e.g., \citet{LiDoSoYa25}), or consideration of brownout mitigation in rotor design (see, e.g., \citet{HuCa26}).  To improve accuracy in the current model, a more detailed description of the primary fluid flow in the helicopter brownout problem could be introduced (as provided in, e.g., \citet{Nathan_Green_2009}).  As such, in \S\ref{sec:CFD} we outline how our model could potentially be incorporated into a Computational Fluid Dynamics (CFD) platform that provides an accurate representation of the single-phase fluid velocity field in the absence of a particle bed.  Such an integration could prove to be a powerful tool in enabling a clearer understanding of particle-cloud formation for a wider range of applications.

Finally, in \S\ref{sec:conc}, we present some conclusions, and a discussion of potential future research.

\section{The mathematical model}
\label{sec:model}
We begin by giving a brief review of the model developed in \citet{NL24} in the context of the helicopter brownout problem. Specifically, for the two-phase flow we introduce a continuum voidage field measuring fluid volume per unit spatial volume of the two-phase flow, and state the advection/diffusion partial differential equation (PDE) that it satisfies in the fluidised region (derived in~\cite{NL24} from conservation of mass and momentum in the fluid and particle phases), with the condition at the boundary with the particle bed as derived in \citet{NL24}.  Throughout this section, we use the same notation as in \citet{NL24}, and though the nonlinear elliptic boundary value problem is stated in full here, for a full explanation of the derivation of the model the present paper should be read in close conjunction with the theory and modelling developed in \citet[\S1--5]{NL24} and the application of the model to the helicopter cloud problem in \citet[\S6]{NL24}.

For a helicopter hovering steadily, with rotor blades rotating in a horizontal plane, in otherwise still air above a sand bed, which has undisturbed level at $z=0$, the steady problem for the associated voidage field in the fluid flow region can be stated in cylindrical polar coordinates $(R,\vartheta,z)$ (with $z$ measuring dimensionless distance vertically upward), with radial symmetry in $\vartheta$ (corresponding to the rotor-induced flowfield in the simple helicopter model being represented by a revolution-averaged induced flowfield), as the elliptic boundary value problem:
\begin{eqnarray}
  & E\left(E_{RR} + \frac{1}{R} E_R + E_{zz} \right) - \bar{a}(R,z) E_R - \bar{b}(R,z) E_z = 0, \quad (R,z)\in \omega,& \label{eqn:h34} \\
  & E=g(R) \mbox{ on } z=0, \quad R\geq 0, & \label{eqn:h35} \\
  & E_R = 0 \mbox{ on } \displaystyle{\left\{ \begin{array}{ll} R=0, & z\in(0,z_d^-) \cup (z_d^+,\infty) \\ R=\Delta, & z\in[z_d^-,z_d^+], \end{array} \right.}& \label{eqn:h36} \\
  & E_z = 0 \mbox{ on } z= z_d^+,z_d^-, \quad R\in(0,\Delta), & \label{eqn:h37} \\
  & E\rightarrow 1 \mbox{ as } (R^2 + z^2)\rightarrow\infty \mbox{ uniformly in } \omega. & \label{eqn:h38}
\end{eqnarray}
Here $E=E(R,z)\in[E_s,1]$ is the voidage field (representing the volume of fluid per unit spatial volume), with $0<E_s\ll 1$ being the packing voidage of the particles, and the domain $\omega$ is defined by
\[
  \omega = \{ (R,z): \, z>0, \, R>0 \} \backslash \{ (R,z): \, z_d^- \leq z \leq z_d^+, \, 0 < R\leq \Delta \}.
\]
In this simple model, the effect of the helicopter is modelled as generating an incompressible, inviscid fluid flow field represented by a half-space fluid dipole located at $R=0$, $z=z_d>0$ (in dimensionless variables, with $z_d$ measuring the ratio of the hovering height of the helicopter rotor to the rotor blade radius), and a uniform fluid line vortex aligned along the positive z-axis. The flow takes place in the half-space region $z>0$ and exterior to a small cylindrical shell, where $\Delta>0$ represents the dimensionless length and radius of the cylindrical shell which is aligned with the $z$-axis and with the dipole at its centre, and measures the ratio of the helicopter rotor hub to the rotor blade radius. In addition,
\[
  z_d^+ = z_d + \frac{\Delta}{2}, \quad z_d^- = z_d - \frac{\Delta}{2},
\]
and the boundary data is given in terms of the continuous and piecewise smooth function $g:[0,\infty)\rightarrow\R$, which is defined by,
\begin{equation}
  g(R) = \left\{ \begin{array}{ll}
                E_s, & \mbox{for }|\nabla_h \bar{\phi}(R,0)|^2 > (\gamma E_s)^{-1}, \\
                \displaystyle{\gamma^{-1}\left( \frac{9z_d^2 R^2}{4\pi^2(R^2+z_d^2)^5} + \frac{\Omega^2}{4\pi^2 R^2} \right)^{-1}}, & \mbox{for }\gamma^{-1}\leq|\nabla_h \bar{\phi}(R,0)|^2\leq (\gamma E_s)^{-1}, \\
                1, & \mbox{for }|\nabla_h \bar{\phi}(R,0)|^2<\gamma^{-1}. \\
                \end{array} \right.
   \label{eqn:h25}
\end{equation}
Here, again using the notation of~\citet{NL24},
\begin{equation}
  |\nabla_h \bar{\phi}(R,0)|^2 = \frac{9z_d^2 R^2}{4\pi^2(R^2+z_d^2)^5} + \frac{\Omega^2}{4\pi^2 R^2}.
  \label{eqn:h20}
\end{equation}

\[
  \bar{a}(R,z) = \frac{9}{2\alpha} a(R,z), \quad \bar{b}(R,z) = \frac{9}{2\alpha} b(R,z),
\]
for $(R,z)\in \omega$, where
\begin{eqnarray}
  a(R,z) & = & \displaystyle{\frac{3R}{4\pi} \left[\frac{(z+z_d)}{(R^2 + (z + z_d)^2)^{5/2}} - \frac{(z-z_d)}{(R^2 + (z - z_d)^2)^{5/2}} \right]}, \label{eqn:h18} \\
  b(R,z) & = & \displaystyle{\frac{3}{4\pi} \left[\frac{(z+z_d)^2}{(R^2 + (z + z_d)^2)^{5/2}} - \frac{(z-z_d)^2}{(R^2 + (z - z_d)^2)^{5/2}} \right]} \nonumber \\
         &   & \quad + \displaystyle{\frac{1}{4\pi} \left[\frac{1}{(R^2 + (z - z_d)^2)^{3/2}} - \frac{1}{(R^2 + (z + z_d)^2)^{3/2}} \right]}. \label{eqn:h19}
\end{eqnarray}
Full derivation of this boundary value problem from fundamental first principles is described in \citet[\S6]{NL24}, together with a discussion of the physical relevance of each of the six dimensionless parameters appearing in the boundary value problem, namely $E_s$, $\alpha$, $\gamma$, $\Omega$, $z_d$ and $\Delta$.  Additional discussion of the relevance of these parameters to particular helicopter operating velocity scales such as the downwash velocity and rotor blade tip velocity can be found in \citet[\S II]{LN25}.  As discussed in~\citet[\S6]{NL24}, and using typical helicopter parameters as recorded, for example, in \citet{GLG,PoMiPaPeVe20,TaGaBaLiZhHu21,TGZWCW,TaYoHeYuWa22,LSXW,LXJHL}, the physical relevance and typical values of these parameters are as follows.  

Whilst in the helicopter example and the drone examples of~\S\ref{sec:gravity}, $\Delta$ has a specific physical meaning (the ratio of the helicopter rotor hub to the rotor blade radius), in some of the other modelling scenarios (bathysphere (\S\ref{subsec:bathysphere}), lunar plume (\S\ref{subsec:lunar}), dust devil (\S\ref{sec:DD})) $\Delta$ does not have a specific physical meaning but rather denotes simply the size of the shell that we place around the singularity.  Typically for the helicopter example $\Delta\approx 10^{-2}$, but in the other examples we simply require $\Delta\ll \text{min}\{1,z_d\}$.  Here, we take $\Delta=0.04$ in the examples where it is constrained by a physical interpretation (specifically, the drone examples in~\S\ref{sec:gravity}), and for computational efficiency we take the larger values $\Delta=0.1$ (in~\S\ref{sec:no_swirl}) and $\Delta=0.2$ (in~\S\ref{sec:DD}) in the examples where it is not.

As noted above, the parameter $z_d$ measures the ratio of the hovering height of the helicopter rotor to the rotor blade radius, and typically in the helicopter example $z_d\approx10^{-1}$--$10^1$.  In each of our examples below we consider values of $z_d$ within this range, but we note that for the dust devil example just a single value $z_d=5$ is appropriate (see \S\ref{sec:DD} for details).

For the fluid flow, the ratio of the hovering helicopter induced swirl velocity to the induced downwash velocity is measured by $\Omega$, and typically we have $\Omega\approx10^{-1}$--$10^0$.  For the case where there is no swirl, as discussed in \S\ref{sec:no_swirl}, we simply have $\Omega=0$.

The parameter $E_s$ is the particle packing voidage, and for sand in air this is typically of magnitude $10^{-2}$.  We will take this value throughout.

The ratio of the lift force per unit volume to the gravity force per unit volume on particles in the interfacial layer is given by $\gamma$, and for the helicopter example we typically have $\gamma\approx10^{2}$--$10^3$.  For other examples, the value of $\gamma$ may vary significantly, and this is discussed further in \S\ref{sec:no_swirl}, \S\ref{sec:gravity} and \S\ref{sec:DD}.

Finally, as mentioned in~\S\ref{sec:intro}, $\alpha$ measures the ratio of the particle collisional pressure to the drag force in the particle flow phase and we tentatively estimate, for sand fluidized in air, that $\alpha\approx10^{-1}$.  For other examples, we consider a wider range of values of $\alpha$, and in Appendix~\ref{sec:alpha} we consider in particular how very small values of $\alpha$ may affect the flow.

For each example in \S\ref{sec:no_swirl}, \S\ref{sec:gravity} and \S\ref{sec:DD}, representing closely related applications of the general model developed in~\citet{NL24} for generic wind-generated particle-fluid flow fields, we derive the associated reduced forms in each case and illustrate the full solution of the problem (in some cases representing a minor variation to (\ref{eqn:h34})--(\ref{eqn:h19}), as described below), computed using the numerical method introduced in~\citet[Appendix~B]{NL24}.  The numerical discretisation is uniquely defined by four parameters:  $\epsilon>0$ (see \citet[(B7)]{NL24}), defining the tolerance level for the nonlinear solve;  $L_R\geq 1$ and $L_z \geq 1$ (see \citet[(B8)]{NL24}), which control the truncation of the unbounded domain $\omega$ in the horizontal and vertical directions, respectively;  $N\geq 1$ (see \citet[(B9--B12)]{NL24}), which controls the mesh size.  To enable precise reproducability of our results, for each example considered below we state the values of $\epsilon$ ($\epsilon\leq10^{-6}$ in each example), $L_R, L_z$ and $N$, and the resulting total number of degrees of freedom (given by \citet[(B14)]{NL24}, at least four million in each example), used in our computations.

\section{A measure of particle cloud formation potential}
\label{sec:cloud}
Before moving on to specific applications, in this section we introduce a simple useful measure of the tendency of a primary fluid flow field above an otherwise static bed of particles to generate a significant particle cloud in the associated fluid flow field. To achieve this, we utilize the concept of incipient fluidization of particles in a thin interfacial layer that accommodates the transmission of particles between the static particle bed and the fully fluidized flow region above, as described in~\S\ref{sec:intro}. The dynamics and in particular the localised voidage field within this layer were fully developed and analysed in \citet{NL24}, and for ease of reference the notation that we use here is adopted directly from \citet{NL24}. To keep things simple at the outset, we will assume that the otherwise undisturbed surface of the static particle bed, and the thin, incipiently fluidized, interfacial layer above this, are horizontal, and occupy an open and simply connected region $\mathcal{A}\subset\mathbb{R}^2$ whose boundary $\partial\mathcal{A}$ is a simple smooth closed curve (the closure of this region is denoted by $\overline{\mathcal{A}}$). Exterior to this region, the surface is assumed to be immobile. In the notation of \citet{NL24} we write the tangential primary fluid velocity field at the interfacial layer as $\mathbf{u}_h(\mathbf{r}_h)$ with $\mathbf{r}_h \in \overline{\mathcal{A}}$. Here $\mathbf{u}_h:\overline{\mathcal{A}} \to\mathbb{R}^2$ is taken to be a smooth two-dimensional vector field (except possibly at a finite number of interior points, at which $|\mathbf{u}_h(\mathbf{r}_h)|\to \infty$ as each such point is approached) on $\overline{\mathcal{A}}$. It is convenient to introduce the associated non-negative scalar field $\chi:\overline{\mathcal{A}} \to [0,\infty)$, given by
\begin{equation}
    \chi(\mathbf{r}_h) =
    \begin{cases}
    \frac{1}{\Lambda}~~\text{when}~~|\mathbf{u}_h(\mathbf{r}_h)|^2 \le \frac{1}{\Lambda},\\
     |\mathbf{u}_h(\mathbf{r}_h)|^2~~\text{when}~~ \frac{1}{\Lambda}< |\mathbf{u}_h(\mathbf{r}_h)|^2 < \frac{1}{E_s\Lambda},\\
     \frac{1}{E_s\Lambda}~~\text{when}~~|\mathbf{u}_h(\mathbf{r}_h)|^2 \ge \frac{1}{E_s\Lambda}. \label{eqne1}
    \end{cases}
\end{equation}
Here, both $\Lambda$ and $E_s$ are the physical parameters introduced in \citet[\S3]{NL24}, with $\Lambda$ given by \citet[(3.25)]{NL24} and $E_s\in(0,1)$ denoting the packing voidage of the particles (see also~\S\ref{sec:model}), with each regarded as fixed.  This scalar field $\chi$ is henceforth referred to as the \emph{admissible} scalar field associated with the tangential fluid velocity field $\mathbf{u}_h$. We readily observe that $\chi$ is continuous, and bounded, with in particular,
\begin{equation}
    \frac{1}{\Lambda} \le \chi{(\mathbf{r}_h)} \le \frac{1}{E_s\Lambda},~~\forall~~\mathbf{r}_h \in \overline{\mathcal{A}}.\label{eqne2}
\end{equation}
Furthermore, if two tangential fluid velocity fields $\mathbf{u}_h^1$ and $\mathbf{u}_h^2$ are such that
\begin{equation}
    |\mathbf{u}_h^1(\mathbf{r}_h)|^2 \le |\mathbf{u}_h^2(\mathbf{r}_h)|^2,~~\forall~~\mathbf{r}_h\in \overline{\mathcal{A}},
\end{equation}
then $\mathbf{u}_h^1$ is said to be a \emph{decrease} of $\mathbf{u}_h^2$.  Correspondingly, the associated admissible scalar field $\chi_1$ is said to be a \emph{decrease} of the associated admissible scalar field $\chi_2$. We now follow the theory in \citet{NL24} to introduce the incipient voidage field in the interfacial layer (from equation~(4.6) in \citet{NL24}), associated with the admissible scalar field, as
\begin{equation}
E(\chi(\mathbf{r}_h)) = \frac{1}{\Lambda \chi(\mathbf{r}_h)},~~\forall~~\mathbf{r}_h \in \overline{\mathcal{A}}. \label{eqne3}
\end{equation}
It is worth noting here that it is from this incipient voidage field in the interfacial layer that the boundary condition~(\ref{eqn:h35}) in the elliptic boundary value problem (\ref{eqn:h34})--(\ref{eqn:h19}) is ultimately derived (again, see \citet[\S5--6]{NL24} for details).

This enables us to introduce the functional
\begin{equation}
\mathcal{PC}:\{\chi \in C(\overline{\mathcal{A}}): \chi ~\text{is an admissible scalar field}\} \to \overline{\mathbb{R}}^+
\end{equation}
as
\begin{equation}
    \mathcal{PC}(\chi) =\frac{1}{|\overline{\mathcal{A}}|}\int\!\!\int_{\overline{\mathcal{A}}}{(1 - E(\chi(\mathbf{r}_h)))}d\mathcal{A} \label{eqnf}
\end{equation}
which gives a simple measure of the average particle volume that is fluidized in the interfacial layer and, as such, immediately available for particle cloud formation in the bulk fluidized flow region.  Key properties of this functional are established in Appendix~\ref{sec:PC}. In particular, it is determined that the image of $\mathcal{PC}$ is precisely the closed interval $[0,1-E_s]$, and that it is sufficiently regular to be Lipschitz continuous on its domain and to have a Fr\'{e}chet derivative at each point in its domain, whilst being decreasing on pairs of decreasing points in its domain. The minimum value of zero is achieved if and only if $\chi\equiv \Lambda^{-1}$, while the maximum value of $(1-E_s)$ is achieved if and only if $\chi\equiv (\Lambda E_s)^{-1}$.

This measure can be used as a useful tool in rapidly assessing the particle cloud forming potential of a given primary fluid flow field which, say, has $M$ free parameters denoted as $\mathbf{\mu}\in \Sigma \subset \mathbb{R}^M$. This then leads to a parameterised admissible scalar field $\chi = \chi(\mathbf{r}_h; \mathbf{\mu})\equiv \chi_{\mathbf{\mu}}$, and substitution into (\ref{eqnf}) results in a parameterisation $\mathcal{PC} = \mathcal{PC}(\chi_{\mu})\equiv f(\mathbf{\mu})$
where $f:\Sigma\to [0,1-E_s]$ is now a prescribed real-valued function of the parameters $\mathbf{\mu} \in \Sigma$. The effects on cloud forming potential through the parameters can now be investigated by direct analysis of this scalar function. It should be noted here that the above theory relies on applications for which the domain $\mathcal{A}$ is bounded on the plane. In applications where $\mathcal{A}$ is formally unbounded, we can still use this theory by choosing a suitable fixed finite representation of $\mathcal{A}$ based on the finite length scale and location of the flow geometry in the associated fluidized region. For example, in the model we have introduced for the helicopter brownout application, a suitable choice for $\mathcal{A}$ would be, in terms of the dimensional coordinates used in the model, the circular disc in the plane $z=0$, with centre at $R=0$, and having radius $\sim 10l$.  In order to enable direct comparison between each of the applications considered in the sections below, and in particular to enable inclusion of the lunar plume model (see \S\ref{subsec:lunar}), we choose the disc to have radius $12l$ consistently across all examples.

As an illustration, we consider here the helicopter cloud problem, as described in~\citet[\S6-7]{NL24}.  For this and the other examples considered in \S\ref{sec:no_swirl}, \S\ref{sec:gravity} and \S\ref{sec:DD} below, $\mathcal{PC}$ depends only on $\Omega$, $\gamma$, $E_s$ and $z_d$, as it is a function solely of boundary data which depends only on those parameters and not on $\Delta$ or $\alpha$ (recall~(\ref{eqn:h25})).  For each example, here and in the subsequent sections, we can use~(\ref{eqnf}) to predict particle cloud formation potential, and then compare with key features of the voidage field calculated through numerical solution of (\ref{eqn:h34})--(\ref{eqn:h19}). In the dimensionless coordinates used throughout (with lengths scaled against the applications' geometric length scale $l$) we take $\mathcal{A}$ to be the circular disc on the plane $z=0$, with centre at the origin and radius $\hat{R}=12$ (as described above). The formula~(\ref{eqnf}) then becomes 
\begin{equation}
  \mathcal{PC}(\chi_{\mu})\equiv f(\mathbf{\mu})=\frac{1}{\pi\hat{R}^2}\int_0^{2\pi}\int_{0}^{\hat{R}}(1-g(R))R\, \rd R\, \rd \theta = \frac{2}{\hat{R}^2}\int_{0}^{\hat{R}}(1-g(R))R\, \rd R,
  \label{eqn:PC}
\end{equation}
and this integral can be easily evaluated using a standard quadrature formula.  For the helicopter cloud problem, as described in~\citet[\S6-7]{NL24}, we fix $E_s=0.01$, and investigate how $f(\mathbf{\mu})$ depends on $\Omega\in[0.1,1]$, $\gamma\in[100,1000]$, and $z_d\in[0.5,5]$, noting that the broad conclusions of the numerical results in~\citet[\S7]{NL24} (computed with $\Delta=0.04$ and $\alpha=0.1$, though those values do not affect the value of $f(\mathbf{\mu})$) were that the particle cloud formation potential was greater for smaller values of $z_d$, and larger values of $\Omega$ and $\gamma$.

\begin{figure}
   \centerline{
   \includegraphics[width=0.33\linewidth]{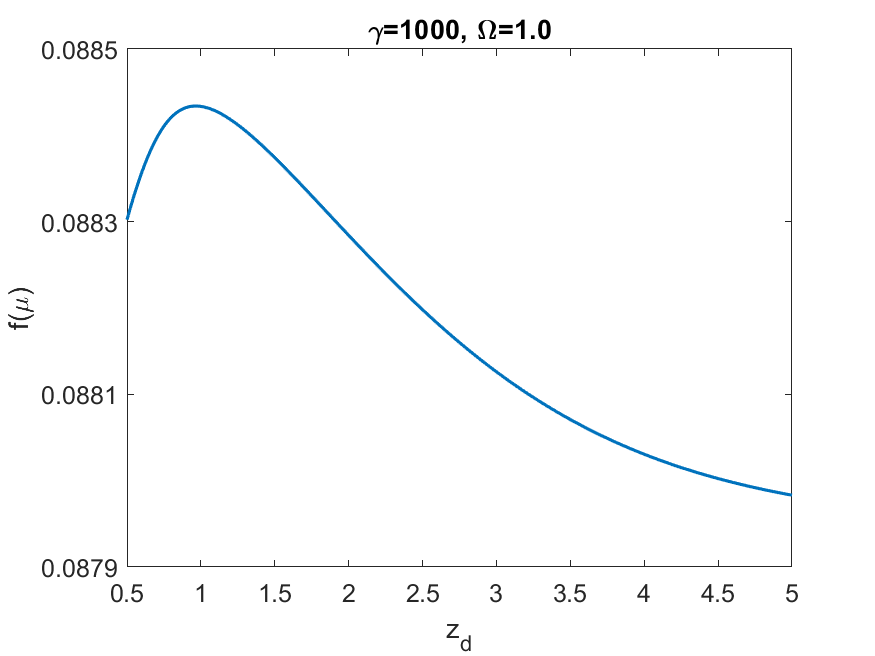}
   \includegraphics[width=0.33\linewidth]{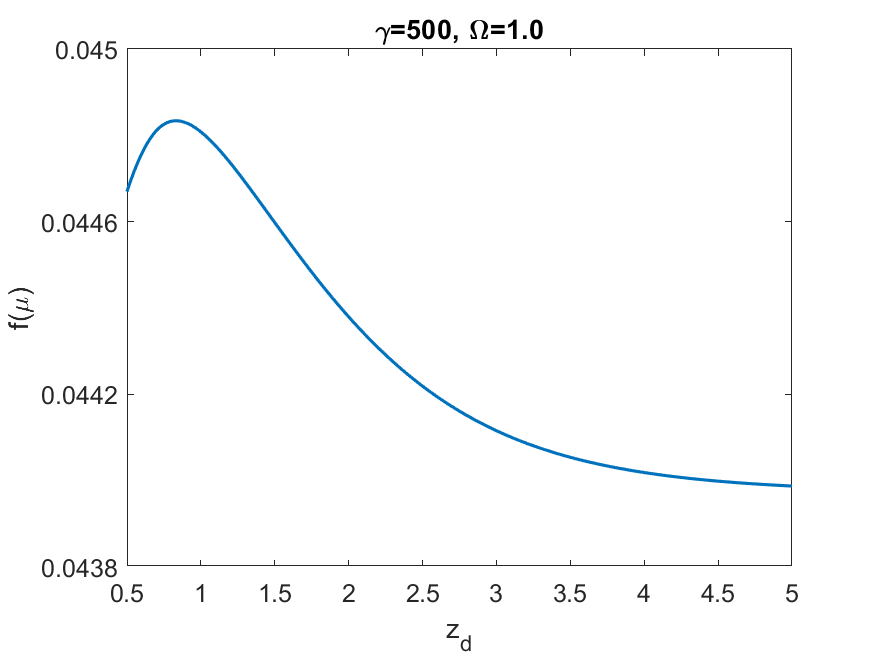}
   \includegraphics[width=0.33\linewidth]{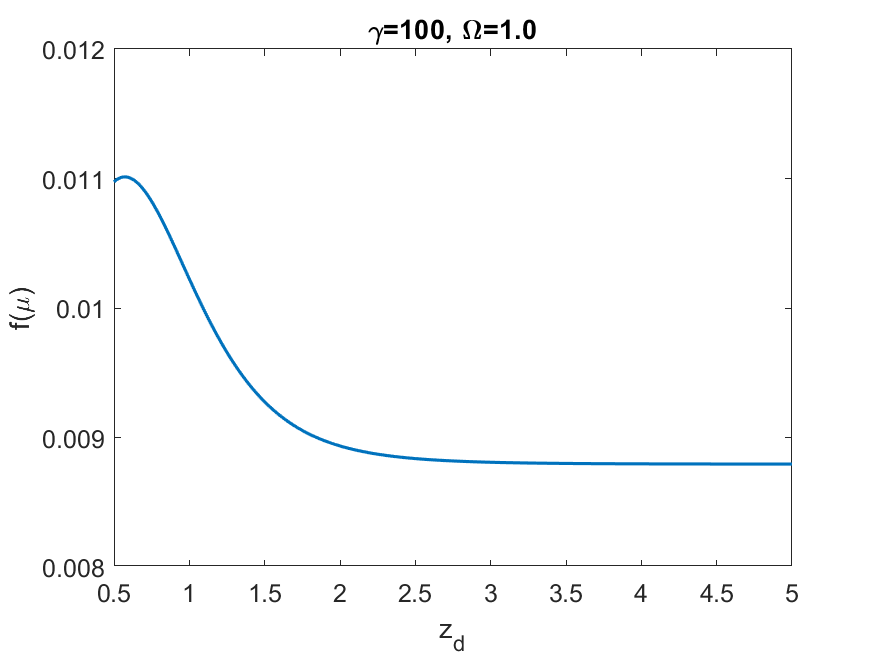}}
   \centerline{
   \includegraphics[width=0.33\linewidth]{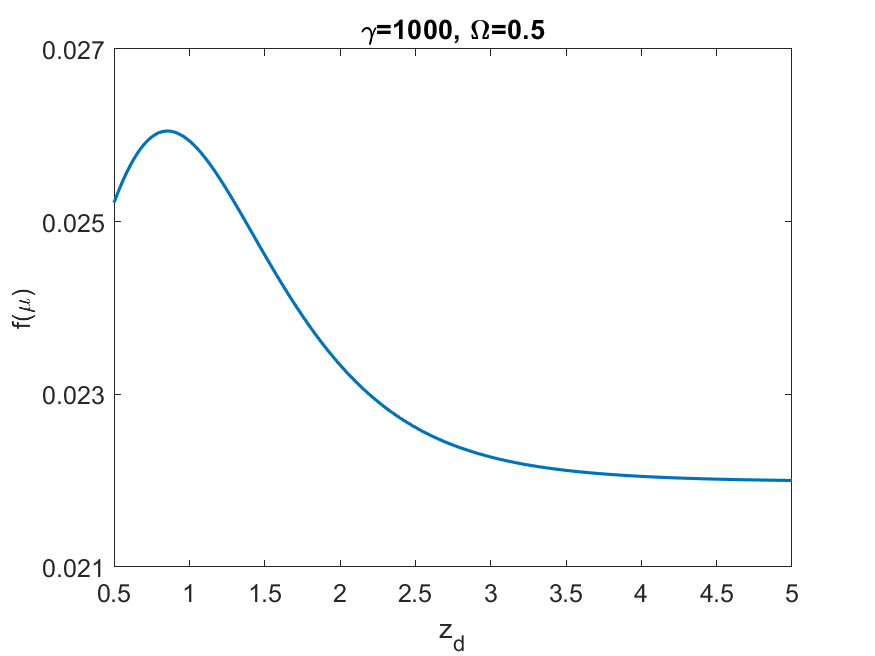}
   \includegraphics[width=0.33\linewidth]{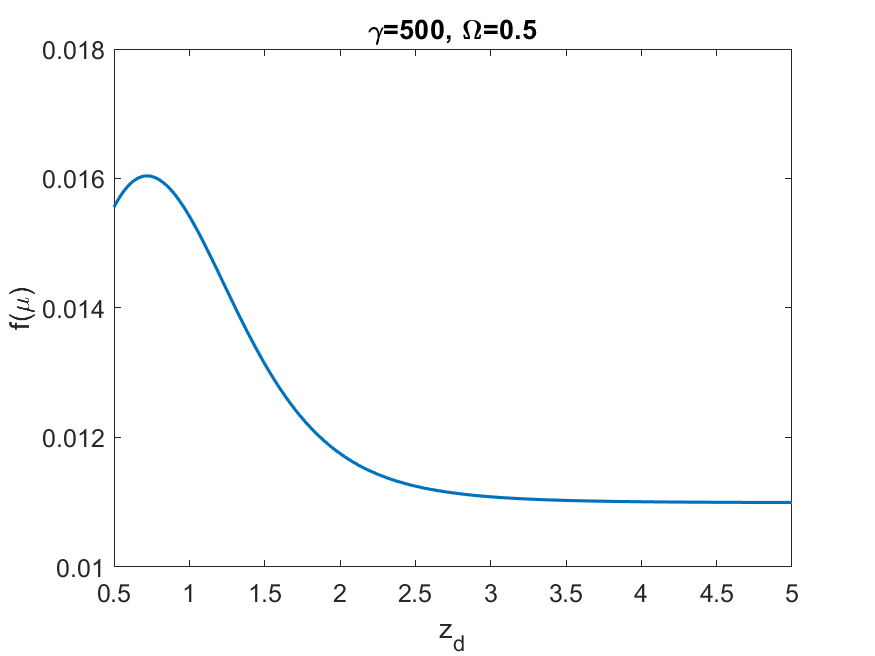}
   \includegraphics[width=0.33\linewidth]{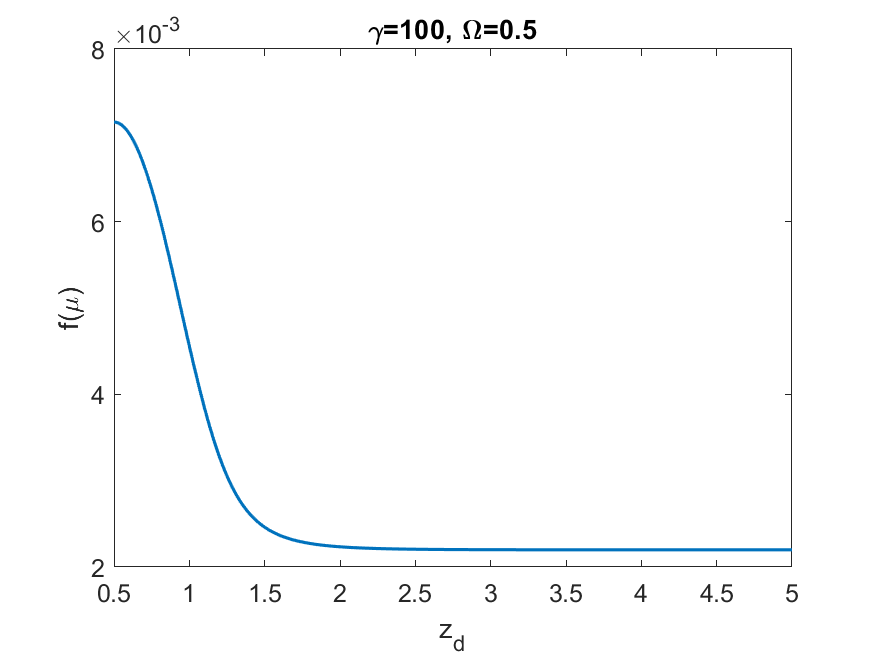}}
   \centerline{
   \includegraphics[width=0.33\linewidth]{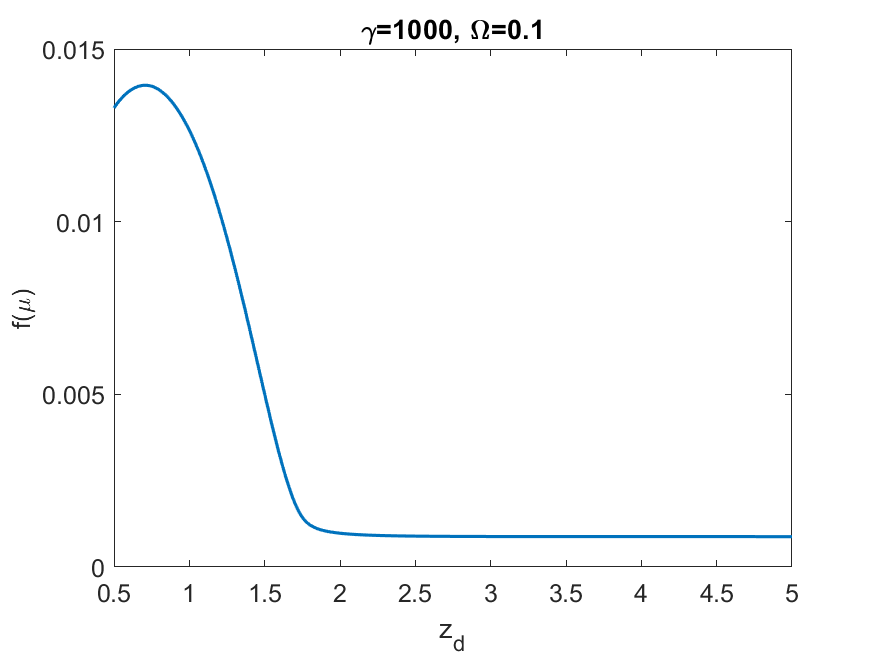}
   \includegraphics[width=0.33\linewidth]{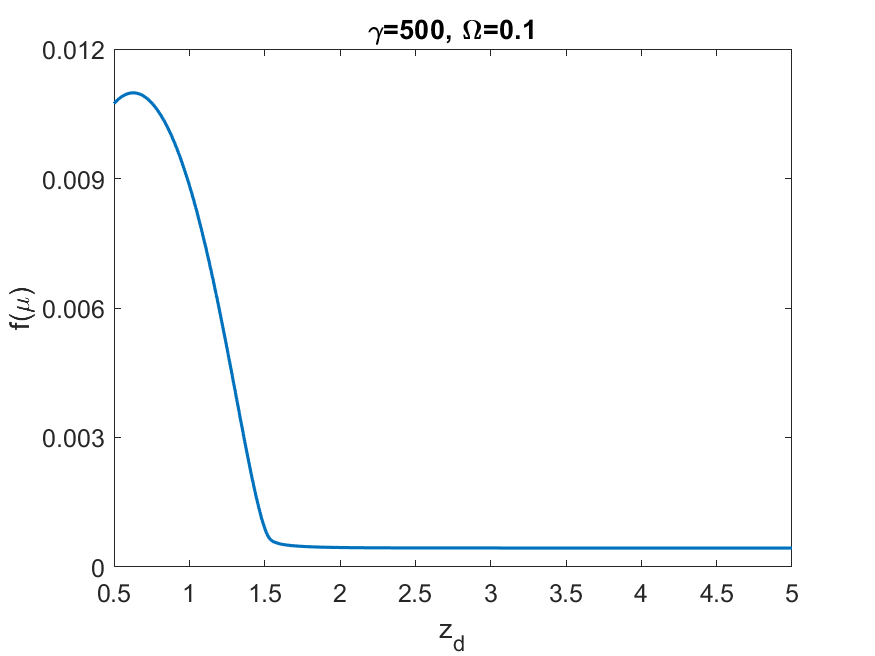}
   \includegraphics[width=0.33\linewidth]{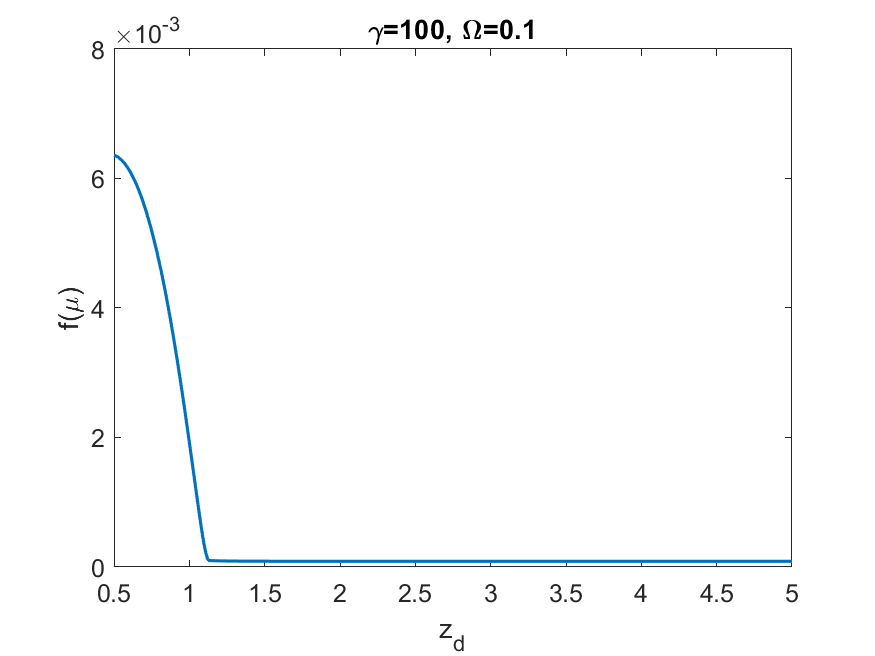}}
   \caption{Plots showing how $f(\mathbf{\mu})$ (given by (\ref{eqn:PC})) varies with $z_d$, for $\gamma=1000$ (left), $\gamma=500$ (middle), $\gamma=100$ (right) and $\Omega=1.0$ (top), $\Omega=0.5$ (middle), and $\Omega=0.1$ (bottom).}
  \label{fig:vary_zd}
\end{figure}

First, in figure~\ref{fig:vary_zd} we show how $f(\mathbf{\mu})$ varies as $z_d$ changes, for each combination of $\gamma=1000,500,100$ and $\Omega=1.0, 0.5, 0.1$.
In each case, we see that $f(\mathbf{\mu})$ is indeed greater for larger values of $\Omega$ and $\gamma$.  In most cases, $f(\mathbf{\mu})$ increases as $z_d$ decreases, but this is not uniformly true - in particular, when $\gamma$ and $\Omega$ are sufficiently large, and $z_d$ is sufficiently small, $f(\mathbf{\mu})$ can in fact decrease (slightly) as $z_d$ decreases.

\begin{figure}
   \centerline{
   \includegraphics[width=0.33\linewidth]{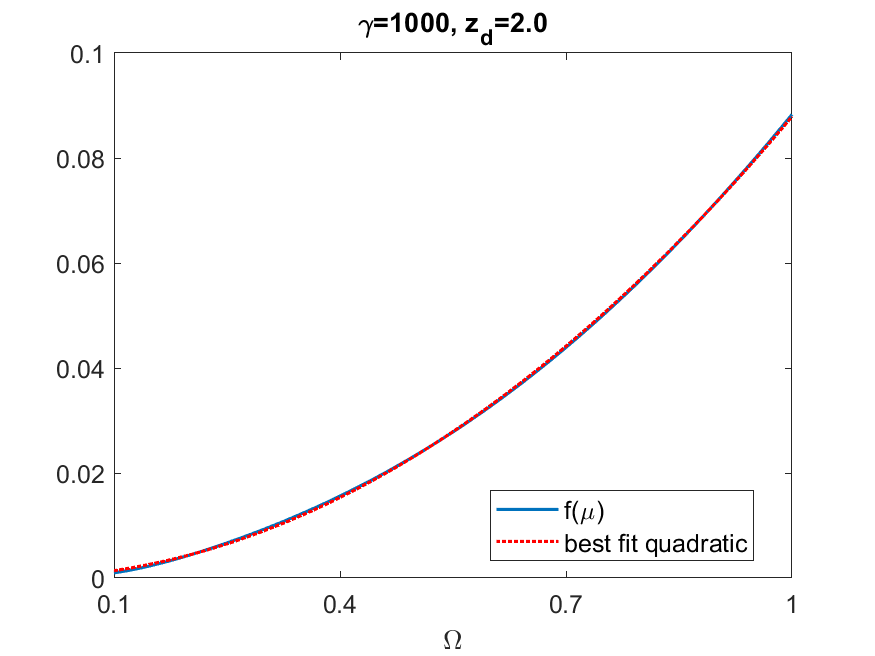}
   \includegraphics[width=0.33\linewidth]{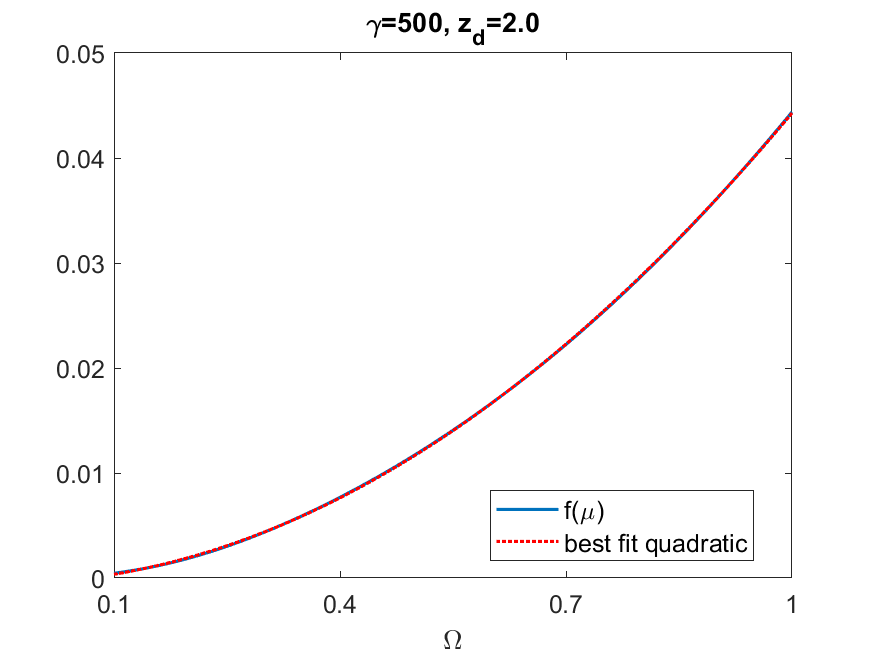}
   \includegraphics[width=0.33\linewidth]{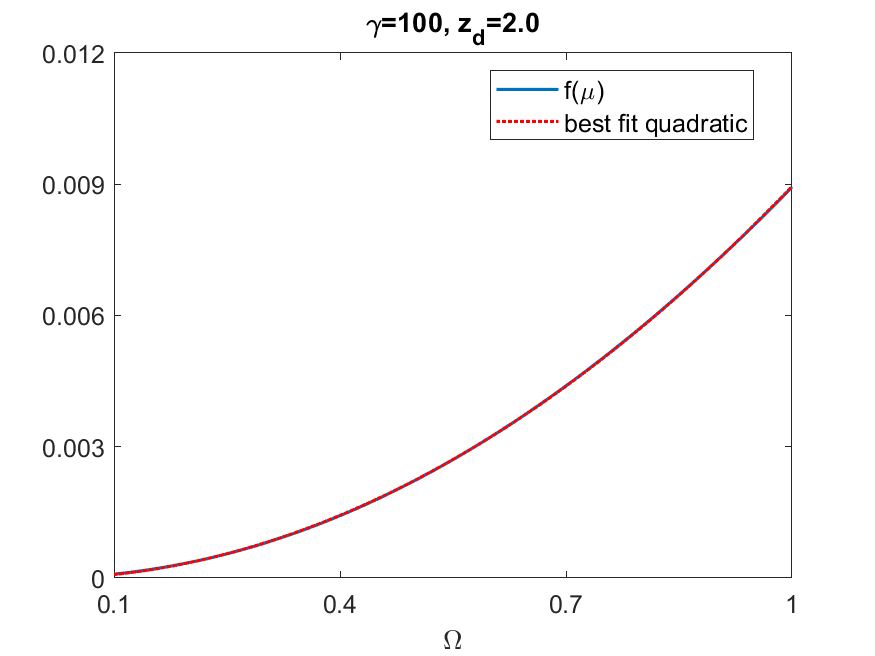}}
   \centerline{
   \includegraphics[width=0.33\linewidth]{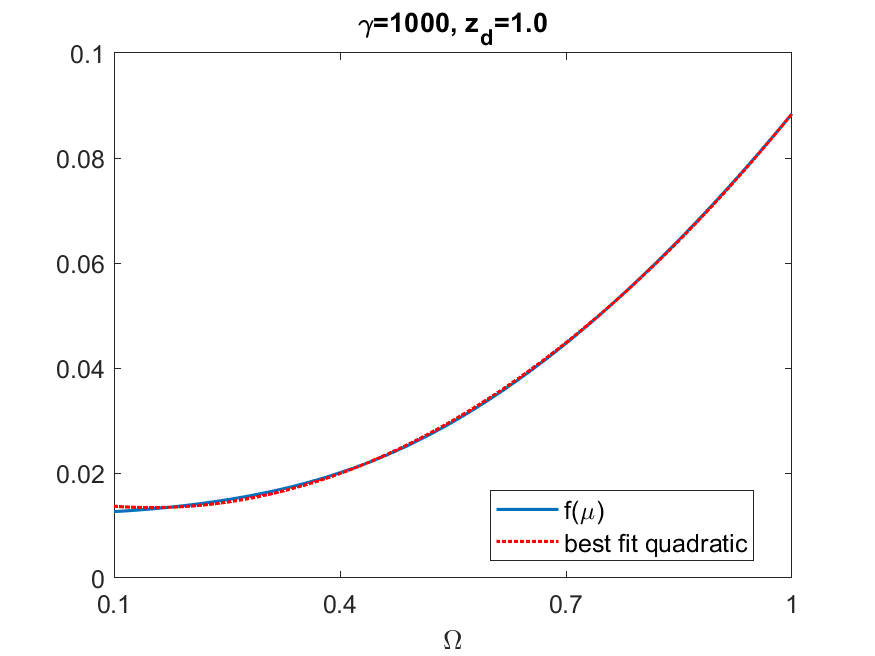}
   \includegraphics[width=0.33\linewidth]{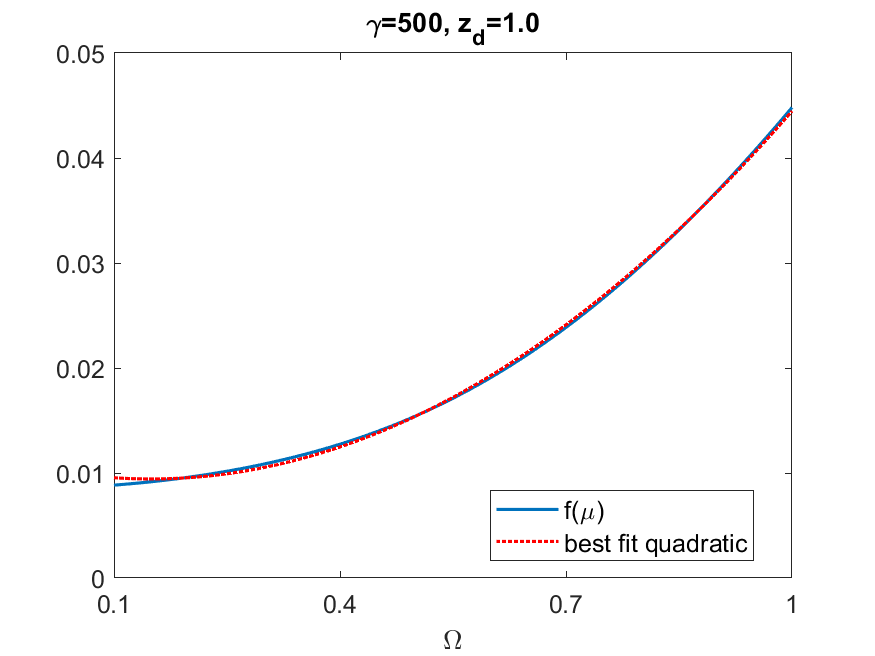}
   \includegraphics[width=0.33\linewidth]{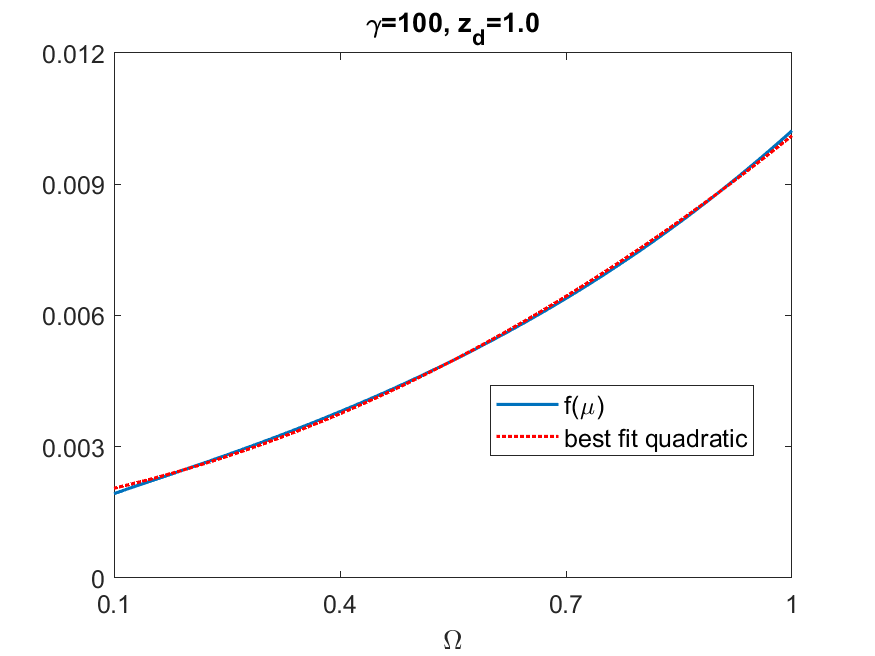}}
   \centerline{
   \includegraphics[width=0.33\linewidth]{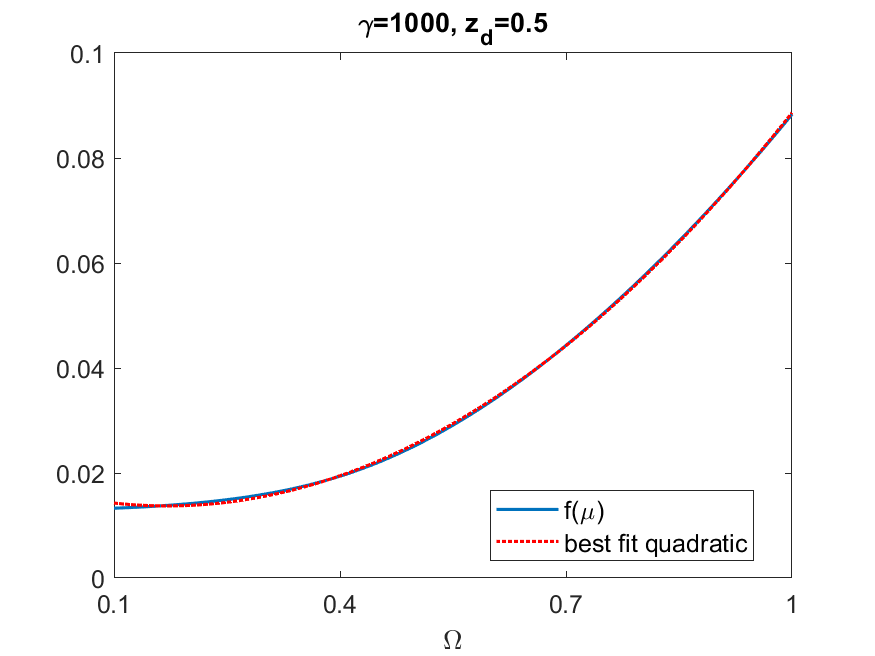}
   \includegraphics[width=0.33\linewidth]{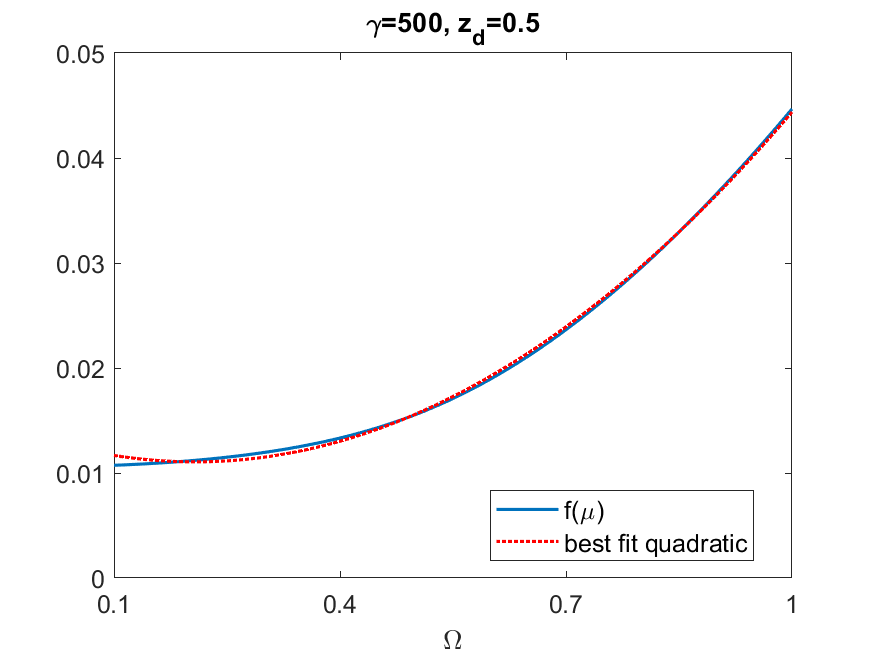}
   \includegraphics[width=0.33\linewidth]{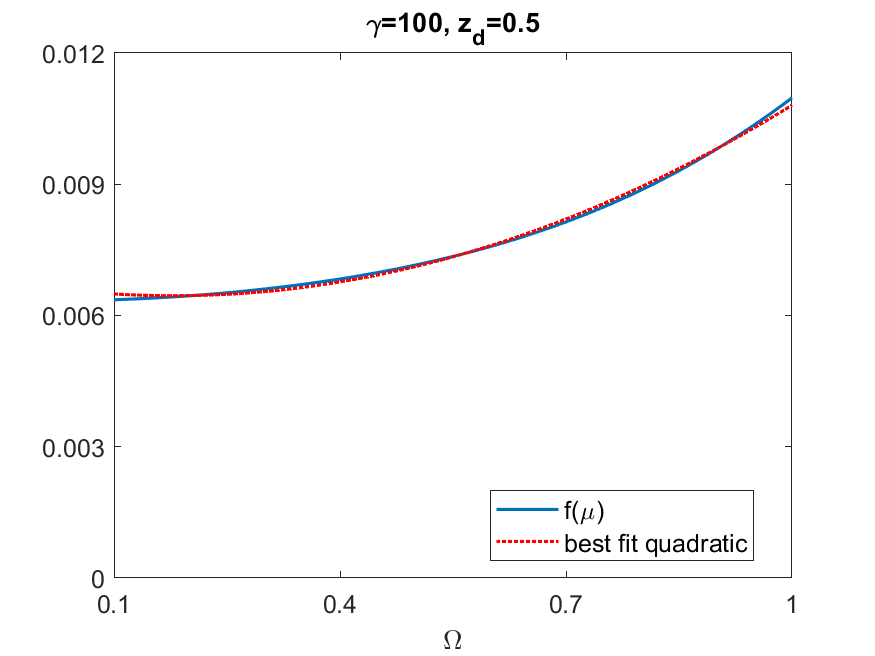}}
   \caption{Plots showing how $f(\mathbf{\mu})$ (given by (\ref{eqn:PC})) varies with $\Omega$, for $\gamma=1000$ (left), $\gamma=500$ (middle), $\gamma=100$ (right) and $z_d=2.0$ (top), $z_d=1.0$ (middle), and $z_d=0.5$ (bottom).}
  \label{fig:vary_omega}
\end{figure}

Next, in figure~\ref{fig:vary_omega} we explore how $f(\mathbf{\mu})$ varies as $\Omega$ changes, for each combination of $\gamma=1000,500,100$ and $z_d=2.0, 1.0, 0.5$.
In each case, $f(\mathbf{\mu})$ is greater for larger values of $\Omega$, and in fact it appears that $f(\mathbf{\mu})$ is a quadratic function of $\Omega$.  For comparison, on each plot we also show the best fitting quadratic (in a least  squares sense).  The quadratic curve is a better fit to the data when $z_d$, $\Omega$ and $\gamma$ are larger.

\begin{figure}
   \centerline{
   \includegraphics[width=0.33\linewidth]{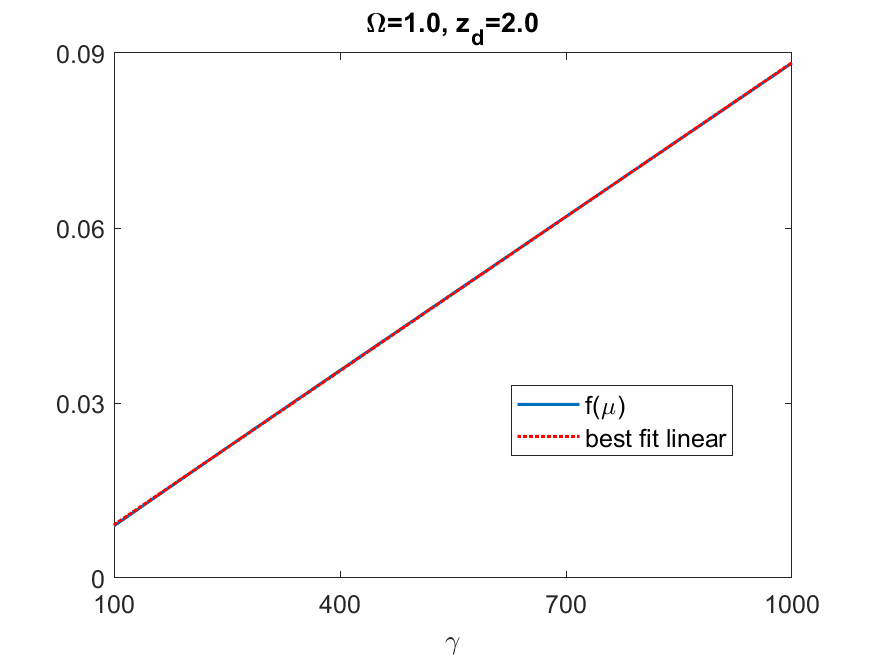}
   \includegraphics[width=0.33\linewidth]{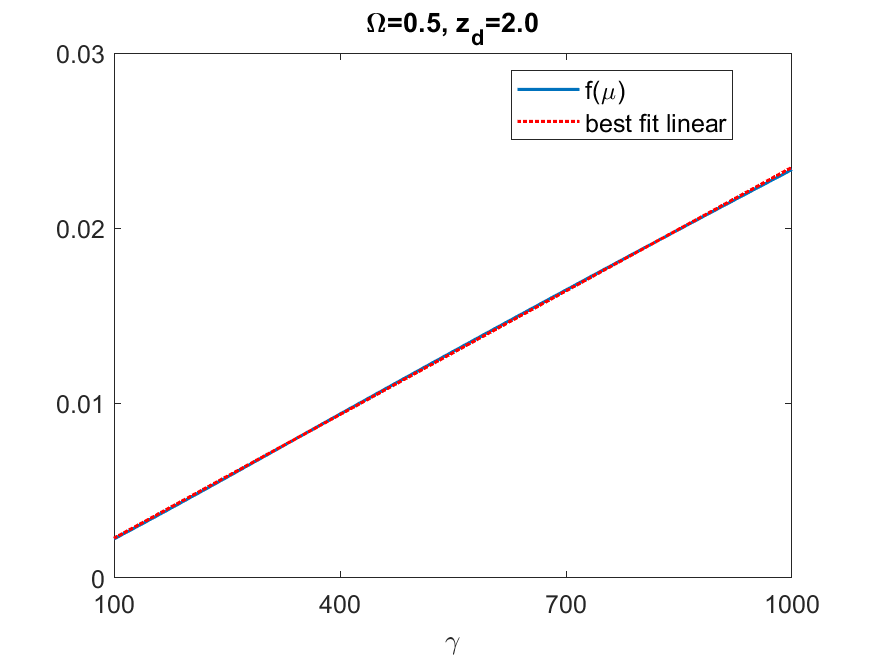}
   \includegraphics[width=0.33\linewidth]{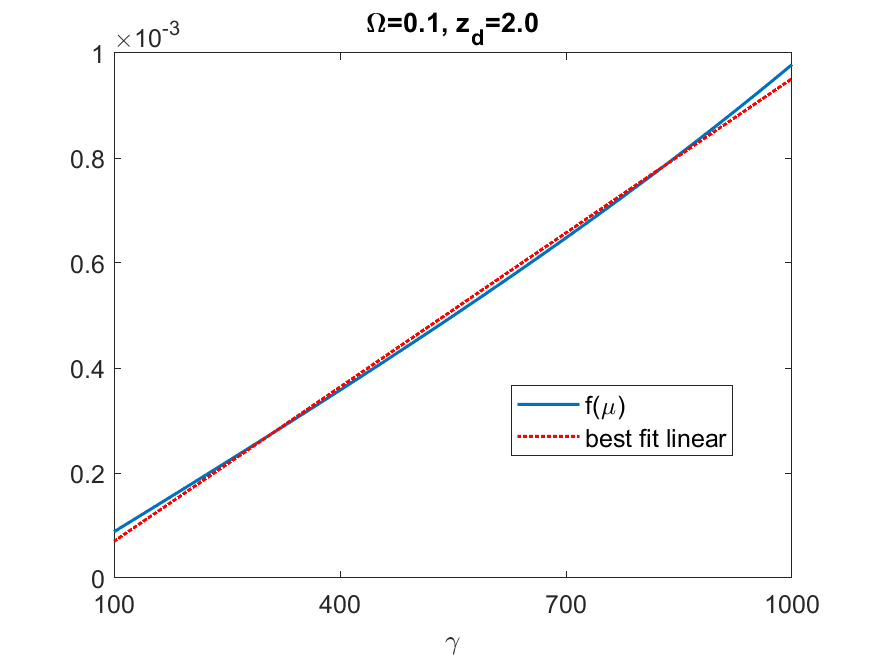}}
   \centerline{
   \includegraphics[width=0.33\linewidth]{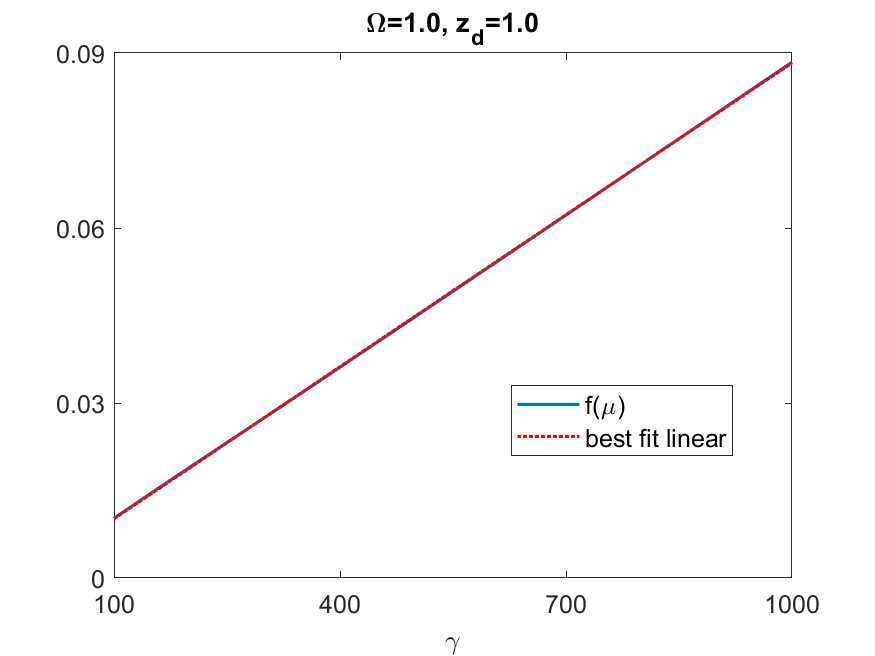}
   \includegraphics[width=0.33\linewidth]{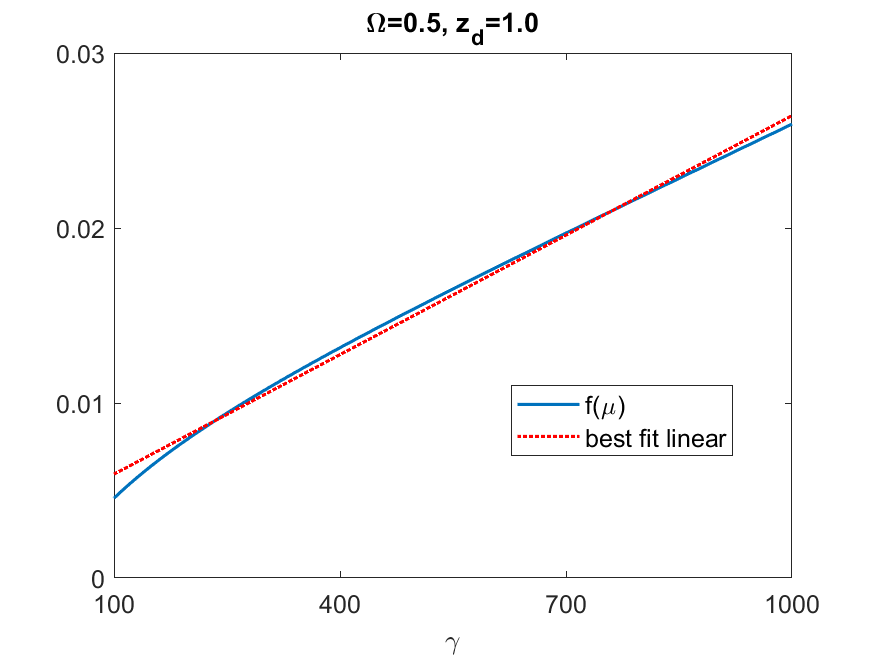}
   \includegraphics[width=0.33\linewidth]{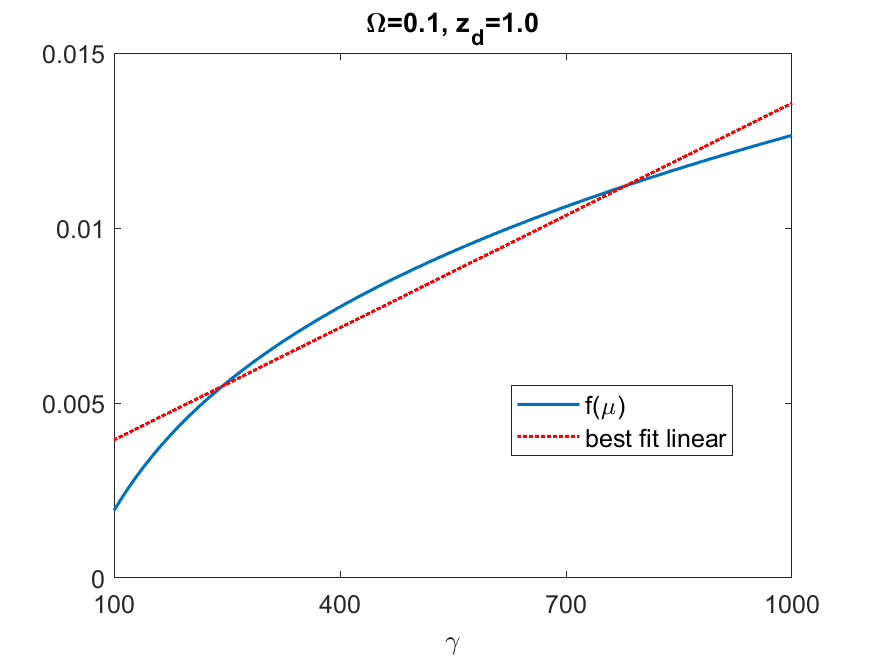}}
   \centerline{
   \includegraphics[width=0.33\linewidth]{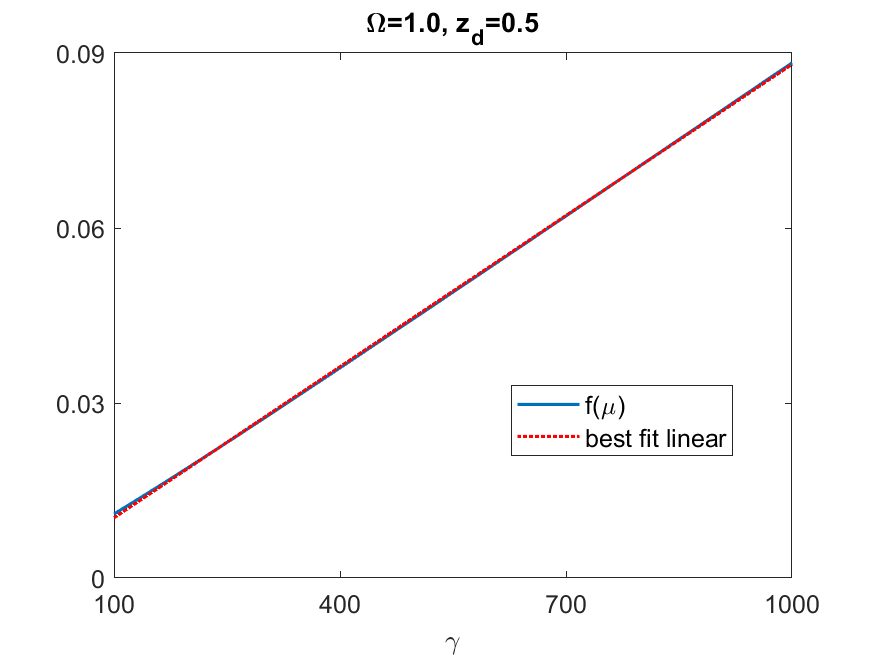}
   \includegraphics[width=0.33\linewidth]{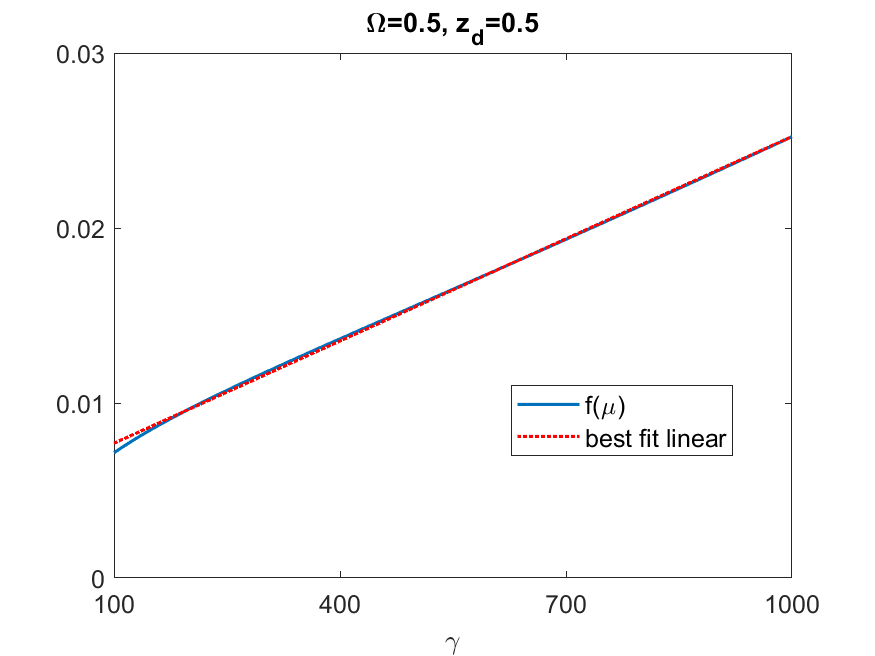}
   \includegraphics[width=0.33\linewidth]{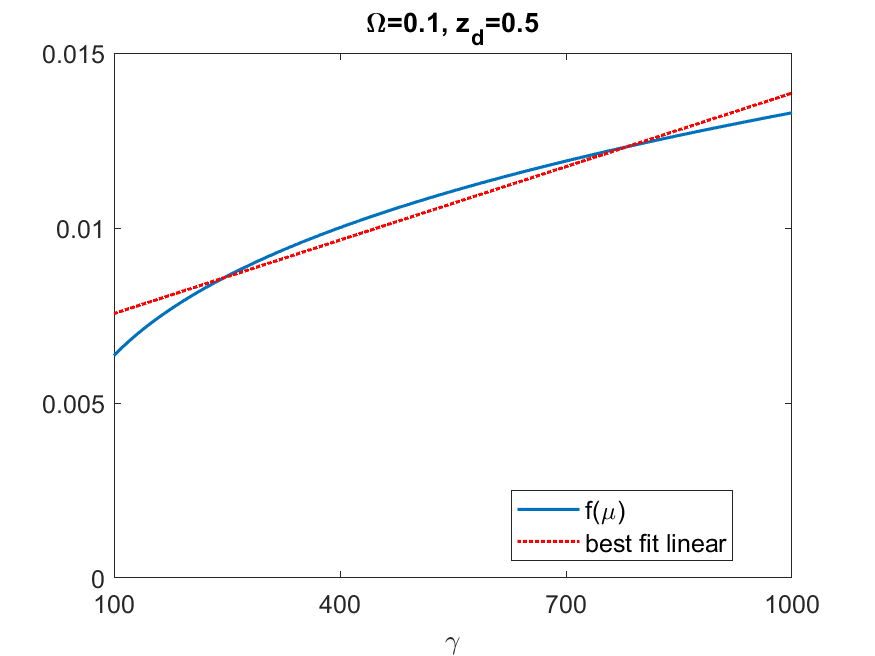}}
   \caption{Plots showing how $f(\mathbf{\mu})$ (given by (\ref{eqn:PC})) varies with $\gamma$, for $\Omega=1.0$ (left), $\Omega=0.5$ (middle), $\Omega=0.1$ (right) and $z_d=2.0$ (top), $z_d=1.0$ (middle), and $z_d=0.5$ (bottom).}
  \label{fig:vary_gamma}
\end{figure}

Finally, in figure~\ref{fig:vary_gamma} we explore how $f(\mathbf{\mu})$ varies as $\gamma$ changes, for each combination of $\Omega=1.0,0.5,0.1$ and $z_d=2.0, 1.0, 0.5$.
In each case, $f(\mathbf{\mu})$ is greater for larger values of $\gamma$, and in fact in several cases it appears that $f(\mathbf{\mu})$ has a linear dependence on $\gamma$.  For comparison, on each plot we also show the best fitting straight line (in a least  squares sense).  The linear plot is, in general, a better fit to the data when $z_d$, $\Omega$ and $\gamma$ are larger.  For  $\Omega=0.1$, and $z_d=1.0$ or $z_d=0.5$, the curve is clearly not linear, and in fact in these cases it appears that the dependence of $f(\mathbf{\mu})$ on $\gamma$ follows a fractional power law.

The nature of the results obtained in this section regarding particle cloud formation potential will be discussed in the context of each application in the subsequent sections.

\section{Applications with no swirl}
\label{sec:no_swirl}

\begin{figure}
   \centerline{
   \includegraphics[width=0.6\linewidth]{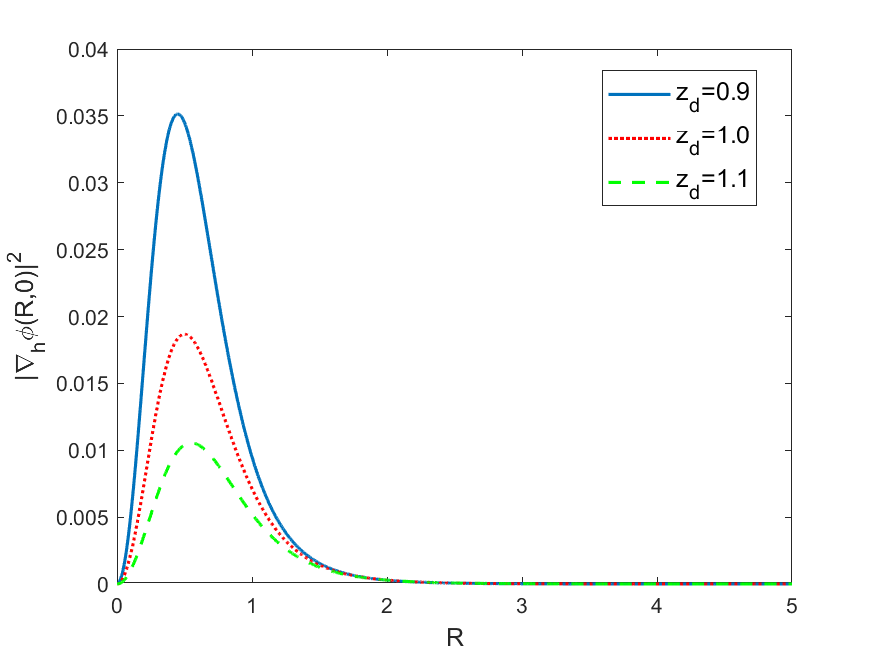}}
   \caption{Plot of $|\nabla_h \bar{\phi}(R,0)|^2$ (given by (\ref{eqn5.4})), for $z_d=0.9, 1.0, 1.1$.}
  \label{fig:no_swirl_fig1}
\end{figure}

In this section we consider situations in which there is no swirl component driving the primary fluid flow; that is, the case in which the parameter $\Omega$, which measures the ratio of the induced swirl velocity to the induced downwash velocity, can be taken as zero.  In this case, the revised form of $|\nabla_h \bar{\phi}(R,0)|^2$ (see~(\ref{eqn:h20})) is simply
\begin{equation}
  |\nabla_h \bar{\phi}(R,0)|^2 = \frac{9z_d^2 R^2}{4\pi^2(R^2+z_d^2)^5},
  \label{eqn5.4}
\end{equation}
for $R\ge0$, and parameter values $z_d>0$.  A graph of equation (\ref{eqn5.4}), for a range of values of $z_d$ is shown in figure~\ref{fig:no_swirl_fig1}.
The local maximum is located at
\begin{equation}
    R = R_m(z_d)\equiv \frac{1}{2}z_d,
\end{equation}
and the maximum value is given by
\begin{equation}
   |\nabla_h \bar{\phi}(R_m(z_d),0)|^2= \frac{576}{3125 \pi^2} z_d^{-6} \approx 0.0187 z_d^{-6}.
\end{equation}
The qualitatively similar shape of $|\nabla_h \bar{\phi}(R,0)|^2$ for each value of $z_d$ can be clearly seen in figure~\ref{fig:no_swirl_fig1}.

The absence of the swirl component in the primary fluid flow leads to a significant change in the nature of the boundary function $g:\mathbb{R}\to\mathbb{R}$ in this application compared to the corresponding form for the helicopter brownout application (as considered in~\citet[\S6]{NL24}, see in particular~\citet[Figure~7]{NL24} and the discussion immediately after~\citet[(6.38)]{NL24}).  A key new feature is that since in this case $|\nabla_h \bar{\phi}(R,0)|^2$ is bounded, which is not the case for $\Omega>0$ (see~\citet[(6.32)]{NL24}), it follows that at heights above the static particle bed corresponding to
\begin{equation}
    z_d^6 \ge \frac{576}{3125\pi^2}\gamma \equiv (z_d^c)^6
    \label{eqn:zdc}
\end{equation}
we have $g(R)\equiv1$, and the corresponding voidage field exterior to the vehicle generating the flow then has $E(R,Z)\equiv 1$, so no particle cloud appears until the height of the vehicle above the static particle bed has reduced below this critical value.  This raises the speculative thought that helicopter brownout could potentially be reduced significantly if the ratio of swirling flow to downwash could be sizeably reduced towards the final stages of landing.

We now consider two specific applications for which the omission of the swirl component of the flow seems reasonable, namely the bathysphere and the lunar plume problems.  For heights below the critical value $z_d^c$, given by~(\ref{eqn:zdc}) we present typical examples of computed voidage fields.

\subsection{Ocean-bed descending bathysphere cloud model}
\label{subsec:bathysphere}
In this application we consider the situation when a rigid spherical bathysphere slowly approaches a horizontal ocean bed composed of saturated particulate sand augmented with fine particulate biological debris, which we anticipate will form a quasi-steady particulate cloud in the locally surrounding (otherwise static) ocean water field. This phenomena has been observed by oceanic bathysphere descents which have taken place since the mid-1930's pioneering descents of William Beebe and Otis Barton (see \citet{Beeb}). Here we will take the flow resulting from the descending bathysphere (very simplistically) as that due to a vertically aligned dipole at the centre of the slowly descending bathysphere, and contained within a small cylindrical shell as introduced earlier in the brownout model. 

We denote the descending velocity of the bathysphere as $u_s$ and the radius of the bathysphere as $l$. The balance between buoyancy, gravity, and the macroscopic drag on the bathysphere then gives the following estimate:
\begin{equation}
    u_s^2\sim \frac{8(\lambda-1)lg}{3c_d}\sim 10^2 \times(\lambda - 1)~ms^{-1}, \label{eqn5.1}
\end{equation}
where we have written the mean density of the bathysphere as $\lambda \rho_f$, with $\rho_f$ being the oceanic water density, the dimensionless parameter $\lambda > 1$ and the total drag force on the bathysphere has been taken as $F_d = (1/2)\pi l^2\rho_fu_s^2c_d$, with $c_d$ being the drag coefficient for the bathysphere descending in oceanic water.  The appropriate strength of the dipole is then given by $M_s = u_sl^3$, and the height of the bathysphere centre above the ocean bed (at instant $t$) is represented by $lz_d$. When the speed of descent of the bathysphere is sufficiently slow, so that we may adopt a quasi-steady approximation for the voidage field, the model determining the voidage field in the fluid flow region local to the bathysphere is again given by equations (\ref{eqn:h34})-(\ref{eqn:h19}), but now with the parameter $\Omega$ set to be identically zero (as described above, simply reflecting that the fluid flow due to the slowly descending bathysphere generates no swirl component). In addition, the two remaining parameters now have estimated values, using (\ref{eqn5.1}) and the respective physical properties of sea water and granular sand, 
\begin{equation}
    \gamma = \frac{3}{16}.\frac{l^2}{a^2}.\frac{\rho_f}{\rho_s}.\frac{(\lambda - 1)\rho_f^2ga}{\mu^2}\sim 10^7\times(\lambda-1),
\end{equation}
and
\begin{equation}
    \alpha = \frac{9}{2}.\frac{a^2}{l^2}.\frac{p_0}{(\lambda-1)\rho_fgl}\sim p_0 \times(\lambda-1)^{-1}\times10^{-8}.
\end{equation}
Here $a$ represents the mean radius of sand grains forming the ocean bed, $\rho_s$ is the density of sand and $\mu$ is the dynamic viscosity of the oceanic water. In this representative case, we select $\lambda$ to give a settling velocity of the bathysphere of $2~ms^{-1}$, which requires the choice $\lambda = 1 + 1/(5\times10^5)$ after which we obtain the estimates,
\begin{equation}
    \gamma\sim10^5~~\text{and}~~\alpha\sim 10^{-4}. \label{gam}
\end{equation}
We observe that in this application each of these parameters is at least an order of magnitude larger and smaller, respectively, than in the helicopter brownout application, which is a consequence of the lower macroscopic flow velocity scale and the higher density and dynamic viscosity of water over air. 

For these parameter estimates, we then obtain $z_d^c \approx 3.5$, (recall~(\ref{eqn:zdc})) so that the centre of the bathysphere is at a height of $3.5$ times its radius above the ocean bed when a local sand cloud begins to form.  We now present some numerical examples which confirm and illustrate that a local sand cloud forms only for $z_d<z_d^c$, and is absent when $z_d\geq z_d^c$.  For this example, following (\ref{gam}), we set $\gamma=10^{5}$, and then, from (\ref{eqn:zdc}), we have, precisely,
\[ z_d^c = \left(\frac{576}{3125\pi^2}\gamma\right)^{1/6} \approx 3.509232, \]
suggesting that we should expect a sand cloud for $z_d\leq 3.5$, but no sand cloud for $z_d>3.51$.  

\begin{figure}
   \centerline{
   \includegraphics[width=0.6\linewidth]{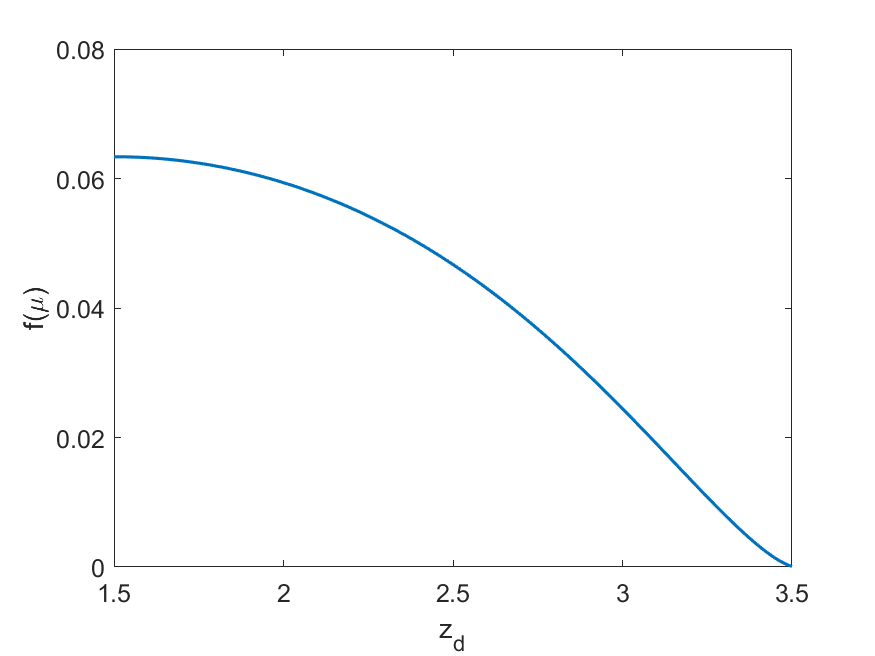}}
   \caption{Plot showing how $f(\mathbf{\mu})$ (given by (\ref{eqn:PC})) varies with $z_d$, for the bathysphere example.}
  \label{fig:bathysphere_PC}
\end{figure}

With the dimensional height of the bathysphere centre above the ocean bed being $lz_d$, where $l$ is the radius of the bathysphere (as described above), we restrict attention to the case $z_d\geq 1.5$ (i.e., the case where the gap distance between the bathysphere and the ocean bed is at least half the radius of the bathysphere), as this seems like a reasonable range for which the current simple fluid flow assumptions should remain reasonable, but below which the fluid dynamics assumptions may lose accuracy.  In figure~\ref{fig:bathysphere_PC} we plot the particle cloud formation potential $\cal{PC}(\chi_{\mu})\equiv$ $ f(\mathbf{\mu})$, given by~(\ref{eqn:PC}), for $\gamma=10^{5}$, $\Omega=0$, $E_s=0.01$ and $z_d\in[1.5,3.51]$.
We see that $f(\mathbf{\mu})$ is indeed zero for $z_d\approx3.51$, and that $f(\mathbf{\mu})$ increases as $z_d$ decreases from 3.5 to 1.5.

\begin{figure}
   \centerline{
   \includegraphics[width=0.5\linewidth]{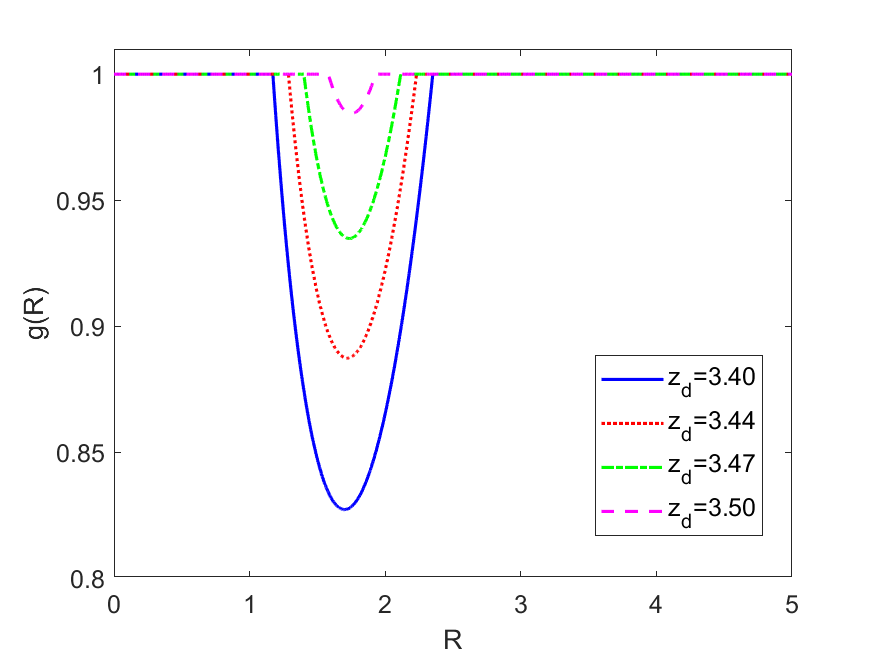}
   \includegraphics[width=0.5\linewidth]{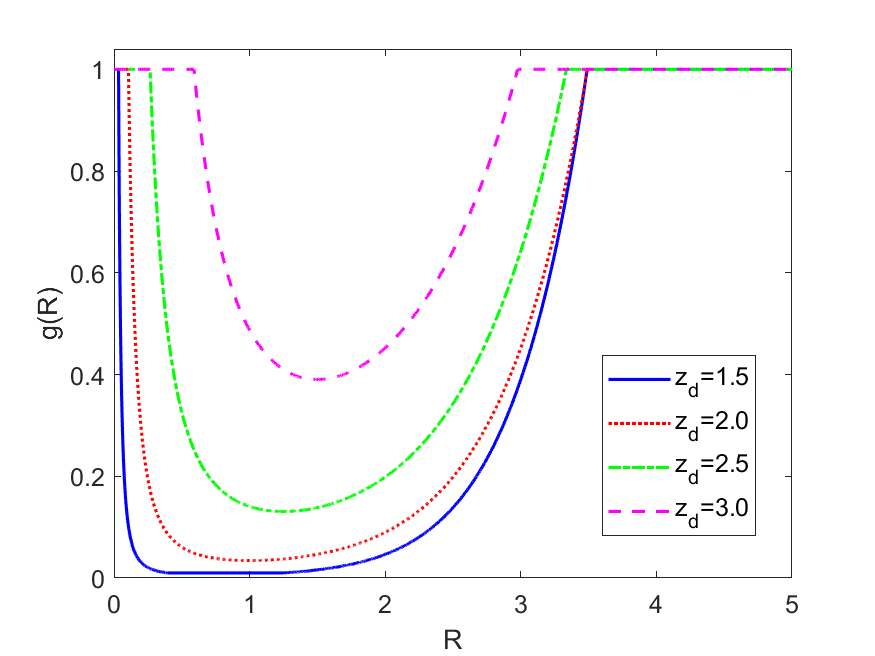}}
   \caption{Boundary data $g(\cdot)$, $z_d\in[3.40,3.50]$ (left) and $z_d\in[1.5,3.0]$ (right)}
  \label{fig:bs_bdy}
\end{figure}

To illustrate this more clearly, in figure~\ref{fig:bs_bdy} we plot the boundary data $g(\cdot)$ (given by (\ref{eqn:h25})), for a small range of $z_d$ close to $z_d^c$ (on the left of figure~\ref{fig:bs_bdy}), and for a wider range of values of $z_d$ (on the right of figure~\ref{fig:bs_bdy}).  Note the differing scales on each plot.  For $z_d\geq3.51$ we simply have $g(R)=1$ for all $R\geq 0$.

We then solve (\ref{eqn:h34})-(\ref{eqn:h19}), with the same parameter choices as those listed above, and with $\alpha=10^{-4}$ and $\Delta=0.1$, using the numerical method described in \citet[Appendix~B]{NL24}, with discretization parameters $\epsilon=10^{-6}$, $L_R=320$, $L_z=256$, and $N=8$ (as described in \S\ref{sec:model}).  The precise number of degrees of freedom that we use in our numerical scheme varies with $z_d$, with more degrees of freedom for greater $z_d$, as in each case we require more degrees of freedom in the mesh below the bathysphere when $z_d$ is larger, and in each case we require the same number of degrees of freedom in our computational mesh above the bathysphere.  Using the discretization parameters listed above, the number of degrees of freedom ranges between 5,575,072 (for $z_d=1.5$) up to 5,985,952 (for $z_d=3.5$), so between approximately 5.5 million and 6.0 million degrees of freedom in our numerical scheme for each example.

\begin{figure}
   \centerline{
   \includegraphics[width=0.6\linewidth]{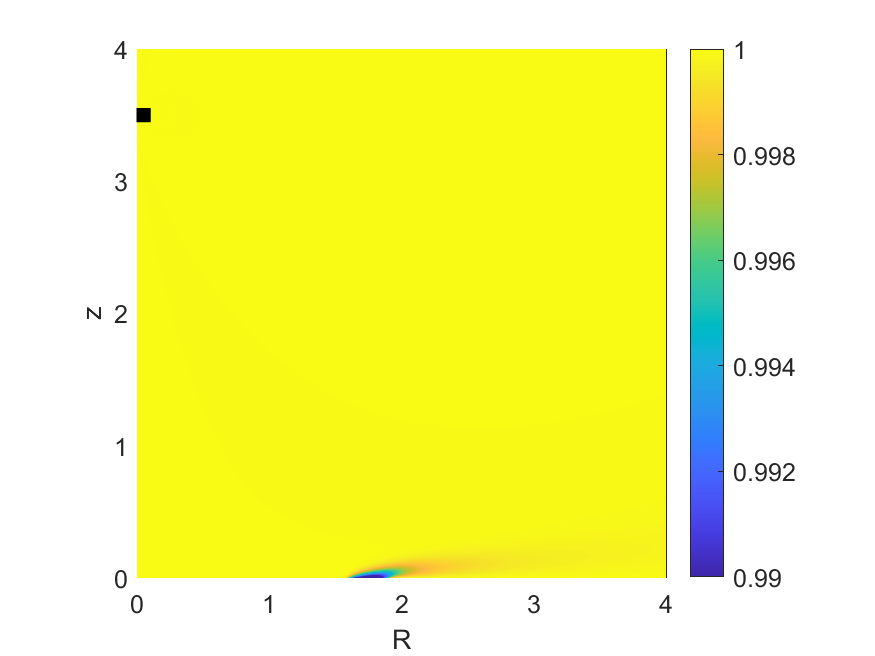}}
   \caption{Voidage field $E(R,z)$, $z_d=3.50$}
  \label{fig:bs_zd_35}
\end{figure}

In figure~\ref{fig:bs_zd_35} we show the voidage field $E(R,z)$ for $z_d=3.50$.  For $z_d=3.51$ (not shown here), we simply have $E(R,z)=1$ for all $(R,z)\in\omega$, but then as $z_d$ decreases we start to see a sand cloud beginning to form.  In order that the formation of the cloud can be seen, the colour scale has been selected such that we have plotted over only a small range of values from $E=0.99$ to $E=1$, with the blue region highlighting all points at which $E\leq 0.99$.  We have also only plotted over a small range of values of $R$ and $z$ - the actual computational domain is much larger (determined by the discretization parameters stated above).

\begin{figure}
   \centerline{
   \includegraphics[width=0.33\linewidth]{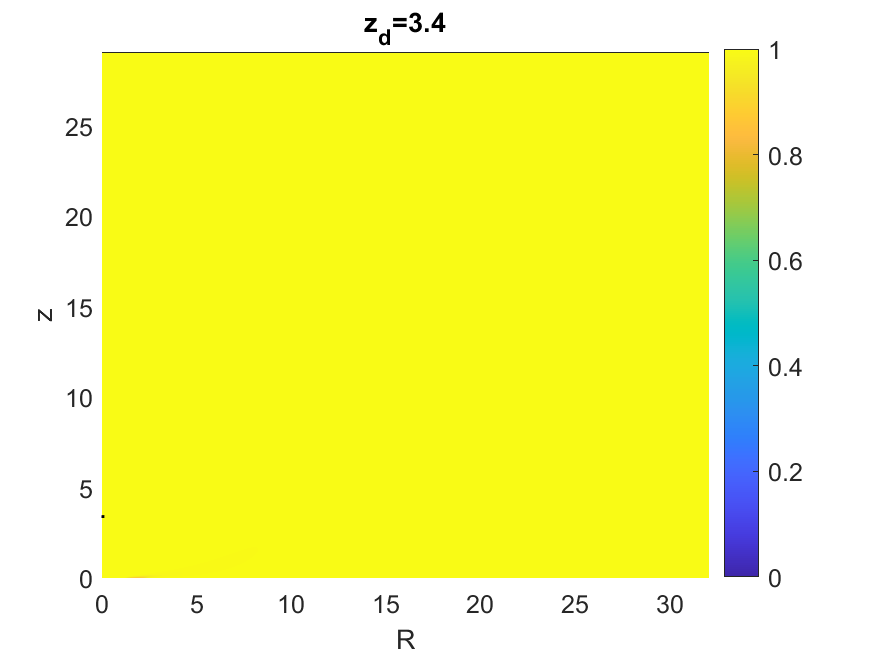}
   \includegraphics[width=0.33\linewidth]{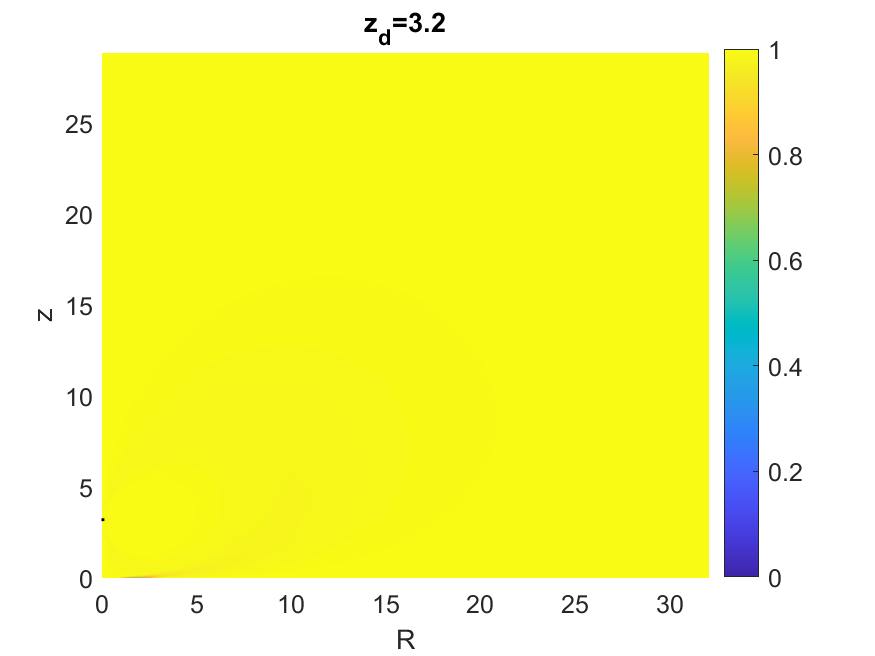}
   \includegraphics[width=0.33\linewidth]{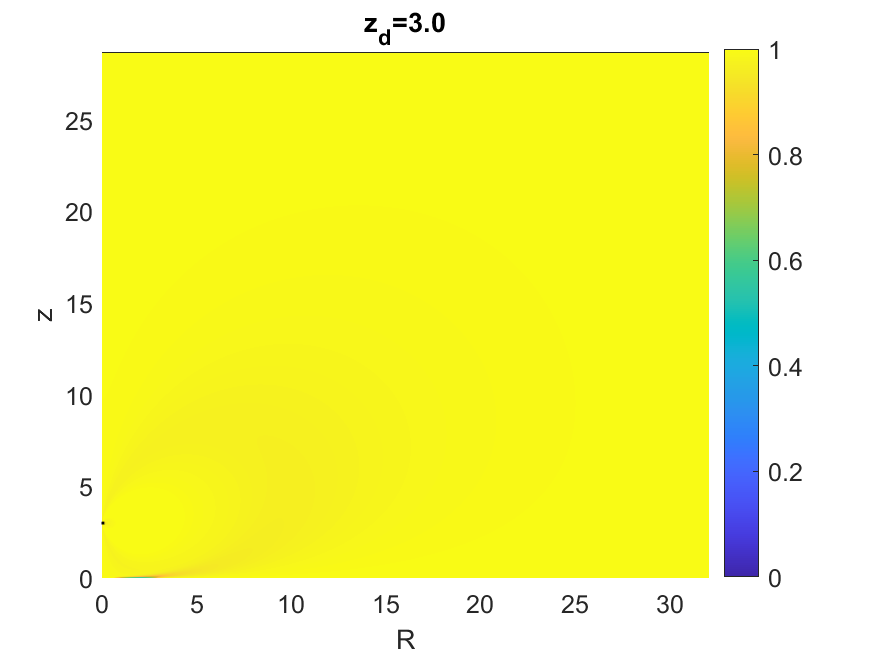}}
   \centerline{
   \includegraphics[width=0.33\linewidth]{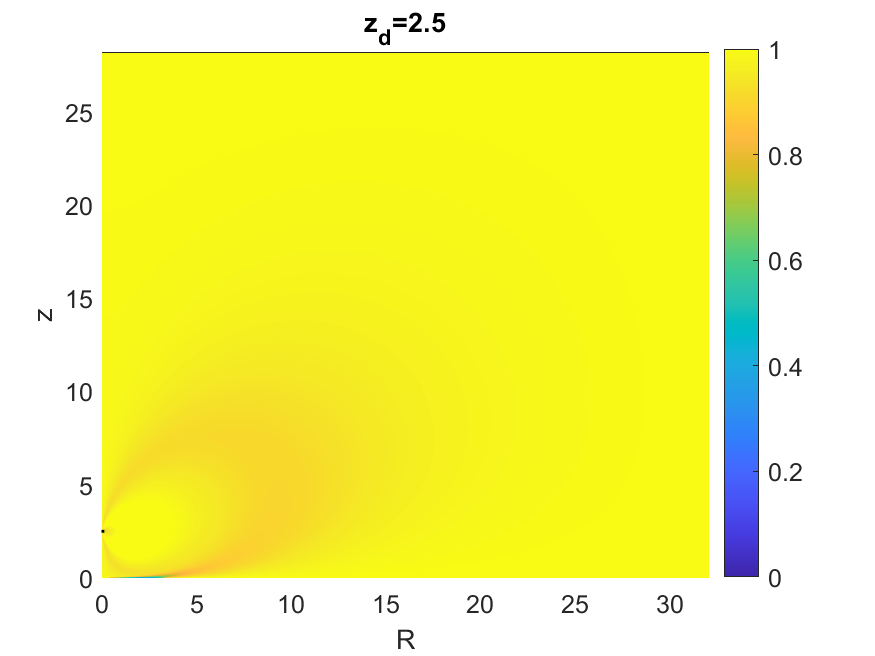}
   \includegraphics[width=0.33\linewidth]{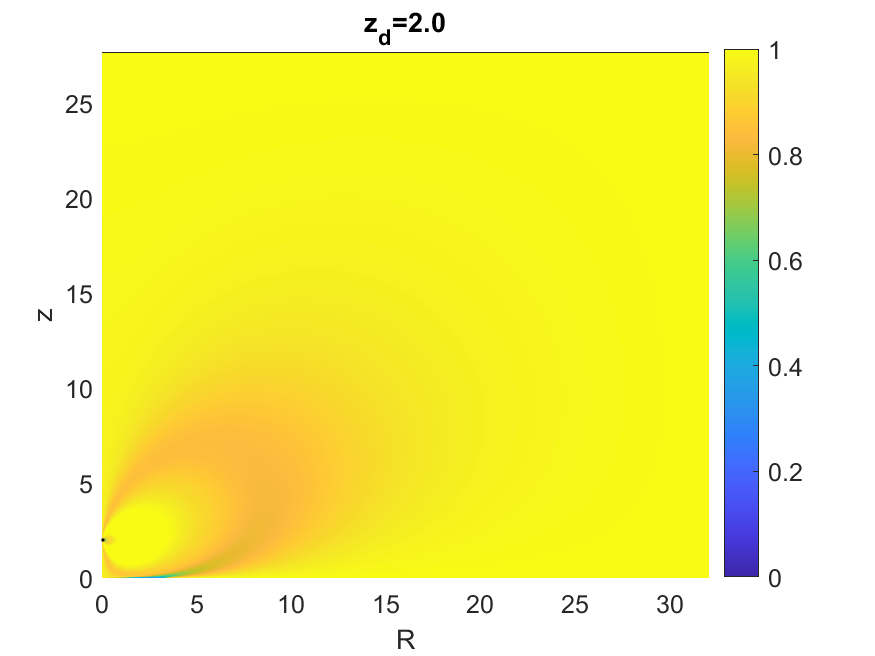}
   \includegraphics[width=0.33\linewidth]{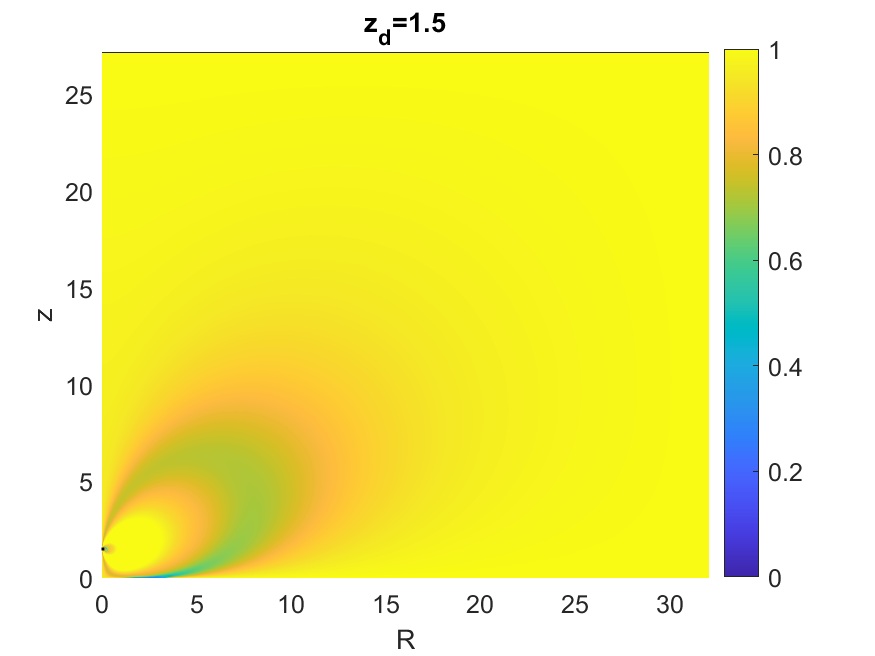}}
   \caption{Bathysphere, voidage field $E(R,z)$, various $z_d$}
   \label{fig:bs_various_zd}
\end{figure}

\begin{figure}
   \centerline{
   \includegraphics[width=0.33\linewidth]{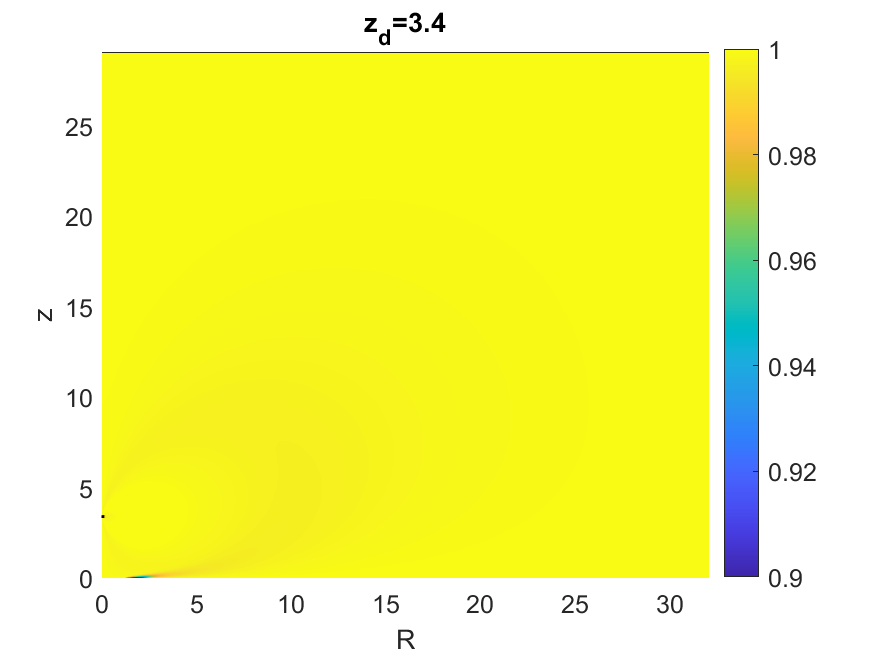}
   \includegraphics[width=0.33\linewidth]{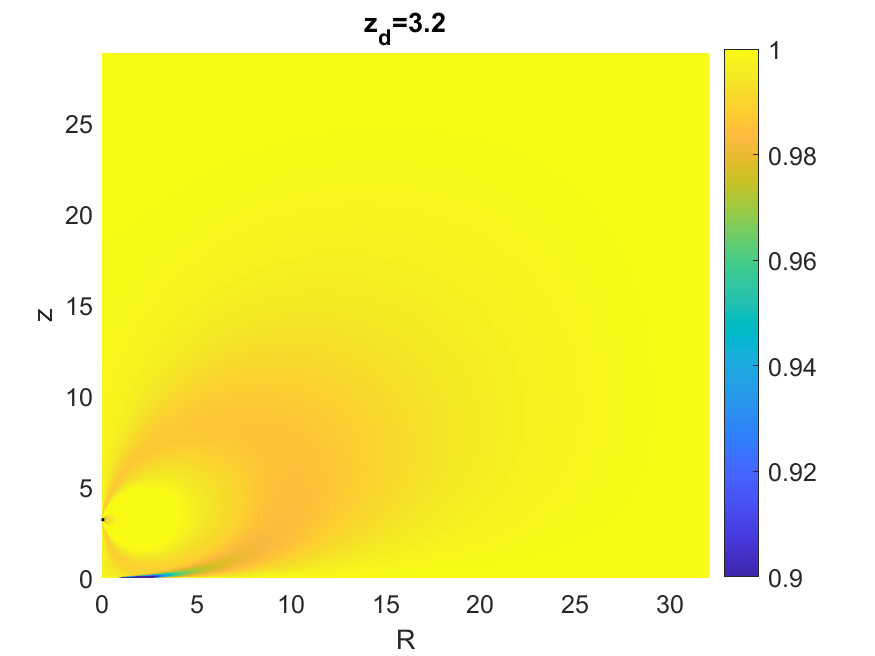}
   \includegraphics[width=0.33\linewidth]{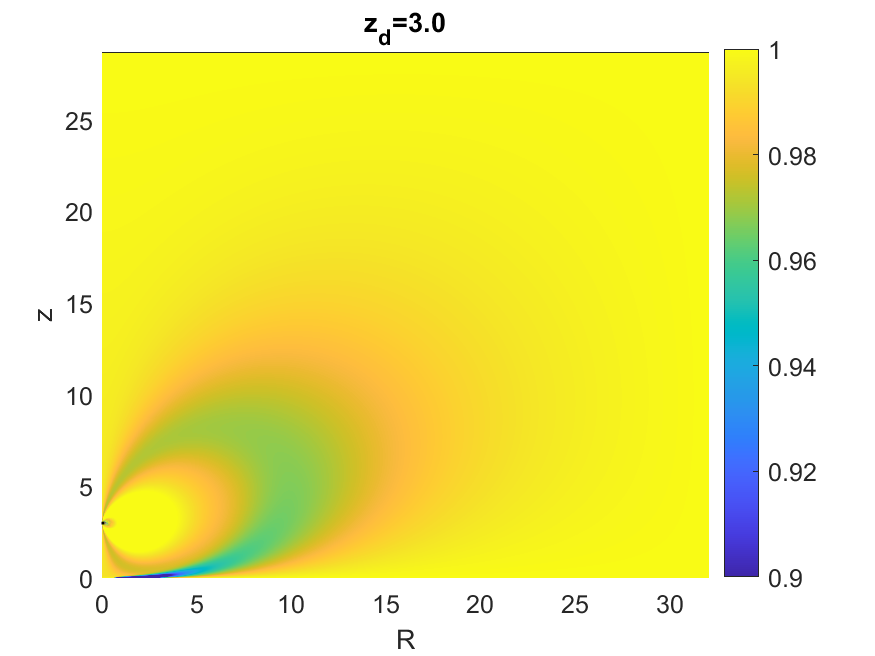}}
   \centerline{
   \includegraphics[width=0.33\linewidth]{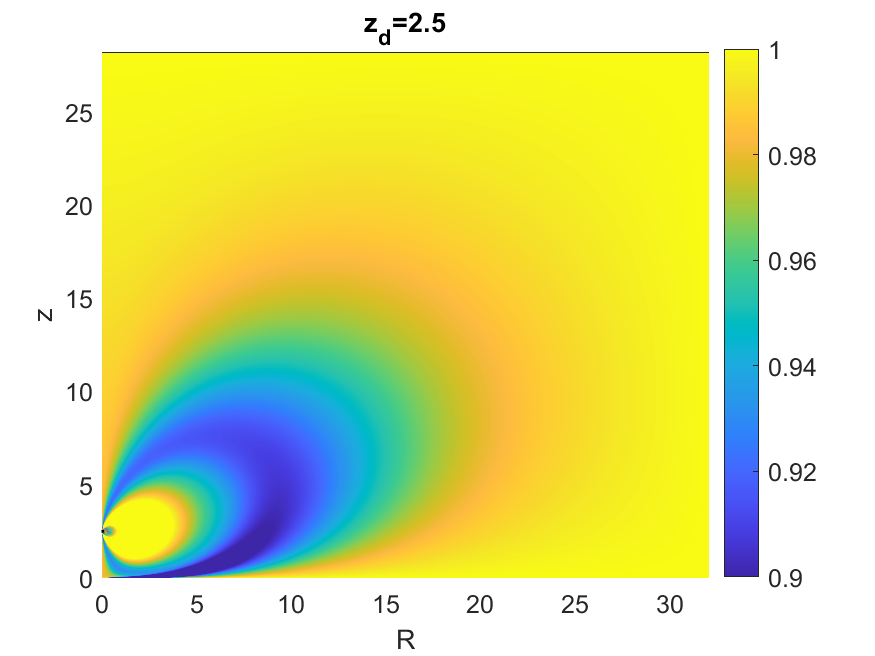}
   \includegraphics[width=0.33\linewidth]{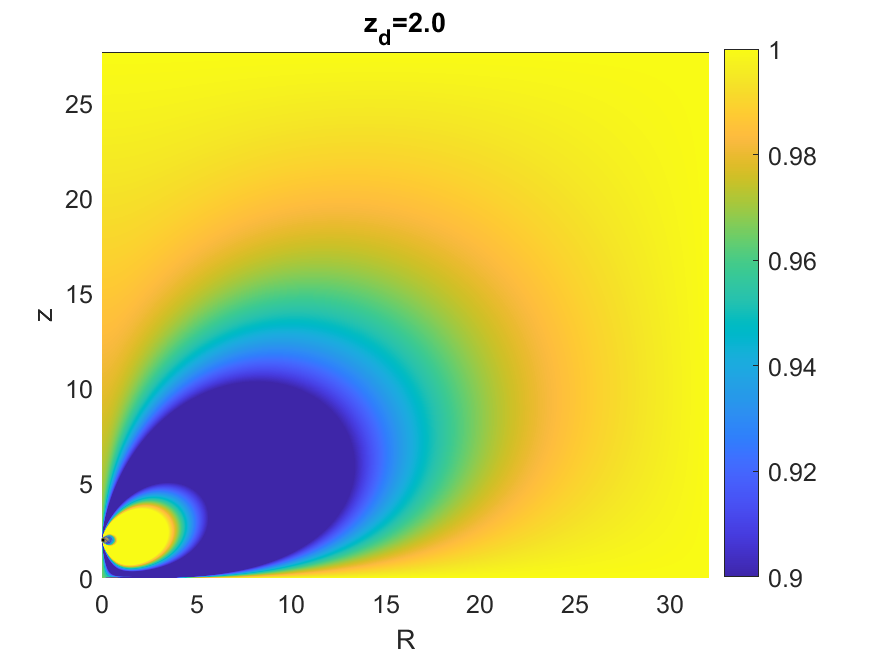}
   \includegraphics[width=0.33\linewidth]{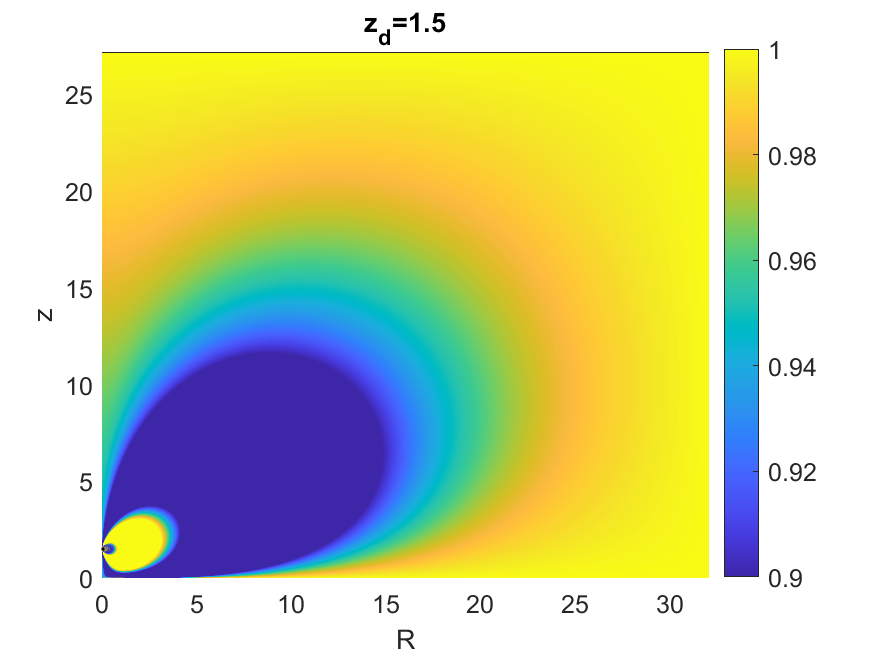}}
   \caption{Bathysphere, voidage field $E(R,z)$, various $z_d$, reduced colour scale}
   \label{fig:bs_various_zd_reduced_clim}
\end{figure}

In figures~\ref{fig:bs_various_zd} and figure~\ref{fig:bs_various_zd_reduced_clim} we then plot the voidage field $E(R,z)$ for $z_d=3.4$, $z_d=3.2$, $z_d=3.0$, $z_d=2.5$, $z_d=2.0$, $z_d=1.5$, using the full colour scale range from zero to 1.0 (recalling that $E\in[E_s,1]$) in figure~\ref{fig:bs_various_zd}, and a reduced colour scale from 0.9 to 1.0 in figure~\ref{fig:bs_various_zd_reduced_clim}, so that the patterns formed by the particle cloud can be more clearly seen, particularly for $z_d=3.4$ and $z_d=3.2$.  As $z_d$ decreases we start to see a sand cloud beginning to form.  Note that the blue region in the latter plot (figure~\ref{fig:bs_various_zd_reduced_clim}) highlights all points at which $E\leq 0.9$.  These plots show more clearly the formation of a sand cloud when $z_d\leq 3.5$, with an increasing amount of sand in the water as $z_d$ decreases.

The supplementary file Langdon-movie1-bathysphere.mp4 shows in a video format how the voidage field changes as $z_d$ is reduced gradually from $z_d=3.5$ to $z_d=1.5$, using the same parameter choices, discretization parameters and colour scale as for the results shown in figure~\ref{fig:bs_various_zd}. 

\subsection{Lunar landing plume model}
\label{subsec:lunar}
In this application we consider the situation in which a lander descends towards a celestial body, such as the Moon or Mars, with its speed of approach regulated by the expulsion of a downward high speed fluid jet that may displace dust, gravel and rocks on the surface during landing.  This problem has received much attention in the recent literature from each of an experimental (see, for example, \cite{WAC20,CJC25}), modelling (see, for example, \citet{WAC20,JC22,ZCCW23,JKJKBJL25}) and CFD (see, for example, \citet{WAC20,ZCCW23,JKJKBJL25}) view point. However, much of this work focuses upon the fine details of the exhaust flow, and relies purely on phenomenological mechanisms for the key aspect in cloud formation associated with the local entrainment and detrainment of particles at the interface between the bulk fluidized region and the otherwise static particle bed. From our point of view we again use the much simplified model of the primary fluid flow characteristics but now coupled with the key interfacial boundary condition which gives a rational representation of the entrainment/detrainment dynamics. In this context, the lunar landing plume can be easily modelled (albeit in a very simplistic and basic form, but one which captures the principal features) within our framework through an appropriate choice of the relevant dimensionless parameters, and the model can be readily used both to determine the voidage field in the fluidized region, and in addition the local lunar surface deformation, following similar calculations for the helicopter brownout problem presented recently in \citet{LN25}. We begin by estimating the magnitudes of the key model parameters $\gamma$ and $\alpha$. In this application the particulate matter is lunar surface dust whilst the fluid is principally the exhaust fluid from the jet landing thrusters. Lunar dust particles have typical radius $a\sim 3\times 10^{-6}~m$ and material density $\rho_s\sim 2700~kgm^{-3}$, whilst typical measurements for the jet exhaust fluid give fluid density $\rho_f \sim 1~kgm^{-3}$ and dynamic viscosity $\mu = 2 \times 10^{-5}~pas$, with typical jet exhaust velocity estimated as $u_s \sim 3 \times 10^2~ms^{-1}$. In addition, the lunar gravitational acceleration is approximated as $1~ms^{-2}$, and we tentatively take the particle collisional pressure coefficient as $p_0 \sim 10^{-2}~pa$. The typical flow length scale $l$ should be taken as that associated with the dimension of the landing vehicle.  The key dimensionless parameters in the model are then computed (using the definitions in \citet{NL24}) from these estimates as having the orders of magnitude
\begin{equation}
    \gamma\sim O(10^6-10^{8})~~\text{and}~~\alpha\sim O(10^{-6}-10^{-10}), \label{gam2}
\end{equation}
and, as may be anticipated, we observe that $\gamma$ and $\alpha$ are, respectively, very much larger and very much smaller than their corresponding values for the descending bathysphere application.

Setting $\gamma=10^{8}$, we obtain (from~(\ref{eqn:zdc}))
\[ z_d^c = \left(\frac{576}{3125\pi^2}\gamma\right)^{1/6} \approx 11.097167. \]

\begin{figure}
   \centerline{
   \includegraphics[width=0.6\linewidth]{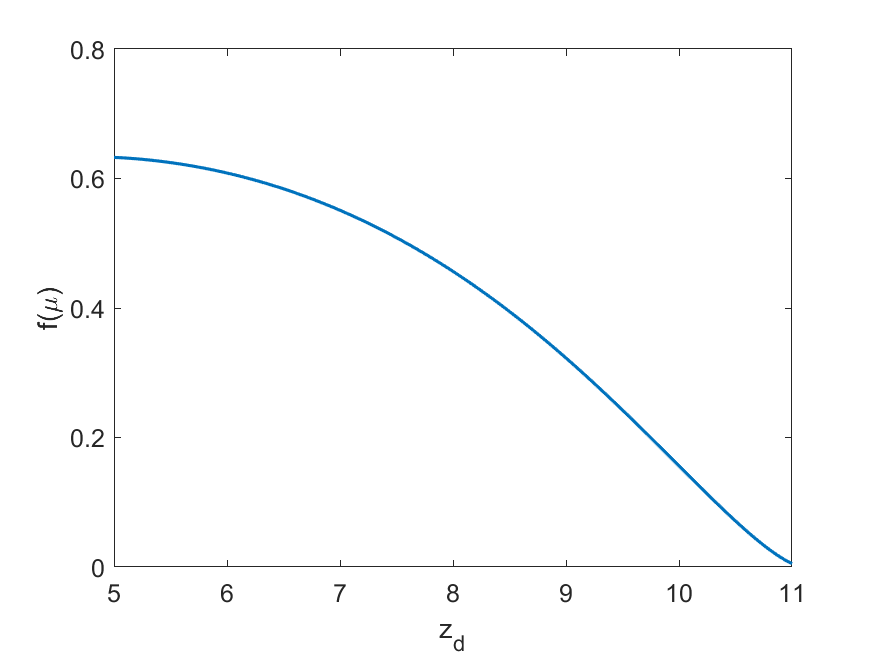}}
   \caption{Plot showing how $f(\mathbf{\mu})$ (given by (\ref{eqn:PC})) varies with $z_d$, for the lunar plume example.}
  \label{fig:lunar_PC}
\end{figure}

In figure~\ref{fig:lunar_PC} we plot the particle cloud formation potential $\cal{PC}(\chi_{\mu})\equiv$ $ f(\mathbf{\mu})$, given by~(\ref{eqn:PC}), for $\gamma=10^{8}$, $\Omega=0$, $E_s=0.01$ and $z_d\in[5,11]$.
We see that $f(\mathbf{\mu})$ is indeed zero for $z_d\approx11$, and that $f(\mathbf{\mu})$ increases as $z_d$ decreases from 11 to 5. Comparing the vertical scale in figure~\ref{fig:lunar_PC} with that in figure~\ref{fig:bathysphere_PC}, we also note the significantly increased particle cloud formation potential of the lunar plume example, reflecting the significantly larger value of the parameter $\gamma$ in this example.

We now present some numerical examples which confirm and illustrate that a dust cloud forms only for $z_d<z_d^c$, and is absent when $z_d\geq z_d^c$.  We solve (\ref{eqn:h34})-(\ref{eqn:h19}), with the same parameter choices as those listed above, and with $\alpha=10^{-6}$ and $\Delta=0.1$, using the numerical method described in \citet[Appendix~B]{NL24}, with discretization parameters $\epsilon=10^{-6}$, $L_R=800$, $L_z=440$, and $N=4$ (as described in \S\ref{sec:model}).  Comparing these values to those used for the bathysphere example, we note that the increased cloud formation potential for the lunar plume leads to the need for a much larger computational domain.  The precise number of degrees of freedom that we use in our numerical scheme varies with $z_d$, with more degrees of freedom for greater $z_d$, as in each case we require more degrees of freedom in the mesh below the lunar lander when $z_d$ is larger, and in each case we require the same number of degrees of freedom in our computational mesh above the lunar lander.  Using the discretization parameters listed above, the number of degrees of freedom ranges between 6,283,032 (for $z_d=5.0$) up to 6,923,832 (for $z_d=10.0$), so between six and seven million degrees of freedom in our numerical scheme for each example.

\begin{figure}
   \centerline{
   \includegraphics[width=0.33\linewidth]{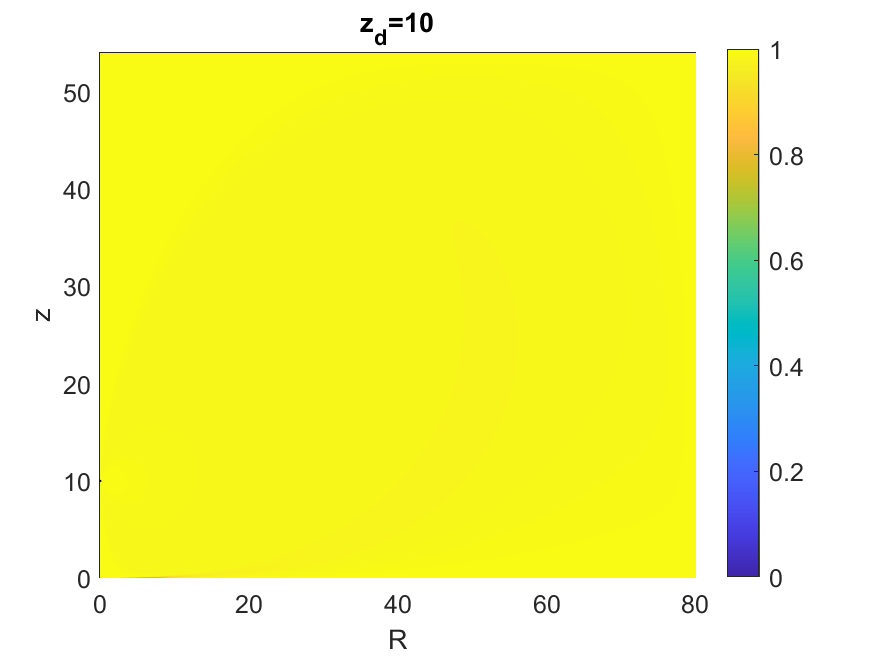}
   \includegraphics[width=0.33\linewidth]{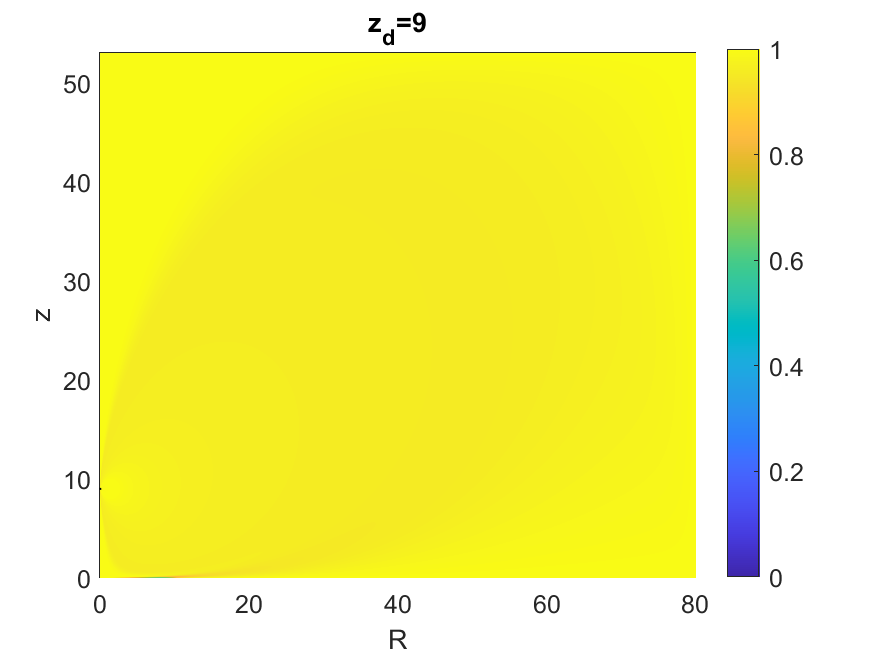}
   \includegraphics[width=0.33\linewidth]{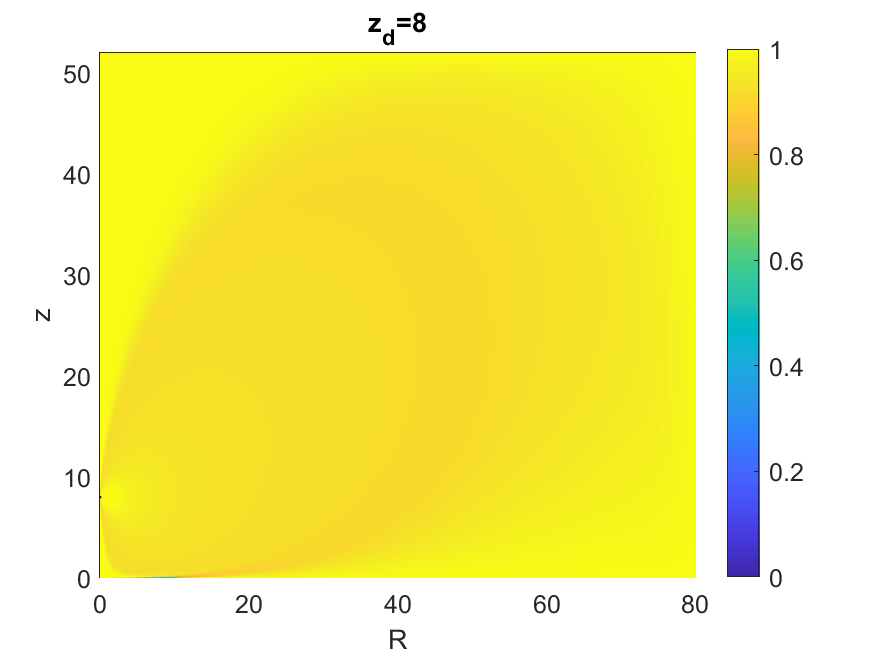}}
   \centerline{
   \includegraphics[width=0.33\linewidth]{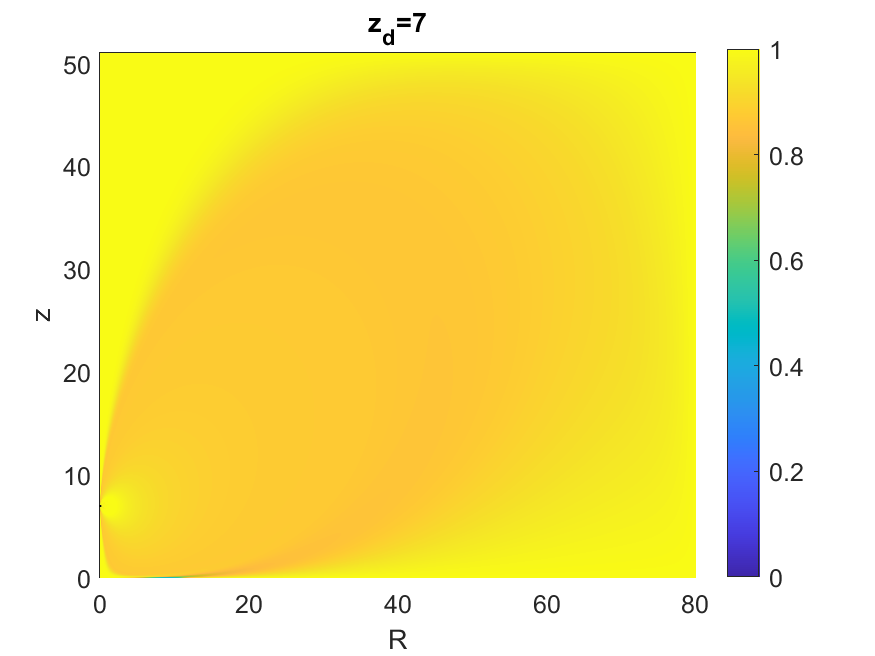}
   \includegraphics[width=0.33\linewidth]{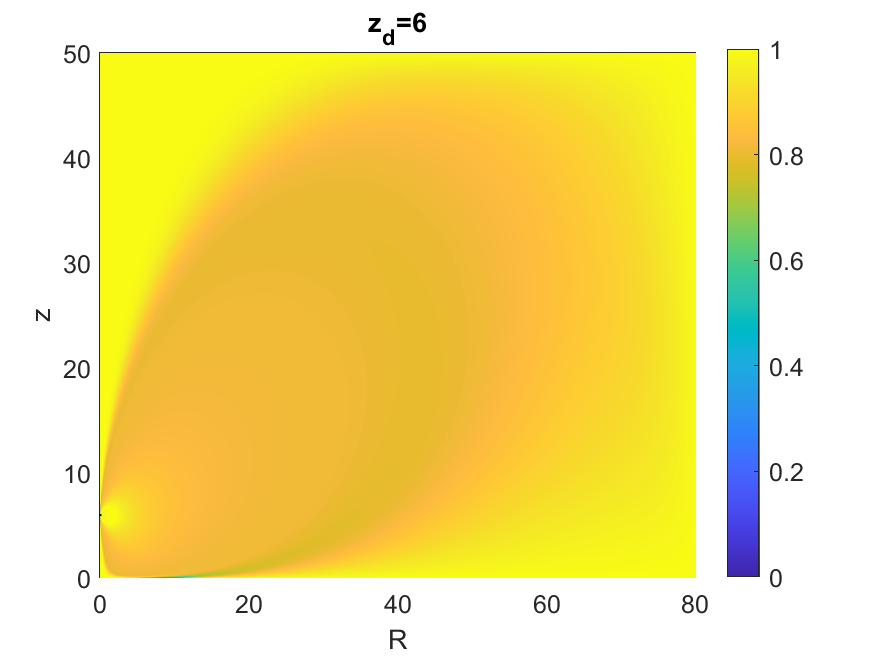}
   \includegraphics[width=0.33\linewidth]{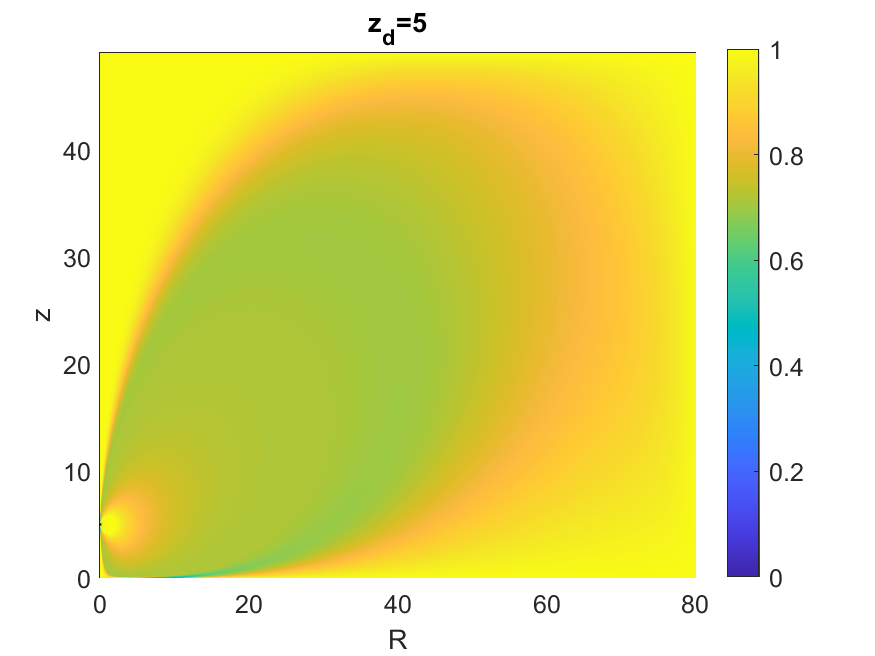}}
   \caption{Lunar landing plume, voidage field $E(R,z)$, $z_d=10, 9, 8, 7, 6, 5$}
  \label{fig:lunar}
\end{figure}

\begin{figure}
   \centerline{
   \includegraphics[width=0.33\linewidth]{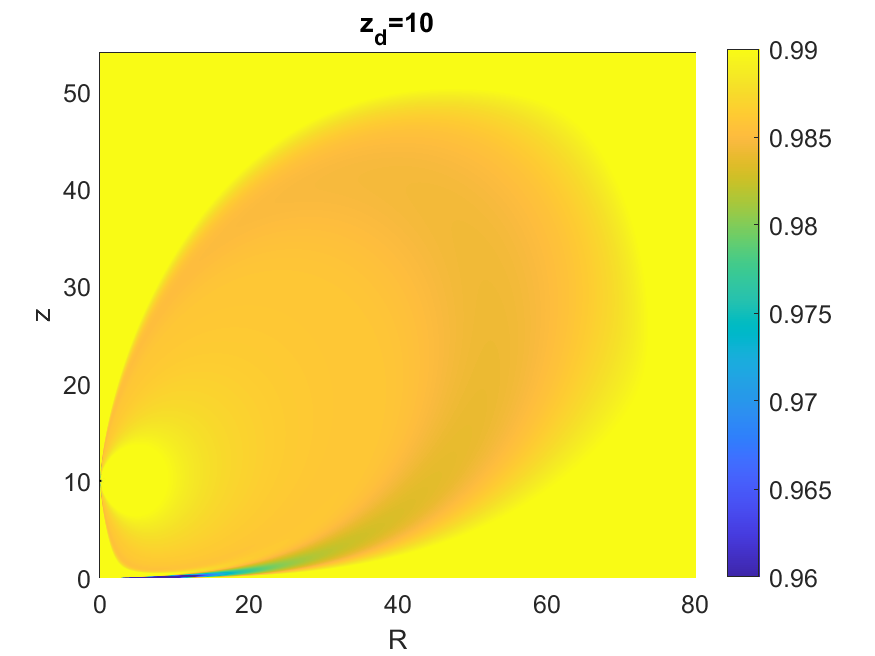}
   \includegraphics[width=0.33\linewidth]{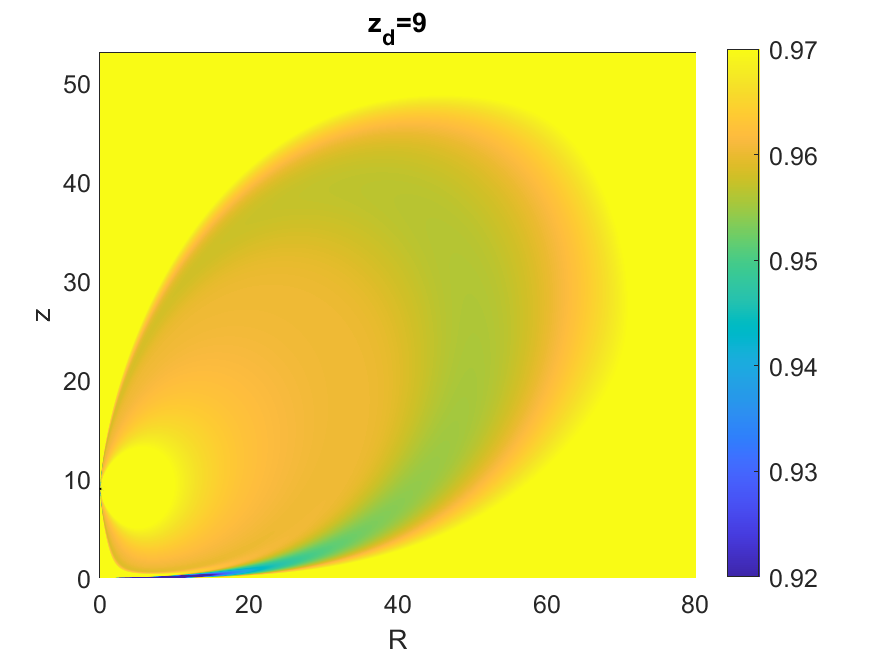}
   \includegraphics[width=0.33\linewidth]{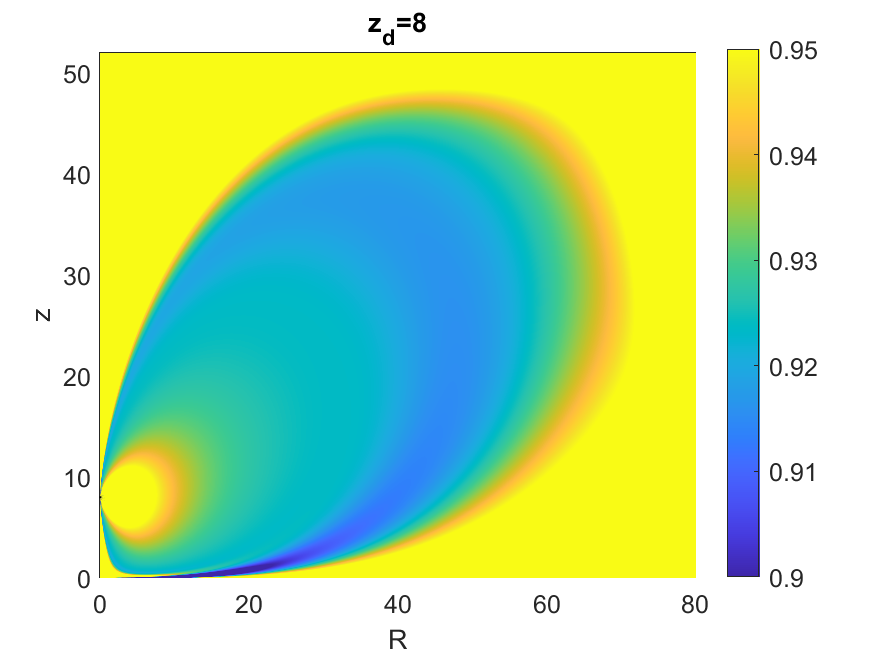}}
   \centerline{
   \includegraphics[width=0.33\linewidth]{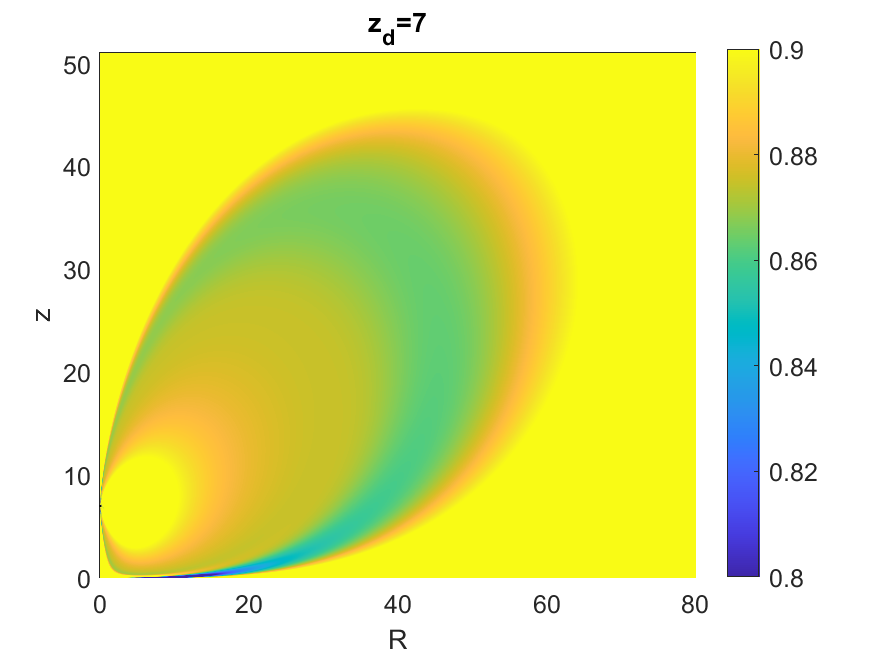}
   \includegraphics[width=0.33\linewidth]{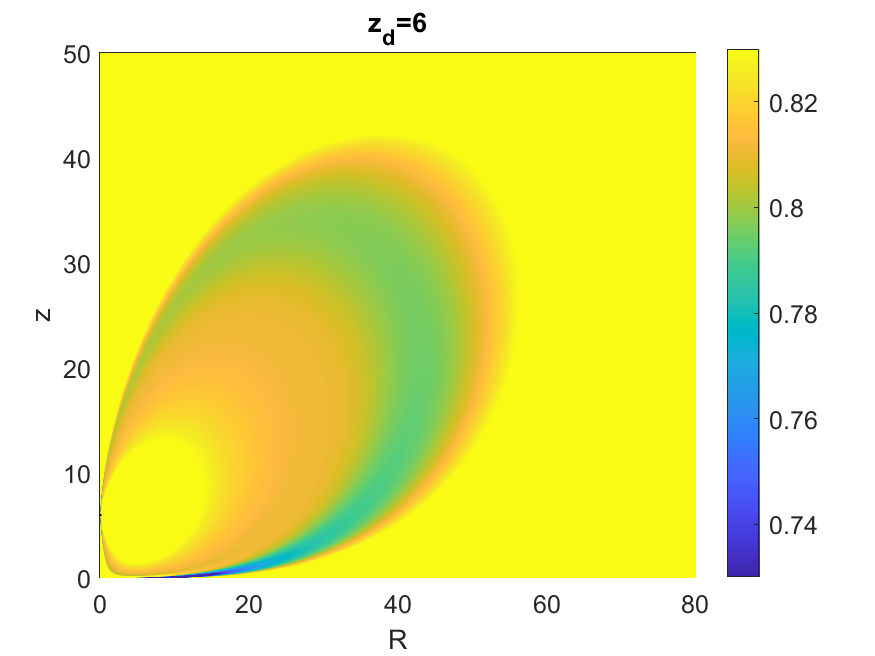}
   \includegraphics[width=0.33\linewidth]{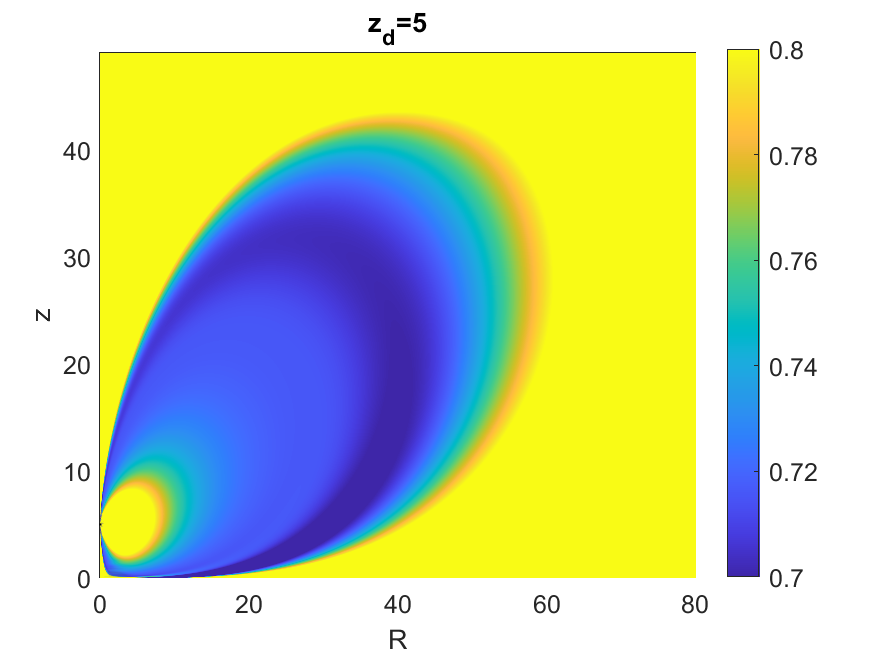}}
   \caption{Lunar landing plume, voidage field $E(R,z)$, $z_d=10, 9, 8, 7, 6, 5$, reduced colour scale.}
  \label{fig:lunar_zoomed}
\end{figure}

In figure~\ref{fig:lunar} we show the voidage field $E(R,z)$ for $z_d=10, 9, 8, 7, 6, 5$.  As $z_d$ decreases, we see the dust cloud grow.
In figure~\ref{fig:lunar_zoomed} we show the voidage field $E(R,z)$ for the same values of $z_d$ as in figure~\ref{fig:lunar}, but with a smaller colour scale range (a different scale for each value of $z_d$), with in each case the blue region representing values of $E$ below the lower end of the plotted range, and the yellow region representing values of $E$ above the upper end of the plotted range.  These plots show more clearly the formation of a dust cloud when $z_d$ is larger, with more dust being raised as $z_d$ decreases, and also illustrate how the qualitative structure of the plume is broadly comparable for each value of~$z_d$.

The supplementary file Langdon-movie2-lunar.mp4 shows in a video format how the voidage field changes as $z_d$ is reduced gradually from $z_d=11.1$ to $z_d=5$, using the same parameter choices, discretization parameters and colour scale as for the results shown in figure~\ref{fig:lunar}. 

\section{Applications requiring the inclusion of gravity}
\label{sec:gravity}
We now show how the mathematical model derived in~\citet{NL24} can be adapted to include gravity.  This may be important in applications where the downwash and swirling flow is less powerful than for a helicopter, such as the case in which the flow field is generated by a much smaller drone.

Following \citet[\S5]{NL24}, the principal PDE governing the structure of the voidage field, now with the inclusion of gravity effects acting on the particle phase flow field, becomes
\begin{equation}
 (\nabla \bar{\phi} + \Gamma E Q'(E)\mathbf{k}) \cdot \nabla E - \alpha E \nabla^2(H(E)) = 0, \quad \br\in D_c, \label{eqn5.6}   
\end{equation}
(compare to \citet[(6.15)]{NL24}), where we recall that $\mathbf{k}$ is the unit vector pointing vertically upward, $D_c$ is the region occupied by the particle cloud and the potential $\bar{\phi}$ is as in \citet{NL24} (equation (6.14) with (6.5)), whilst,
\begin{equation}
     H(E) = \frac{2}{9} (E-E_s).
     \label{eqn:H}
\end{equation}
In deriving this equation, we have followed precisely the same derivation as in \citet[\S5]{NL24}, except that when beginning the derivation (at equations (5.5) and (5.6) in~\citet{NL24}), we have included at the leading order of approximation, in addition to all previously included terms, the term which represents the effect of gravity in the particle phase momentum equation and is premultiplied by the parameter $\rho G$. Specifically, the term $-\rho G(1-E)\bk$ is carried forward from~\citet[(5.6)]{NL24} to~\citet[(5.15)]{NL24}, which becomes
\[ 0 = \alpha\nabla E + (1-E)\bar{\beta}(E)(\bu-\bv) -\rho G (1-E)\bk, \]
so~\citet[(5.16)]{NL24} becomes
\[ \nabla(p-\alpha E) + \rho G (1-E)\bk = 0. \]
This new term introduces the dimensionless parameter
\begin{equation}
    \Gamma = \frac{\rho_sga^2}{\mu u_s}, \label{G}
\end{equation}
which measures the ratio of the magnitude of the gravitational force to the magnitude of the drag force on the particle phase (with $a,~\rho_s,~\mu,~u_s$ as introduced in \cite{NL24}), and
\begin{equation}
    Q(E) \equiv (1-E) H'(E) = \frac{2}{9} (1 - E).
    \label{eqn:Q}
\end{equation}
The precise estimate for $\Gamma$ will vary according to the nature of each application. We see that the inclusion of the gravity term has generated a nonlinear convective term in the vertical direction, accounting for particle cloud settling under gravity. The elliptic boundary value problem describing the voidage field in the particle cloud region is as in (\ref{eqn:h34})-(\ref{eqn:h19}), except that now the coefficient $b(R,z)$ is augmented by the addition of the term $-(2/9) \Gamma E$,  and hence~(\ref{eqn:h34}) is replaced by
\begin{equation}
  E\left(E_{RR} + \frac{1}{R} E_R + \frac{\Gamma}{\alpha}E_z + E_{zz} \right) - \bar{a}(R,z) E_R - \bar{b}(R,z) E_z = 0, \quad (R,z)\in \omega.
  \label{eqn:h34new}
\end{equation}
In addition to this change in the field equation, there is also a consequent change to the boundary condition at both the upper and lower surfaces of the cylindrical shell encasing the dipole, and this is readily determined by following the development of this boundary condition in \citet[\S5,6]{NL24}; in the present application, this leads us to modify boundary conditions (\ref{eqn:h37}) to
\begin{equation}
    E_z = \frac{\Gamma}{\alpha}(1-E)\mbox{ on } z= z_d^+,z_d^-, \quad R\in(0,\Delta). \label{eqn:h37new}
\end{equation}

We now show how, through an appropriate choice of parameters, we can model two relevant applications: the problem in which a small drone hovers over a sand bed on Earth and the problem in which a small drone, such as the NASA Ingenuity Mars helicopter, is hovering over a particle bed on Mars.

\subsection{Terrestrial drone cloud model}
\label{sec:terrestrial}
This application considers the situation in which a small-scale drone hovers above a horizontal sand bed. As such, the modelling situation is the same as in \citet{NL24} for the helicopter brownout application, except that there is now a significant reduction in the macroscopic length scale (from helicopter blade radius to drone blade radius) and in the macroscopic fluid velocity scale.  The key change is that the parameter $\rho G$ introduced in \citet{NL24} (which measures the relative strength of the gravity term to the drag term in the particle phase momentum equation) is significantly increased (now $10^{-1}-10^{-2}$), which leads us to include this gravity term in the principal model, as described above.  In addition to the inclusion of this new term, we also observe that $\alpha$ (now $10^{-1}-10^0$) increases compared to the parameter values estimated for the helicopter brownout application, while $\Omega$ and $\gamma$ remain broadly similar, although typically smaller in general.  From~(\ref{G}) we estimate that $\Gamma$ is of $O(10^{-1})$.  In our numerical examples below, we thus take the following parameter choices:
\begin{equation}
    \Gamma=0.1, \quad \gamma=500, \quad \alpha =0.1, \quad \Omega=0.5, \quad E_s=0.01, \quad \Delta=0.04.
    \label{eqn:terrestrial_param}
\end{equation}

\begin{figure}
   \centerline{
   \includegraphics[width=0.6\linewidth]{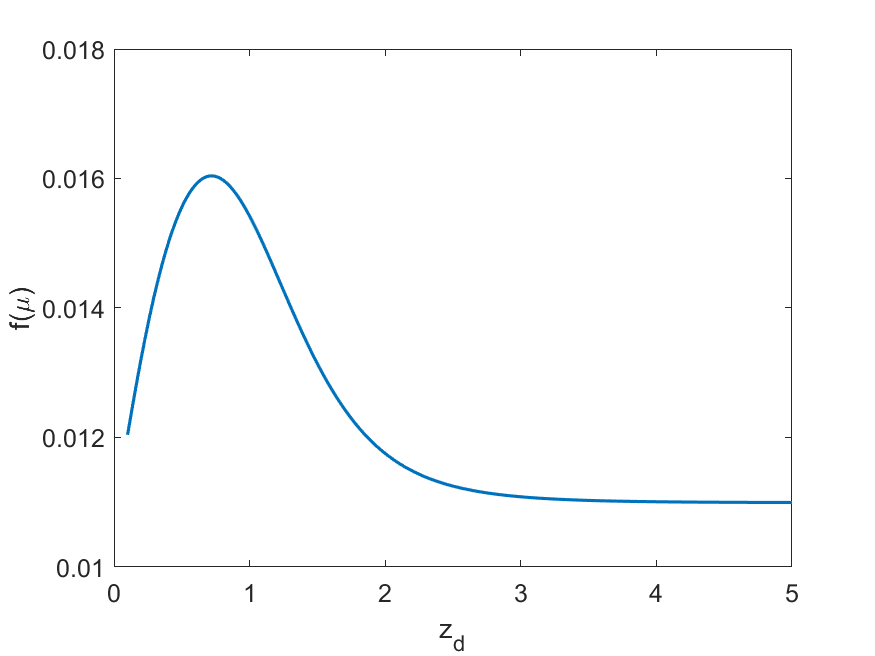}}
   \caption{Plot showing how $f(\mathbf{\mu})$ (given by (\ref{eqn:PC})) varies with $z_d$, for the terrestrial drone example.}
  \label{fig:terrestrial_PC}
\end{figure}

In figure~\ref{fig:terrestrial_PC} we plot the particle cloud formation potential $\cal{PC}(\chi_{\mu})\equiv$ $ f(\mathbf{\mu})$, given by~(\ref{eqn:PC}), for $\gamma=500$, $\Omega=0.5$, $E_s=0.01$ and $z_d\in[0.1,5]$.
As for the helicopter cloud example (see the central plot of figure~\ref{fig:vary_zd}) we see that $f(\mathbf{\mu})$ increases as $z_d$ decreases from 5 to approximately $z_d=1$, before then decreasing as $z_d$ decreases further.

\begin{figure}
   \centerline{
   \includegraphics[width=0.33\linewidth]{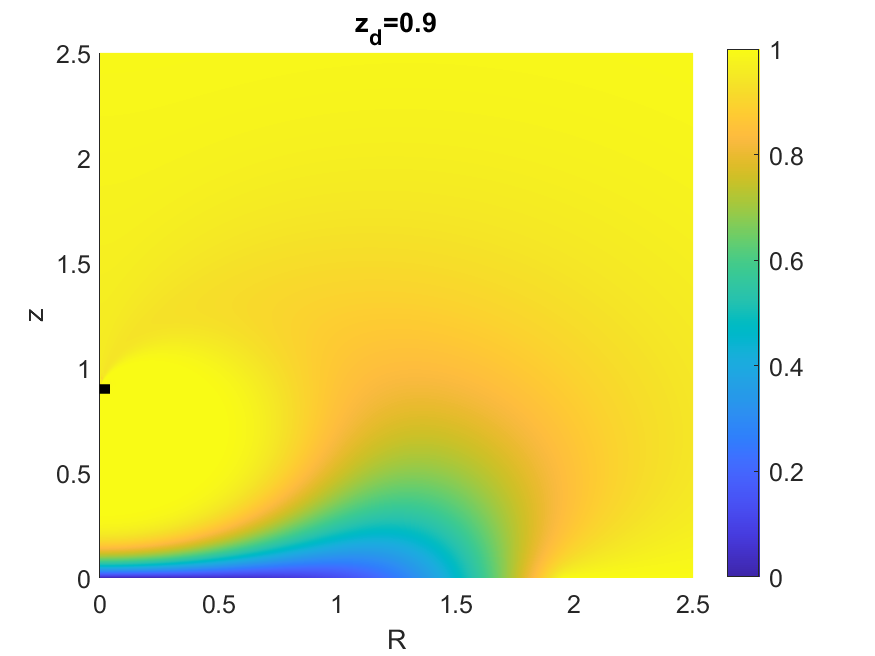}
   \includegraphics[width=0.33\linewidth]{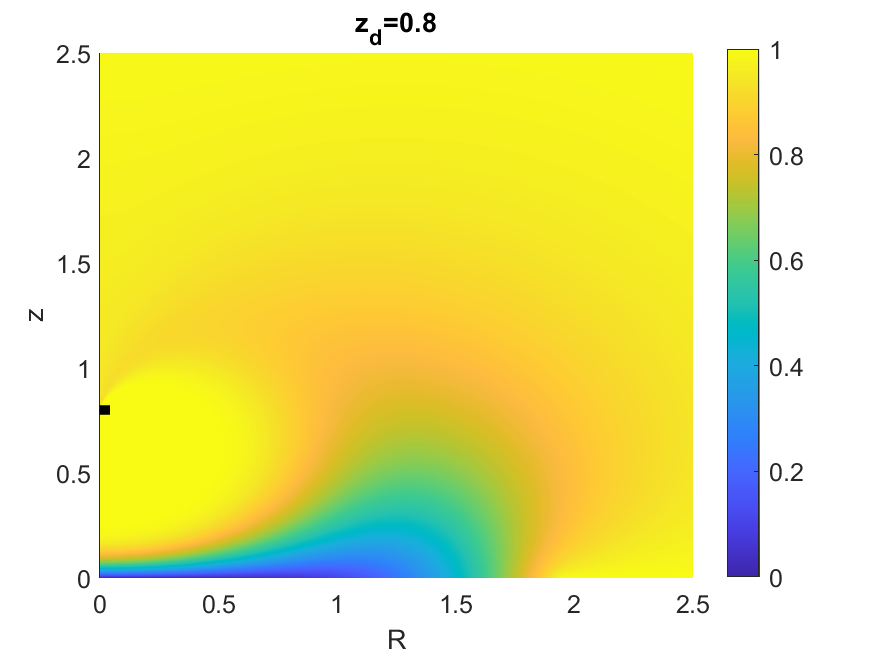}
   \includegraphics[width=0.33\linewidth]{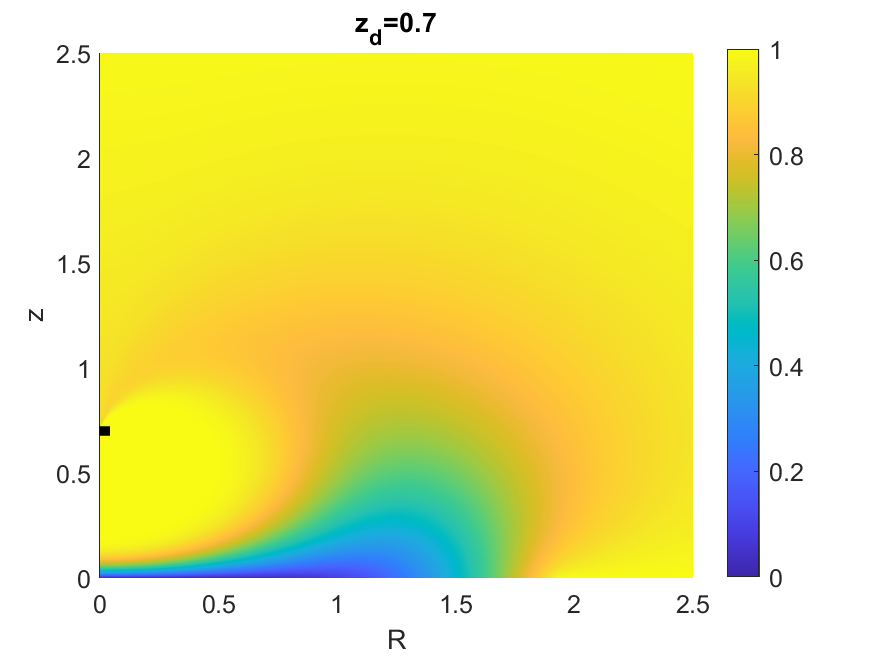}}
   \centerline{
   \includegraphics[width=0.33\linewidth]{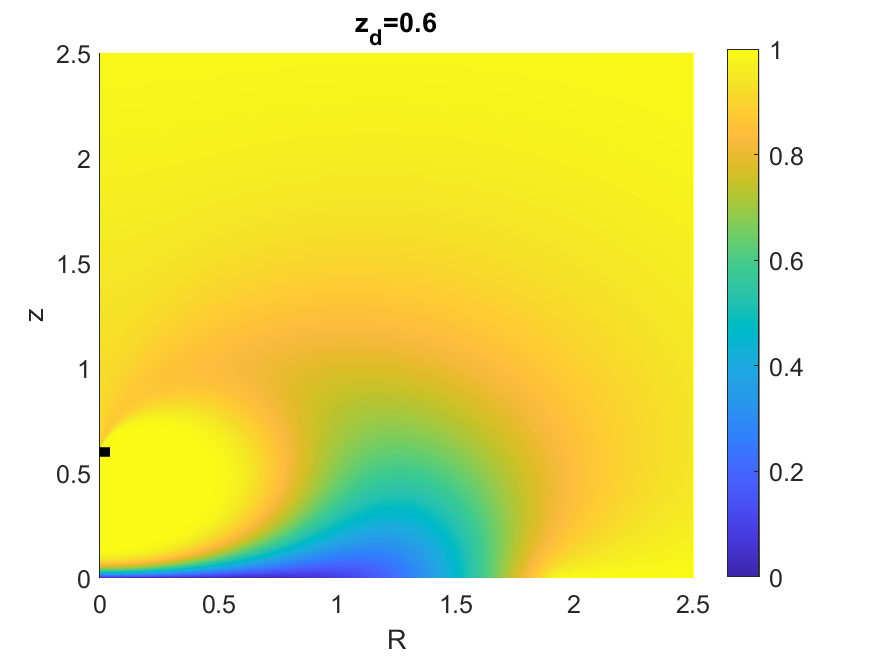}
   \includegraphics[width=0.33\linewidth]{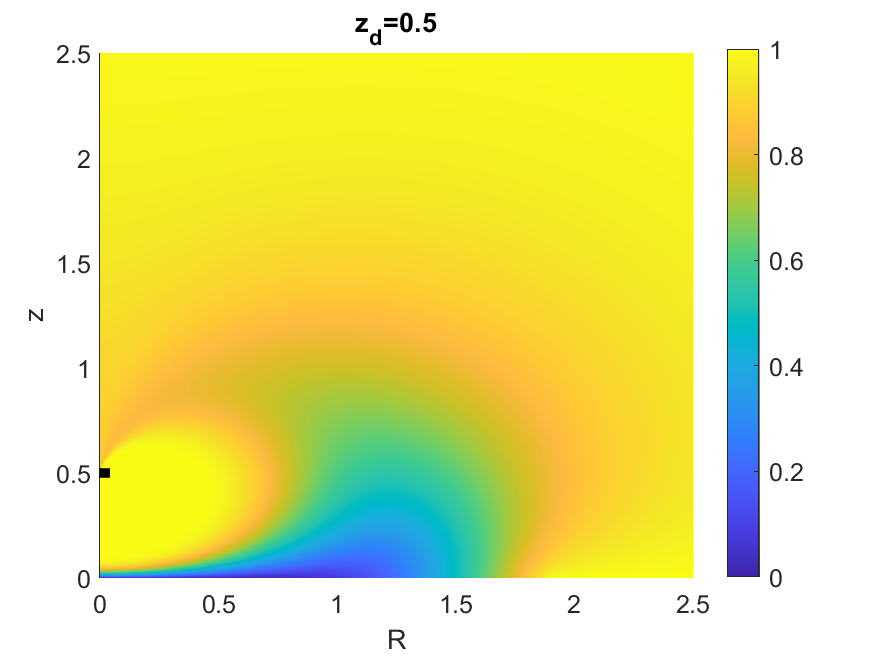}
   \includegraphics[width=0.33\linewidth]{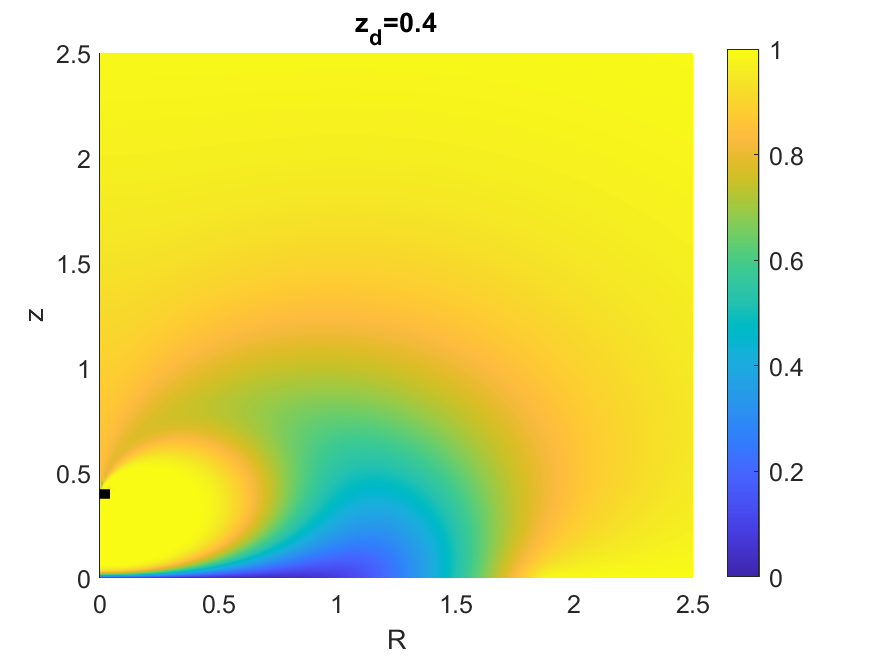}}
   \centerline{
   \includegraphics[width=0.33\linewidth]{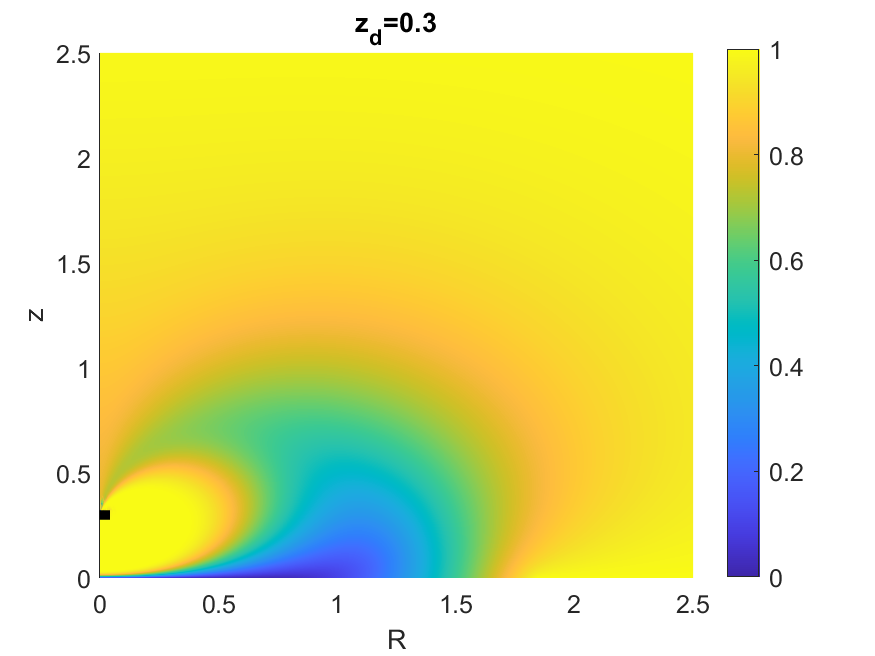}
   \includegraphics[width=0.33\linewidth]{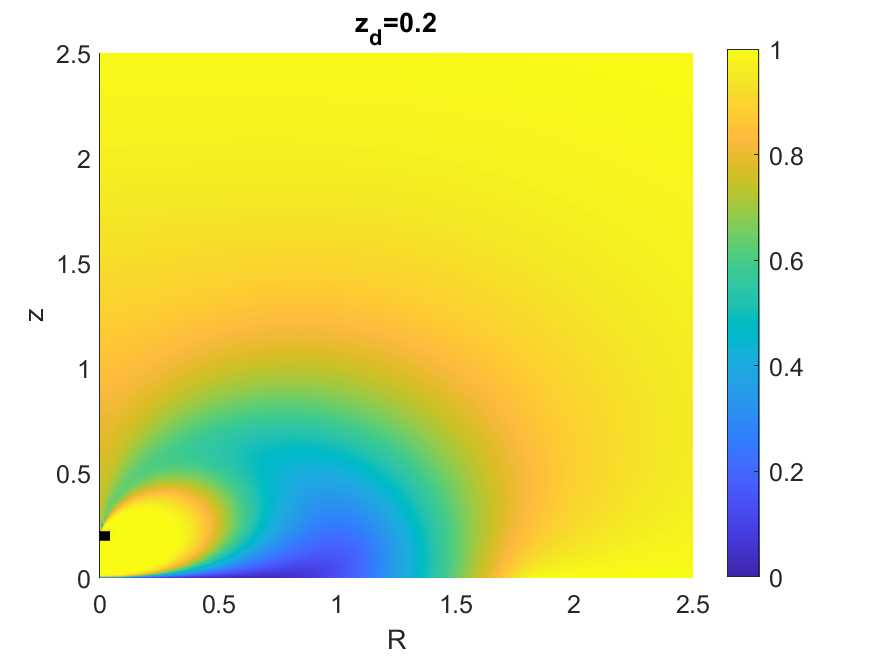}
   \includegraphics[width=0.33\linewidth]{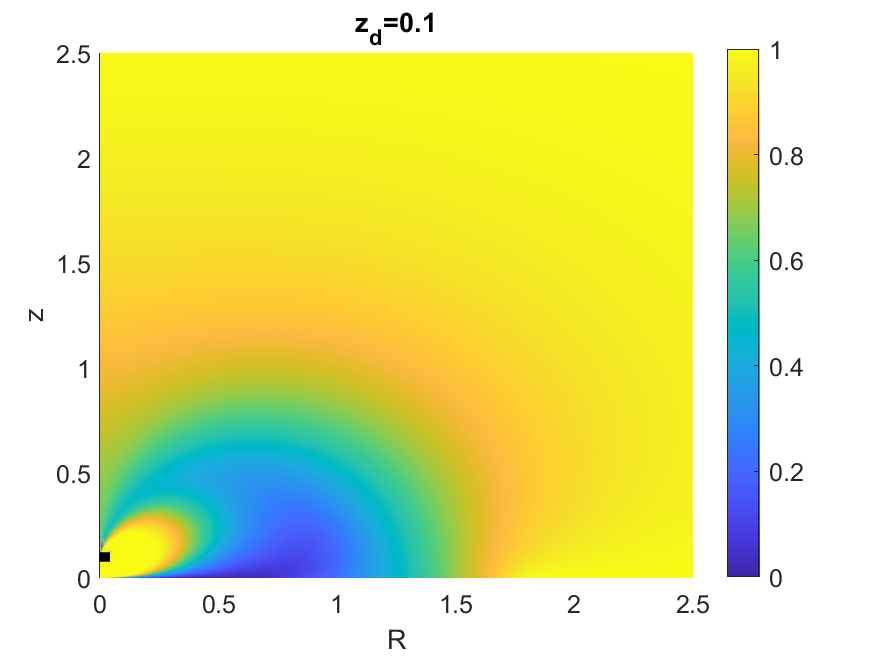}}
   \caption{Voidage field $E(R,z)$, terrestrial drone, $z_d=0.9$ (top left), $z_d=0.8$ (top middle), $z_d=0.7$ (top right), $z_d=0.6$ (middle left), $z_d=0.5$ (middle), $z_d=0.4$ (middle right), $z_d=0.3$ (bottom left) $z_d=0.2$ (bottom middle), $z_d=0.1$ (bottom right)}
  \label{fig:td_various_zd}
\end{figure}

We solve (\ref{eqn:h34new}), (\ref{eqn:h35}), (\ref{eqn:h36}), (\ref{eqn:h37new}), together with (\ref{eqn:h38})-(\ref{eqn:h19}), using the numerical method described in \citet[Appendix~B]{NL24}, with discretization parameters $\epsilon=10^{-8}$, $L_R=256$, $L_z=64$, and $N=16$ (as described in \S\ref{sec:model}).  Using these discretization parameters, the number of degrees of freedom ranges between 4,403,712 (for $z_d=0.1$) up to 5,719,552 (for $z_d=0.9$), so between four and six million degrees of freedom in our numerical scheme for each example.  In Figure~\ref{fig:td_various_zd} we plot the voidage field $E(R,z)$ for $z_d=0.9$, $z_d=0.8$, $z_d=0.7$, $z_d=0.6$, $z_d=0.5$, $z_d=0.4$, $z_d=0.3$, $z_d=0.2$ and $z_d=0.1$.  
Each of these plots is plotted on the same scale, for ease of comparison, but in each case the actual domain of computation was much larger (as determined by the discretization parameters stated above).  By plotting over a smaller range for $R$ and $z$, we hope to make it easier to see the finer detail of the field in the vicinity of the drone.
As $z_d$ decreases, we see the formation of a horn-shaped plume in the voidage field, with particles being drawn into the horn and enveloping the drone, when $z_d$ is sufficiently small.  For this example, we take $z_d$ much lower than we did for the helicopter examples presented in~\citet{NL24}, noting that the ratio of rotor blade length to rotorcraft body is in general much lower for drones than it is for helicopters.  Despite the inclusion of gravity and the different parameter choices considered in this section, the manner in which the horn-shaped plume forms as $z_d$ decreases is broadly comparable to the results for the helicopter presented in~\citet{NL24}, and aligns well with experimental evidence as seen in, e.g., \citet[Figures 6,7,8]{Nathan_Green_2009}.  

The supplementary file Langdon-movie3-terrestrial.mp4 shows in a video format how the voidage field changes as $z_d$ is reduced gradually from $z_d=2.48$ to $z_d=0.12$, using the same parameter choices, discretization parameters and colour scale as for the results shown in figure~\ref{fig:td_various_zd}. 

\subsection{Martian drone cloud model}
\label{sec:Martian}
This application considers the situation of a helicopter drone hovering steadily over a dusty Martian surface.  Simulating dust cloud formation in this scenario has been considered by \citet{RaLoSlWa21} (who considered how terrestrial simulations might be scaled appropriately) and \citet{CaBaDiYo25} (who modelled entrainment of dust particles through the physical process of saltation, see, e.g., \citet{KPMK}).  With the NASA-developed Martian helicopter ``Ingenuity'' relying on vision-based navigation and, therefore, particularly susceptible to dust clouds generated by hovering at low altitudes (see, e.g., \citet{BaCoBrDe19} and \citet[\S3 ``Dust on Optics'']{GaScMeGi26}), understanding these processes is particularly relevant to this application.

Again, the modelling situation is the same as that in \citet{NL24} for the helicopter brownout application, except that there is again a significant reduction in the macroscopic length scale (from helicopter blade radius to drone blade radius) and in the macroscopic fluid velocity scale. Furthermore, the gravitational acceleration is significantly reduced and the atmospheric density is much smaller than that on Earth. With drone operating parameters based on those associated with the NASA Ingenuity helicopter (see \citet{NASA_ingenuity}) the dimensionless parameters for this application are estimated as
\begin{equation}
    \Gamma\sim 1.5 \times 10^{-2},~\gamma\sim  2\times 10^3,~\alpha \sim 10^{-3},~\Omega\sim 10^0.
    \label{eqn:martian_param}
\end{equation}
As in the last subsection, the value of the parameter $\Gamma$ is not sufficiently small to suggest that gravitational terms can be neglected, and so this application again leads us to employ the modified elliptic PDE given in (\ref{eqn:h34new}) and the associated modified boundary condition (\ref{eqn:h37new}). We now determine the voidage field for the parameters estimated in~(\ref{eqn:martian_param}) above, also choosing $E_s=0.01$ and $\Delta=0.04$, as for the terrestrial drone example.

\begin{figure}
   \centerline{
   \includegraphics[width=0.6\linewidth]{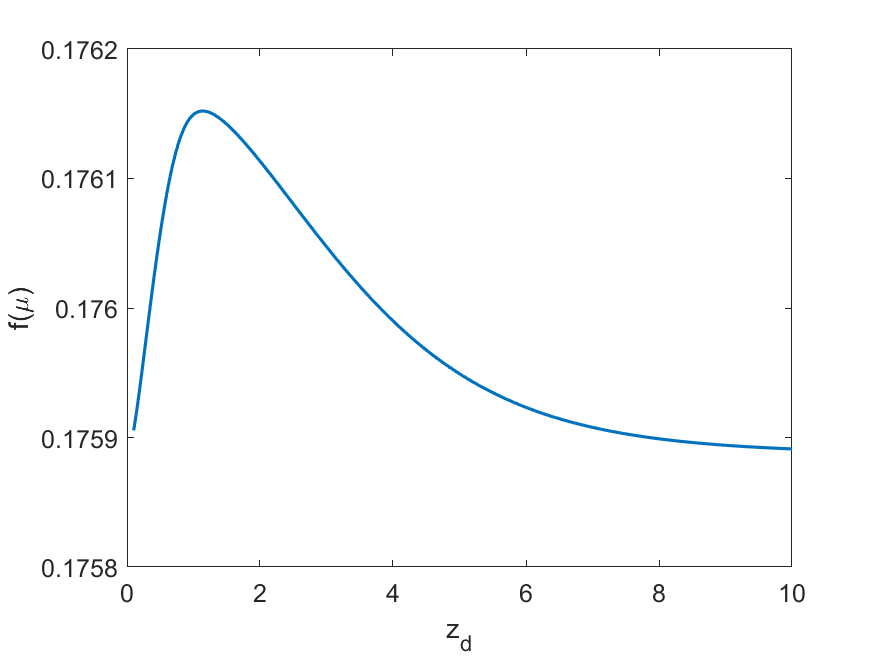}}
   \caption{Plot showing how $f(\mathbf{\mu})$ (given by (\ref{eqn:PC})) varies with $z_d$, for the Martian drone example.}
  \label{fig:martian_PC}
\end{figure}

In figure~\ref{fig:martian_PC} we plot the particle cloud formation potential $\cal{PC}(\chi_{\mu})\equiv$ $ f(\mathbf{\mu})$, given by~(\ref{eqn:PC}), for $\gamma=2000$, $\Omega=1.0$, $E_s=0.01$ and $z_d\in[0.1,10]$.
As for the helicopter cloud example (see the central plot of figure~\ref{fig:vary_zd}) and the terrestrial drone example (see figure~\ref{fig:terrestrial_PC}), we see that $f(\mathbf{\mu})$ increases as $z_d$ decreases from 10 to approximately $z_d=1.5$, before then decreasing as $z_d$ decreases further.  Note also the scale on the vertical axis - for this example, the range of values of $\mathcal{PC}(\chi_{\mu})\equiv f(\mathbf{\mu})$ is extremely narrow.

\begin{figure}
   \centerline{
   \includegraphics[width=0.45\linewidth]{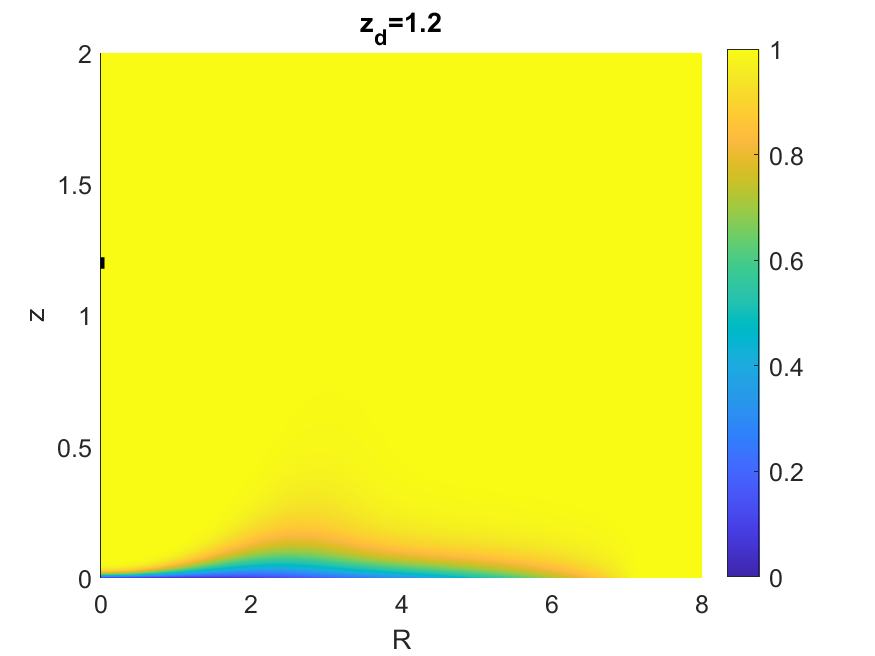}
   \includegraphics[width=0.45\linewidth]{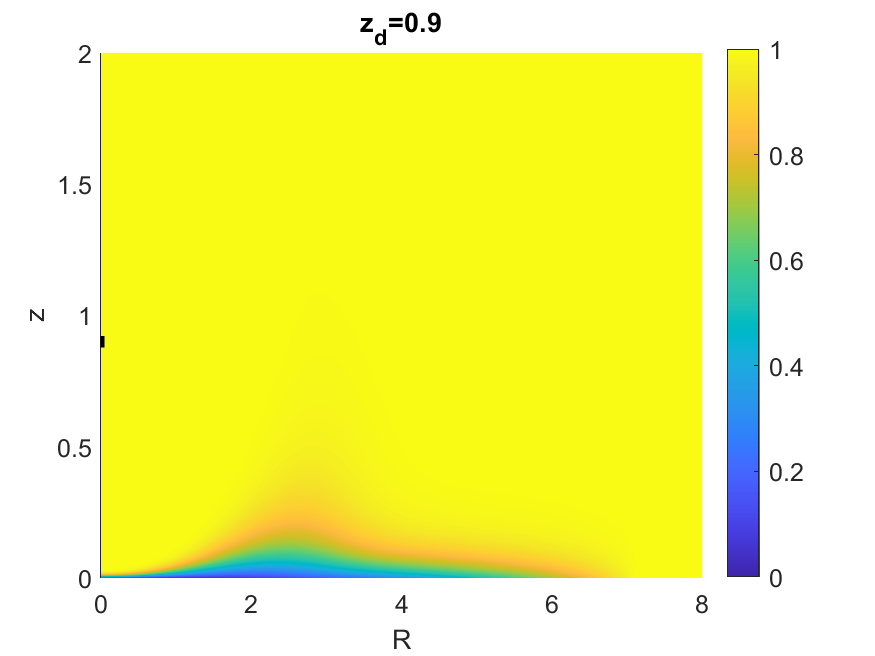}}
   \centerline{
   \includegraphics[width=0.45\linewidth]{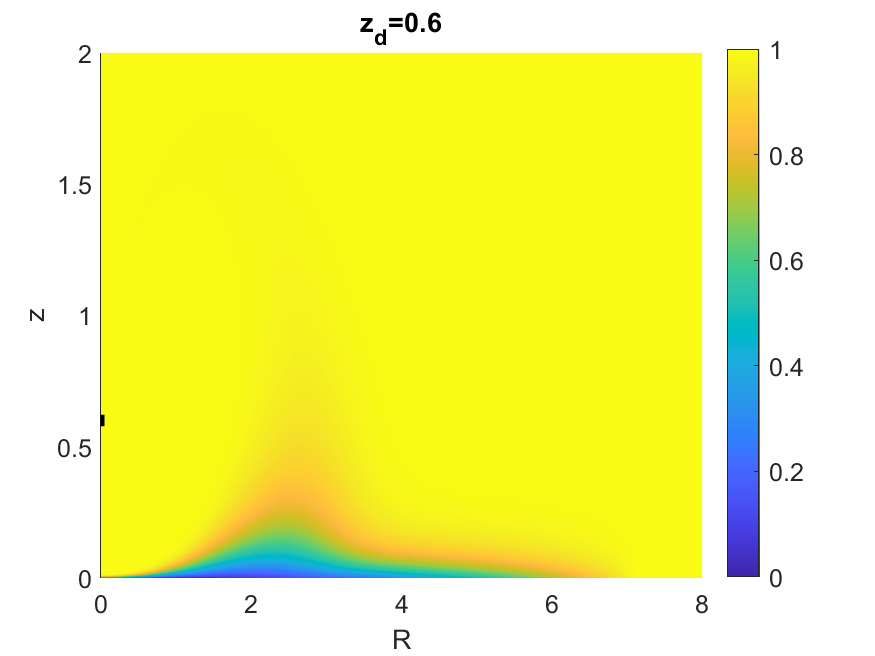}
   \includegraphics[width=0.45\linewidth]{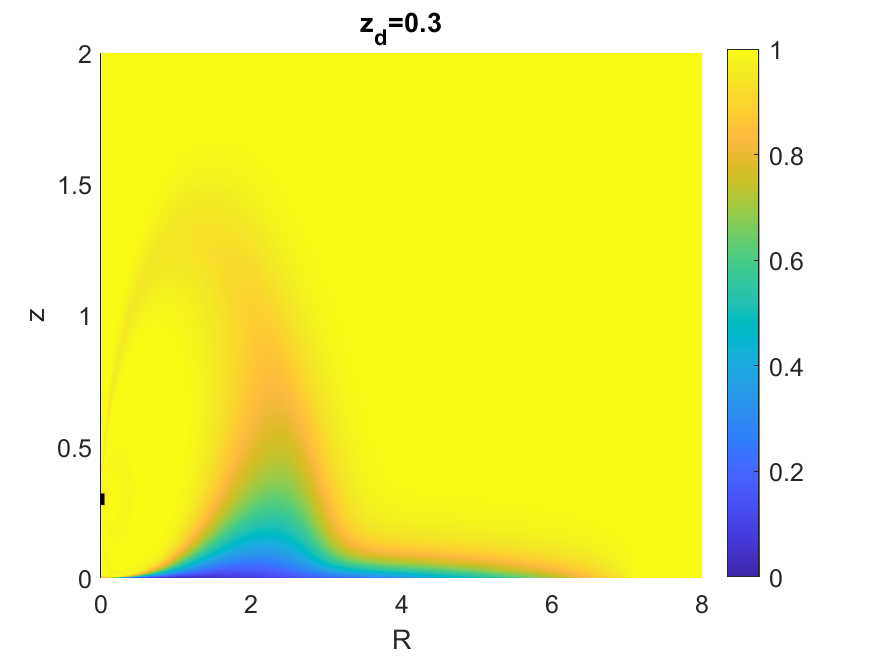}}
   \caption{Voidage field $E(R,z)$, Martian drone, $z_d=1.2$ (top left), $z_d=0.9$ (top right),  $z_d=0.6$ (bottom left), $z_d=0.3$ (bottom right)}
  \label{fig:martian_various_zd}
\end{figure}

We solve (\ref{eqn:h34new}), (\ref{eqn:h35}), (\ref{eqn:h36}), (\ref{eqn:h37new}), together with (\ref{eqn:h38})-(\ref{eqn:h19}), using the numerical method described in \citet[Appendix~B]{NL24}, with discretization parameters $\epsilon=10^{-8}$, $L_R=288$, $L_z=64$, and $N=16$ (as described in \S\ref{sec:model}).  Using these discretization parameters, the number of degrees of freedom ranges between 5,321,984 (for $z_d=0.3$) up to 6,986,624 (for $z_d=1.2$), so between five and seven million degrees of freedom in our numerical scheme for each example.  In Figure~\ref{fig:martian_various_zd} we plot the voidage field $E(R,z)$ for $z_d=1.2$, $z_d=0.9$, $z_d=0.6$ and $z_d=0.3$.  Each of these plots is plotted on the same scale for ease of comparison, but in each case the actual domain of computation was much larger (as determined by the discretization parameters stated above) - by plotting over a smaller range for $R$ and $z$, we hope to make it easier to see the finer detail of the field in the vicinity of the Martian drone.  A comparison with \citet[Figure~14]{CaBaDiYo25} reveals similarities with the horn-shaped plume seen in the ``initial burst of dust'' in their simulation of ``lifted dust during take-off''.

The supplementary file Langdon-movie4-martian.mp4 shows in a video format how the voidage field changes as $z_d$ is reduced gradually from $z_d=2.48$ to $z_d=0.28$, using the same parameter choices, discretization parameters and colour scale as for the results shown in figure~\ref{fig:martian_various_zd}.

\begin{figure}
   \centerline{
   \includegraphics[width=0.45\linewidth]{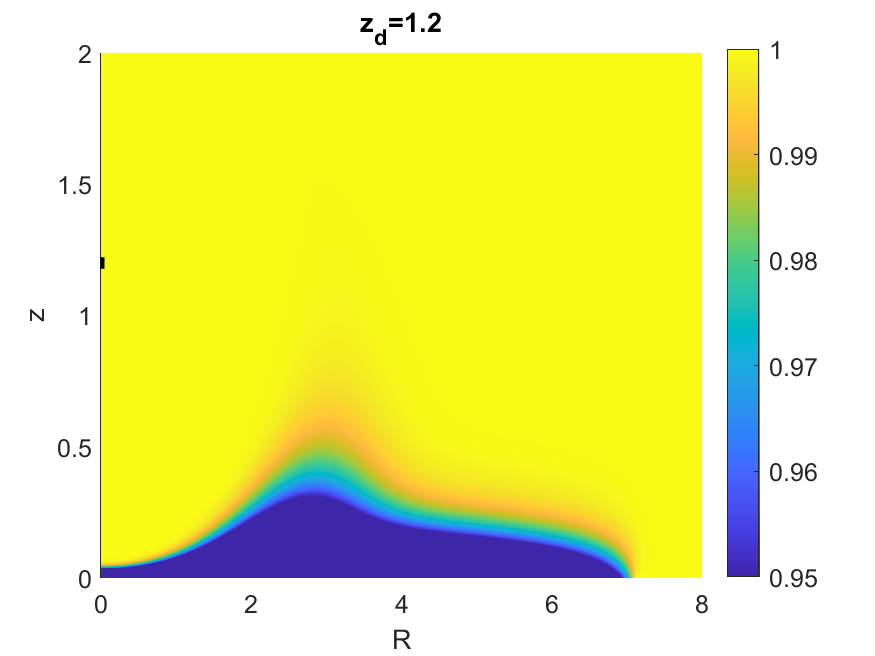}
   \includegraphics[width=0.45\linewidth]{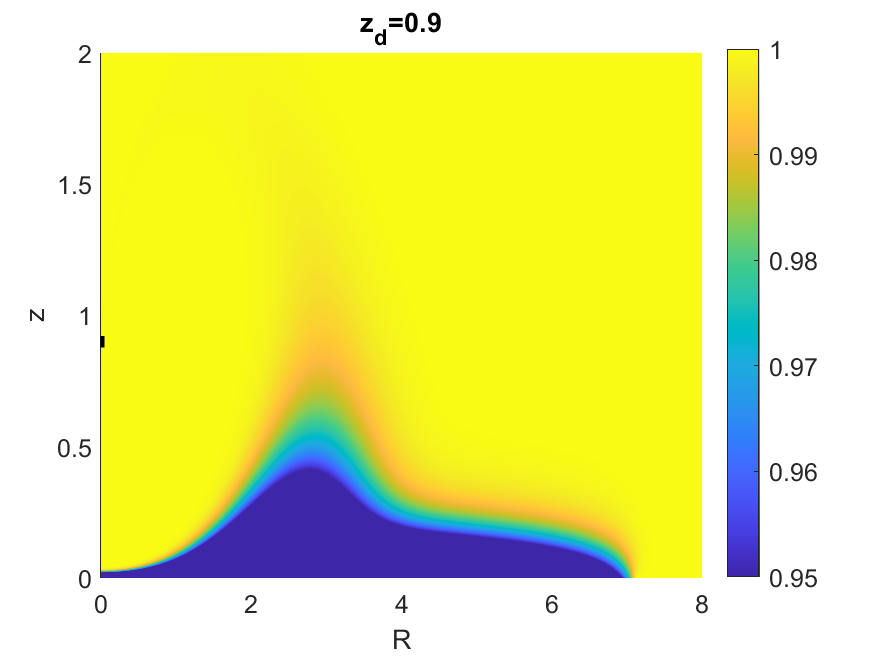}}
   \centerline{
   \includegraphics[width=0.45\linewidth]{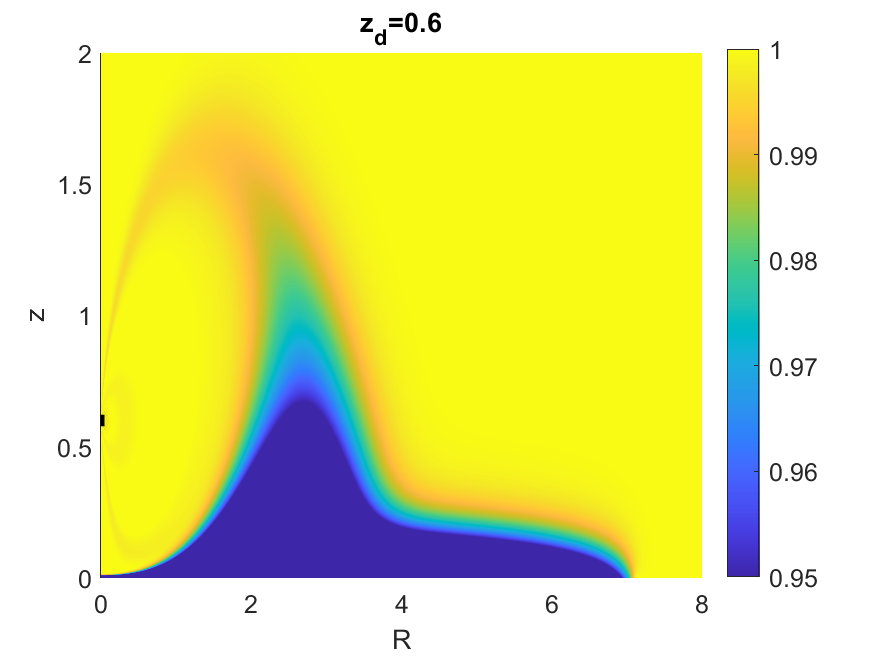}
   \includegraphics[width=0.45\linewidth]{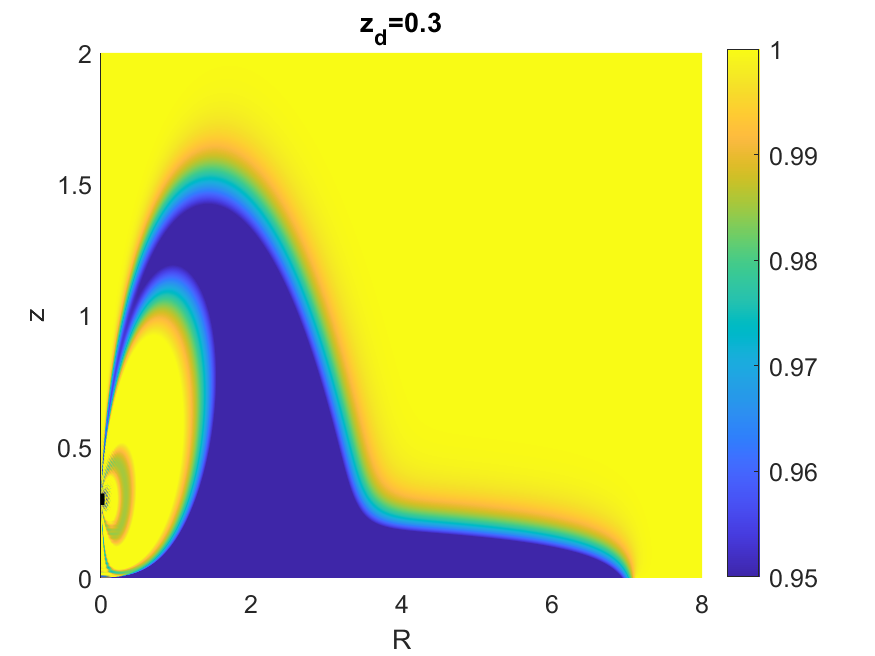}}
   \caption{Voidage field $E(R,z)$, Martian drone, $z_d=1.2$ (top left), $z_d=0.9$ (top right), $z_d=0.6$ (bottom left), $z_d=0.3$ (bottom right), reduced colour scale.}
  \label{fig:martian_various_zd_clim099}
\end{figure}

A new feature that we see in Figure~\ref{fig:martian_various_zd} that was not present in the other results (in this paper, or in~\citet{NL24}), clearer in the figures for the smaller values of $z_d$, is a ``second swirl'', i.e., a closed loop within the outer loop.  To illustrate this more clearly, in Figure~\ref{fig:martian_various_zd_clim099} we show the same results as in Figure~\ref{fig:martian_various_zd}, but now with the colour scale set from 0.95 to 1, rather than from 0 to 1 (as was previously the case).  In this case, if the plot is blue, that means that $E(R,z)\leq 0.95$ in that region.
As noted above, this feature appears to be new for this example, and some computational experimentation suggests that it is primarily driven by the choice of very small values of $\alpha$ in this application, rather than being specific to other parameter choices for the Martian drone example.  We further explore the dependence of this new feature on the parameter choices in Appendix~\ref{sec:alpha}.

\section{Dust devil cloud model}
\label{sec:DD}
In this section we consider an application of our model to what is referred to as a \emph{dust devil}. To describe a dust devil, consider a static dry sand bed with horizontal free surface below a static atmosphere of air at ambient temperature $T_a$ and density $\rho(T_a)$. An approximately circular patch of the surface of the static sand bed, of radius say $r_d (\sim l)$, is heated to a higher temperature $T_d$, and this heats the air just above to temperature $T_d$, and as a consequence, to a reduced density $\rho(T_d)$. This heated air rises in a column through buoyancy, and this requires an inflow of cooler air at the base, and the process continues. In addition, local vortex lines are concentrated and stretched in the core (of radius $r_c<<r_d$) of the rising air column, and this drives a rapid horizontal circulation of the air about the thin vortex core. The rapid horizontal inflow and circulation close to the surface of the otherwise static sand bed entrains sand into the upwardly circulating columnar air flow, and a dust cloud is formed within and around the column. This phenomenon is often referred to as a dust devil. Descriptions and images of dust devils can be found in the review papers by \citet{dust_devil_A}, \citet{dust_devil_B} and \citet{dust_devil_E}, where it should be noted that dust devils have been reported as frequently occurring on the surface of Mars (see also \citet{dust_devil_C}). From a theoretical point of view, the details of the fluid flow field within a dust devil have been addressed recently from a computational standpoint (see, for example, \citet{dust_devil_D}). However, the key aspect of dust cloud formation has either been neglected entirely, or included in a purely phenomenological way. This is due to the absence of a rational mechanistic model to account for the entrainment/detrainment of dust from the surface into the two phase dust cloud. In this section we can remedy this with use of our current theory, whilst adopting a very simplistic representation of the overall fluid flow in the dust devil.

The coherent height of a dust devil (before collapse under cooling), say $h_d$, has been shown \citep{Jackson20} to satisfy the scaling,
\begin{equation}
    h_d \sim O(r_d^{2}). \label{DD1}
\end{equation}
In addition a simple vertical momentum balance in the dust devil rising air column determines, under the Boussinesq approximation, that the rising air in the column has vertical velocity scale,
\begin{equation}
    u_s^2 \sim (1 - \rho(T_d)/\rho(T_a)) g h_d. \label{DD2} 
\end{equation}
Also, with $\Gamma_s$ representing the horizontal circulation of the air flow around the column, then the angular horizontal velocity scale in the column has,
\begin{equation}
    u_d \sim \Gamma_s/2\pi r_d. \label{DD3}
\end{equation}
A simple flow field which captures the principal features of the air flow in the dust devil is obtained by taking exactly the same flow field that we used for the helicopter cloud model reviewed in section 2, \emph{except now we reverse the orientation of the fluid dipole}, and fix its location at a distance  $(1/2) h_d$ vertically above the origin, representing buoyancy balance in the fluid flow occuring approximately halfway up the dust devil column.  With this fluid flow field, we take the dipole strength as,
\begin{equation}
    M_s = u_s r_d^3, \label{DD4}
\end{equation}
with the velocity scale $u_s$ given in (\ref{DD2}), and the circulation of the line vortex as,
\begin{equation}
    \Gamma_s = 2 \pi r_d u_d, \label{DD5}
\end{equation}
with $u_d$ being the horizontal velocity scale for the air flow in the dust devil column. Under this approximation to the airflow in the dust devil, our model for the particle voidage field in the dust devil particle cloud again leads directly to the elliptic boundary value problem (\ref{eqn:h34})-(\ref{eqn:h19}), except that now the signs in front of the  coefficients $\overline{a}$ and $\overline{b}$ are reversed, and in addition, we again choose to include the effect of gravity on the particle phase flow. Hence~(\ref{eqn:h34}) is replaced by
\begin{equation}
  E\left(E_{RR} + \frac{1}{R} E_R + \frac{\Gamma}{\alpha}E_z + E_{zz} \right) + \bar{a}(R,z) E_R + \bar{b}(R,z) E_z = 0, \quad (R,z)\in \omega.
  \label{DD6}
\end{equation}
In addition to this change in the field equation, there is again a consequent change to the boundary condition at both the upper and lower surfaces of the cylindrical shell encasing the dipole due to the inclusion of gravity and  this leads us again to modify boundary conditions (\ref{eqn:h37}) to
\begin{equation}
    E_z = \frac{\Gamma}{\alpha}(1-E)\mbox{ on } z= z_d^+,z_d^-, \quad R\in(0,\Delta). \label{DD7}
\end{equation}

It now remains for us to use the approximations (\ref{DD1})--(\ref{DD5}) to estimate the dimensionless parameters $\Omega$, $\gamma$, $\alpha$ and $\Gamma$ for the dust devil application. Typical measurements from naturally occurring dust devils (see, e.g., \citet{dust_devil_A, dust_devil_E, Jackson20, dust_devil_D}) lead to typical height and radius of the orders of $h_d\sim 100 ~m$ and $r_d\sim 10 ~m$ respectively (which are in accord with the empirical relation (\ref{DD1})), whilst typical measurements of the vertical core air velocity scale are of the order $u_s\sim 10 ~ms^{-1}$, and a similar order of  $u_d \sim 25~ms^{-1}$ for the angular air velocity scale in the core. This leads, via (\ref{DD5}), to a typical circulation scale of $\Gamma_s \sim 1.5\times 10^3~m^2s^{-1}$, whilst the strength of the dipole in the model should be set as $M_s= u_s r_d^3\sim 10^4~m^4s^{-1}$. We note that these scales then predict a typical air density ratio across the length of the dust devil as $\rho(T_d)/\rho(T_a) \sim 0.9$, via (\ref{DD2}), which is in line with typical measurements. We can now use these estimates to compute the dimensionless parameters $\Omega$, $\gamma$, $\alpha$ and $\Gamma$, as introduced in \citet{NL24}, for this dust devil application. We obtain,
\begin{equation}
    \Gamma\sim10^{-1} -10^0, \quad \gamma=2\times(10^1-10^2), \quad \alpha = 3\times 10^{-1}, \quad \Omega=1.5\times(10^1-10^2).
    \label{DDE}
\end{equation}
with the dimensionless height of the dipole location being $z_d \sim 5$.

We now present some numerical examples showing the voidage field generated by the dust devil.  We solve (\ref{DD6}), (\ref{eqn:h35}), (\ref{eqn:h36}), (\ref{DD7}), together with (\ref{eqn:h38})-(\ref{eqn:h19}), with $\gamma=20$, $\Omega=15$, $\alpha=0.3$, $E_s=0.01$, $\Gamma=0.1$, $\Delta=0.2$, and $z_d=5$, using the numerical method described in \citet[Appendix~B]{NL24}, with discretization parameters $\epsilon=10^{-8}$, $L_R=128$, $L_z=480$, and $N=16$ (as described in \S\ref{sec:model}), for which the number of degrees of freedom is 16,691,328.

For these parameter choices, the particle cloud formation potential $\cal{PC}(\chi_{\mu})\equiv$ $ f(\mathbf{\mu})$, given by~(\ref{eqn:PC}), is $ f(\mathbf{\mu}) \approx 0.4$.  Comparing this to the other examples in this paper, we see that the particle cloud formation potential for the dust devil is significantly larger than for all other examples, with the exception of the lunar plume example considered in~\S\ref{subsec:lunar}.

\begin{figure}
   \centerline{
\includegraphics[width=0.6\linewidth]{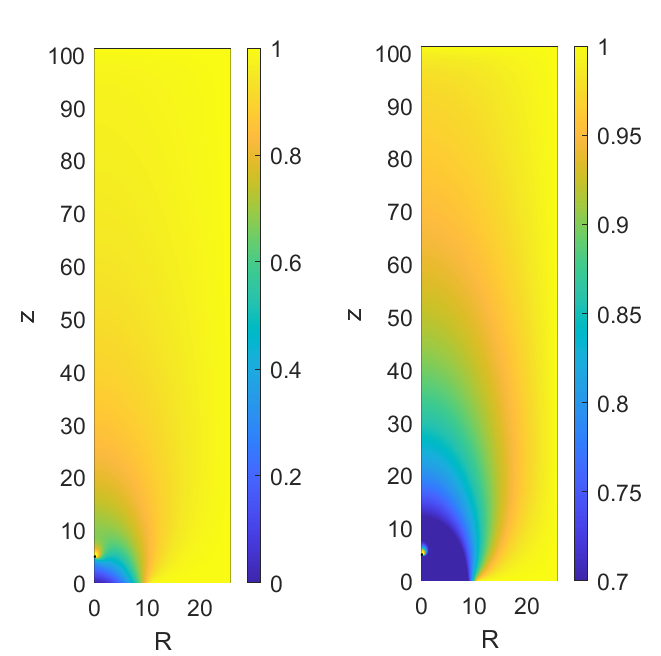}}
   \caption{Dust devil, voidage field $E(R,z)$, two different colour scales}
  \label{fig:dust_devil}
\end{figure}

In figure~\ref{fig:dust_devil} we show the voidage field $E(R,z)$ on two different scales (note the range of the colour bar in each plot). On the left, we plot the voidage field using the full colour scale range from zero to 1.0 (recalling that $E\in[E_s,1]$), and on the right we plot over the range $E\in[0.7,1]$, so that the detail of the pattern can be more clearly seen for $E\approx 1$ (in this plot, blue colour implies that $E$ is less than or equal to 0.7 on the colour scale).  Note also the horizontal and vertical scales for the plot;  the particle cloud for this example extends far higher vertically, compared to all other examples elsewhere in the paper. In regard to overall performance of the model, we see that the dust cloud voidage predictions from the model in figure~\ref{fig:dust_devil} give overall a favourable visual comparison with photographic images, such as \citet[Figure 1]{dust_devil_B}, or the first photographic image on the Wikipedia page ``Dust Devil'', which is an image of a large scale dust devil taken in Arizona, see \citet{DD_wiki}.

\section{Model integration with fluid flow CFD platforms or fluid flow measurement data}
\label{sec:CFD}
In this section we briefly explain how the model developed in \citet{NL24} can be readily incorporated into a CFD platform, or with experimental data, which provides an accurate representation of the primary single-phase fluid velocity field under consideration, when the particle bed is absent and is replaced by a rigid impermeable smooth surface. We denote the rigid (and possibly moving) boundaries in the fluid flow field away from the particle bed location as $\mathcal{M}_u$ and the surface of the particle bed as $\mathcal{M}_l$. We assume that the fluid velocity field has been determined via a CFD platform or data measurements (for a single-phase flow with relative no-slip boundary conditions applied on the whole boundary $\mathcal{M}_u \cup \mathcal{M}_l$). In dimensionless variables (with lengths scaled on the macroscopic geometrical length scale $l_s$ and velocities scaled on the macroscopic fluid velocity scale $u_s$) we write this determined fluid velocity field as $\mathbf{u}(\mathbf{r},t)$ with $\mathbf{r}$ and $t$ being vector position (relative to a suitable origin that is \emph{static relative to the static particle bed})) and time t. Now, referring to \citet[\S4]{NL24} we can estimate the locally tangential fluid velocity in the thin interfacial layer, at any point $p$, located at $\mathbf{r}_p \in \mathcal{M}_l$, as,
\begin{equation}
    \mathbf{u}_h(\mathbf{r}_p,t) = (\mathbf{u} - (\mathbf{u}.\mathbf{\hat{n}})\mathbf{\hat{n}})(\mathbf{r}_p + \epsilon\mathbf{\hat{n}},t),
\end{equation}
where $\epsilon=a/l_s$ and $a$ is the typical particle radius, whilst $\mathbf{\hat{n}}$ is the unit normal at $\mathbf{r}_p$ pointing into the fluid flow region. With $\epsilon$ being small, for an inviscid fluid flow we may approximate this as simply the slip velocity field induced on $\mathcal{M}_l$, whilst for a viscous fluid we may approximate as,
\begin{equation}
      \mathbf{u}_h(\mathbf{r}_p,t) = (\mathbf{u} - (\mathbf{u}.\mathbf{\hat{n}})\mathbf{\hat{n}})_n(\mathbf{r}_p,t) \epsilon, \label{eqna}
\end{equation}
with subscript $n$ representing the normal derivative at $\mathbf{r}_p$ in the direction of $\mathbf{\hat{n}}$. With the determined fluid flow, on the macroscopic scales, generally being of moderately high to very high macroscopic Reynolds number $R_f = u_sl_s/\nu$, then the normal derivative in~(\ref{eqna}) is calculated in the viscous boundary layer for the fluid flow, which determines the dimensionless scale for $\mathbf{u}_h$ as
\begin{equation}
    \mathbf{u}_h \sim \epsilon R_f^{\frac{1}{2}}
\end{equation}
which we formally anticipate to be less than or equal to $O(1)$ in many applications due to the extremely small value of the microscopic to macroscopic length scale ratio $\epsilon$. Having determined $\mathbf{u}_h$ on $\mathcal{M}_l$ via~(\ref{eqna}) we can now immediately formulate the problem which determines the voidage field in the fluid flow region as:
\begin{equation}
E_t +  (\mathbf{u}(\mathbf{r},t)+ \Gamma E Q'(E)\mathbf{k}) \cdot \nabla E - \alpha E \nabla^2(H(E)) = 0, \quad \br \in D(t), ~t>0, \label{eqnE}
\end{equation}
subject to the initial condition,
\begin{equation}
    E(\mathbf{r},0) = E_0(\mathbf{r}),~~\mathbf{r}\in \overline{D}(0),
\end{equation}
and the boundary conditions,
\begin{equation}
   E(\mathbf{r},t) = g(\mathbf{r},t)),~~\mathbf{r} \in \mathcal{M}_l,~t>0,\label{E1}
\end{equation}
\begin{equation}
    E_n(\mathbf{r},t)=0,~~\mathbf{r}\in \mathcal{M}_u,~t>0,\label{E2}
\end{equation}
\begin{equation}
    E(\mathbf{r},t) \to 1~~\text{as}~~|\mathbf{r}|\to \infty ~\text{uniformly on} ~D(t),~t>0.\label{E3}
\end{equation}
Here $D(t)$ represents the open region occupied by the two-phase flow at time $t\ge 0$, the subscript $n$ represents the directional derivative in the direction normal to the relevant surface and pointing into $D(t)$, $\mathbf{k}$ is the unit vector pointing vertically upwards and the functions $Q(E)$ and $H(E)$ are as introduced in~(\ref{eqn:Q}) and~(\ref{eqn:H}), respectively. The prescribed function $g:\mathcal{M}_l \times \overline{\mathbb{R}}^+\to \mathbb{R}$ is given in terms of the determined fluid velocity on $\mathcal{M}_l$, following directly the derivation in \citet[(5.8)]{NL24}, as
\begin{equation}
    g(\mathbf{r},t) =  
    \begin{cases}
    1~~\text{when}~~|\mathbf{u}_h(\mathbf{r},t)|^2 \le \frac{1}{\gamma},\\
    (\gamma|\mathbf{u}_h(\mathbf{r},t)|^2)^{-1}~~\text{when}~~ \frac{1}{\gamma}< |\mathbf{u}_h(\mathbf{r},t)|^2 < \frac{1}{E_s\gamma},\\
     E_s~~\text{when}~~|\mathbf{u}_h(\mathbf{r},t)|^2 \ge \frac{1}{E_s\gamma}, \label{eqng}
    \end{cases}
\end{equation}
for $(\mathbf{r},t) \in \mathcal{M}_l \times \overline{\mathbb{R}}^+$ where $\mathbf{u}_h(\mathbf{r},t)$ is as given in (\ref{eqna}). The function $E_0:\overline{D}(0)\to [E_s,1]$ is the prescribed initial voidage field and is taken to be continuous and satisfying the boundary constraints in (\ref{E1})--(\ref{E3}) at $t=0$. The parameter $\Gamma$ is as given in (\ref{G}), whilst,
\begin{equation}
    \gamma = \frac{3u_s^2\rho_f}{ag\rho_s},~~\alpha = \frac{p_0a^2}{l_s\mu u_s},
\end{equation}
as introduced in \citet{NL24}, with $p_0$ being the interparticle collisional coefficient and $\mu$ the dynamic viscosity of the fluid. We note that it is straightforward to establish that this parabolic evolution problem has a global in time classical solution, which is unique, and depends continuously on the positive parameters $\alpha$ and $\gamma$, and on the initial data function $E_0$. In addition, the Strong Parabolic Maximum/Minimum Principles determine that
\begin{equation}
    E_s < E(\mathbf{r},t) < 1~~\forall~~\mathbf{r}\in D(t),~t>0.
\end{equation}
The parabolic evolution problem (\ref{eqnE})--(\ref{E3}) can be solved numerically using an approach such as that described in \citet[Appendix~B]{NL24}, combined with an appropriate timestepping scheme.

\section{Conclusions}
\label{sec:conc}
In this paper, we have generalised a recently derived model for particle-fluid flow fields generated by static particle bed entrainment, and have applied it to a range of applications.  The key features of our simulated particle clouds resemble those seen in photographs, where those are readily attainable.  It would be interesting to undertake experimental work in order to further validate our model; for the bathysphere application in particular, we anticipate that such an experimental programme should be reasonably straightforward.  For applications involving swirling flows and extraterrestrial applications, such an experimental programme would be more challenging, but the level of this challenge also serves to highlight the value of simulations based on rational mathematical models.

To enable a straightforward comparison of particle cloud formation potential across different applications and parameter choices, we have derived a simple a-priori measure that can be calculated requiring only the knowledge of the tangential primary fluid velocity field at the particle bed surface. The relative values of this measure can be seen to align well with qualitative features of the particle cloud in the full two-phase flow field.

As described in \S\ref{sec:intro}, our model contains a number of simplifying assumptions that enable consideration of the key features of the flow without requiring consideration of the complexity of the fine-scale details of the primary fluid flow field.  To facilitate wider applicable use of our model, we have outlined in \S\ref{sec:CFD} how it could potentially be incorporated into a CFD platform, thereby enabling simulations requiring finer-scale details to be considered.  We anticipate that implementation of such an approach may be a fruitful topic of future research.

\backsection[Funding]{This research received no specific grant from any funding agency, commercial or not-for-profit sectors.}

\backsection[Declaration of interests]{The authors report no conflict of interest.}

\backsection[Data availability statement]{The authors confirm that the data supporting the findings of this study are available within the paper.}

\backsection[Rights retention statement]{For the purposes of open access, the authors have applied a Creative Commons Attribution (CC BY) Licence to any Accepted Author Manuscript version arising from this submission.}

\backsection[Author ORCIDs]{S. Langdon, https://orcid.org/0000-0002-0572-5137;\\ D. J. Needham, https://orcid.org/0000-0002-4958-6976}

\begin{appendix}
\section{Properties of $\mathcal{PC}$} 
\label{sec:PC}
    In this appendix we determine the main properties of the functional $\mathcal{PC}$. In the current setting it is appropriate to work in the Banach space of real, continuous scalar fields on $\overline{\mathcal{A}}$ under the usual sup-norm, which we denote as $B(C(\overline{\mathcal{A}}),||\cdot||_{\infty})$, and we denote by $K(\overline{\mathcal{A}})\subset C(\overline{\mathcal{A}})$ the closed subset of scalar fields in this Banach space which have image in the interval $[\Lambda^{-1}, (\Lambda E_s)^{-1}]$. Clearly this closed subset contains all admissible scalar fields. The functional $\mathcal{PC}$ is now considered as a mapping from $K(\overline{\mathcal{A}})$ to $\overline{\mathbb{R}}^+$ ($\mathcal{PC}: K(\overline{\mathcal{A}}): \to \overline{\mathbb{R}}^+$). It is a straightforward consequence that $\text{Im}(\mathcal{PC}) = [0,1-E_s]$, after also noting that $\chi \equiv c$ is in $K(\overline{\mathcal{A}})$ for each constant $c\in [\Lambda^{-1}, (\Lambda E_s)^{-1}]$. Next, let $\chi_1$ and $\chi_2$ be chosen from $K(\overline{\mathcal{A}})$. It then follows from  (\ref{eqne3}) and the Mean Value Theorem that,
    \begin{equation}
        |E(\chi_2(\mathbf{r}_h)) - E(\chi_1(\mathbf{r}_h))| \le \Lambda |\chi_2(\mathbf{r}_h) - \chi_1(\mathbf{r}_h)|~~\forall~~\mathbf{r}_h \in \overline{\mathcal{A}}.
    \end{equation}
    Consequently from (\ref{eqnf}) we obtain,
    \begin{equation}
        |\mathcal{PC}(\chi_2) - \mathcal{PC}(\chi_1)| \le \Lambda ||\chi_2 - \chi_1||_{\infty},
    \end{equation}
    via the triangle inequality, and we conclude that $\mathcal{PC}$ is Lipschitz continuous on $K(\overline{\mathcal{A}})$, with Lipschitz constant $\Lambda$. Next suppose $\chi_1$ is a \emph{decrease} of $\chi_2$. It then follows from (\ref{eqne3}) that,
    \begin{equation}
        E(\chi_2(\mathbf{r}_h)) - E(\chi_1(\mathbf{r}_h))\le0~~\forall~~\mathbf{r}_h \in \overline{\mathcal{A}},
    \end{equation}
    and so from (\ref{eqnf}),
    \begin{equation}
      \mathcal{PC}(\chi_1) - \mathcal{PC}(\chi_2)  \le 0.
    \end{equation}
    Thus $\mathcal{PC}$ is decreasing on pairs of points in $K(\overline{\mathcal{A}})$ which are decreasing. Since all points $\chi\in K(\overline{\mathcal{A}})$ are continuous and we have $(1-E(\chi)) \equiv (1-(\Lambda \chi)^{-1})$ it also readily follows that,
    \begin{equation}
        \mathcal{PC}(\chi) =0~~\text{if and only if}~~\chi\equiv \Lambda^{-1},
    \end{equation}
    and,
      \begin{equation}
        \mathcal{PC}(\chi) =(1-E_s)~~\text{if and only if}~~\chi\equiv (\Lambda E_s)^{-1}.
    \end{equation}
    Finally we observe that at each point $\chi \in \text{int}(K(\overline{\mathcal{A}}))$ we have the Fr\'{e}chet derivative given by the bounded linear mapping $D\mathcal{PC}_{\chi}:C(\overline{\mathcal{A}}) \to \mathbb{R}$ where,
    \begin{equation}
        D\mathcal{PC}_{\chi}(h) = \frac{1}{|\mathcal{\overline{A}}|} \int\!\!\int_{\overline{\mathcal{A}}}{\frac{h(\mathbf{r}_h)}{\Lambda \chi(\mathbf{r}_h)^2}}d\mathcal{A} ~~\forall~~h\in C(\overline{\mathcal{A}}).\label{eqnF}
    \end{equation}
    We observe that, for each point $\chi \in \text{int}(K(\overline{\mathcal{A}}))$, then, 
    \begin{equation}
        |D\mathcal{PC}_{\chi}(h)| \le \Lambda ||h||_{\infty}~~\forall~~h\in C(\overline{\mathcal{A}}).  
    \end{equation}

\section{Low ratio of particle collision pressure to drag force in the particle flow phase}
\label{sec:alpha}
In this appendix, motivated by our findings in~\S\ref{sec:Martian}, we consider the case in which the parameter $\alpha$, measuring the ratio of particle collision pressure to drag force in the particle flow phase, is very small.  As noted in~\S\ref{sec:Martian}, for the specific parameter choices considered for the Martian drone example, we see a new feature, not present in the other examples, namely a closed loop within the main swirl on the voidage contour colour maps. The key difference between the Martian drone example and the terrestrial drone example is the reduced value of $\alpha$, though the variations in $\Gamma$ complicate matters.  

\begin{figure}
   \centerline{
   \includegraphics[width=0.45\linewidth]{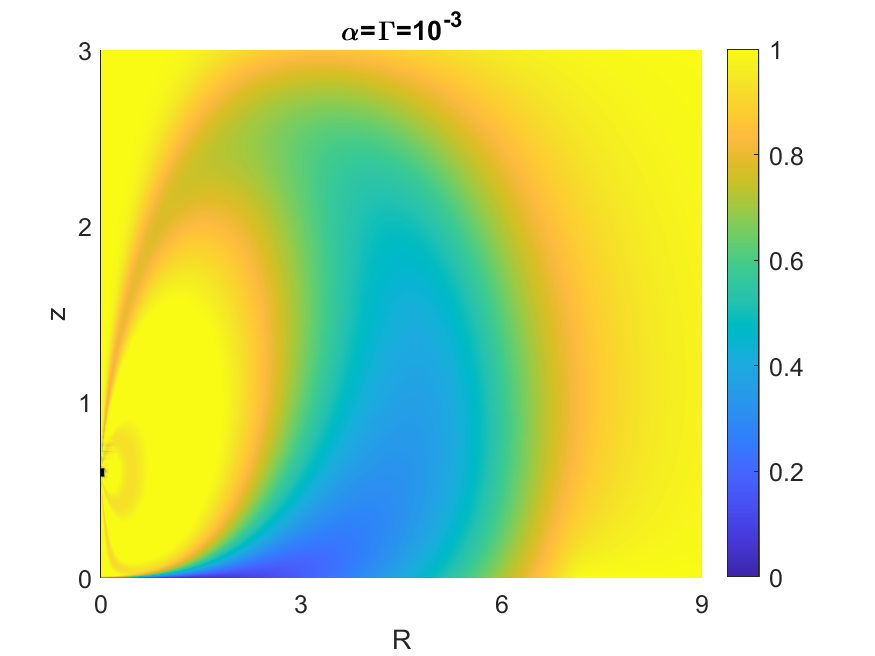}
   \includegraphics[width=0.45\linewidth]{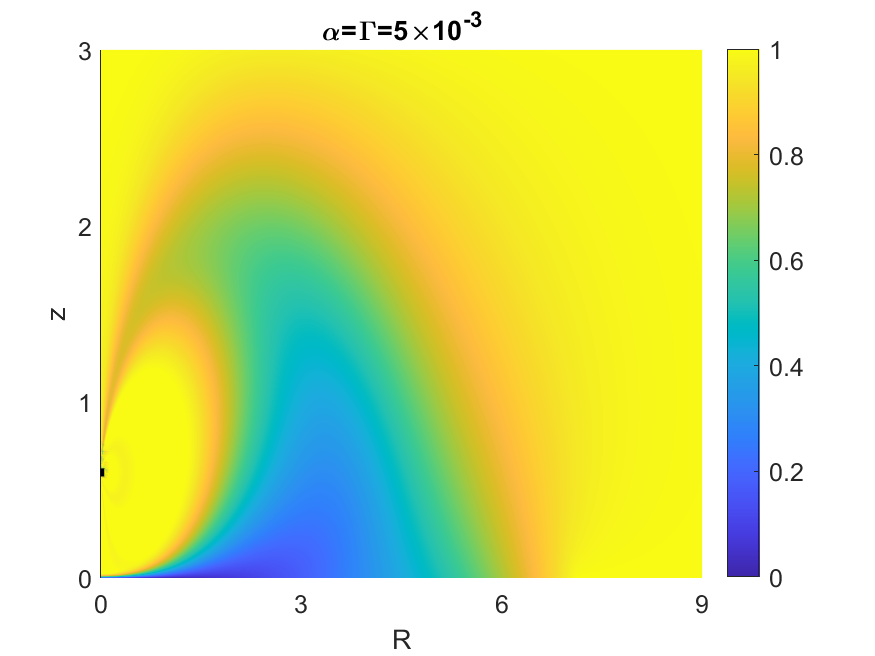}}
   \centerline{
   \includegraphics[width=0.45\linewidth]{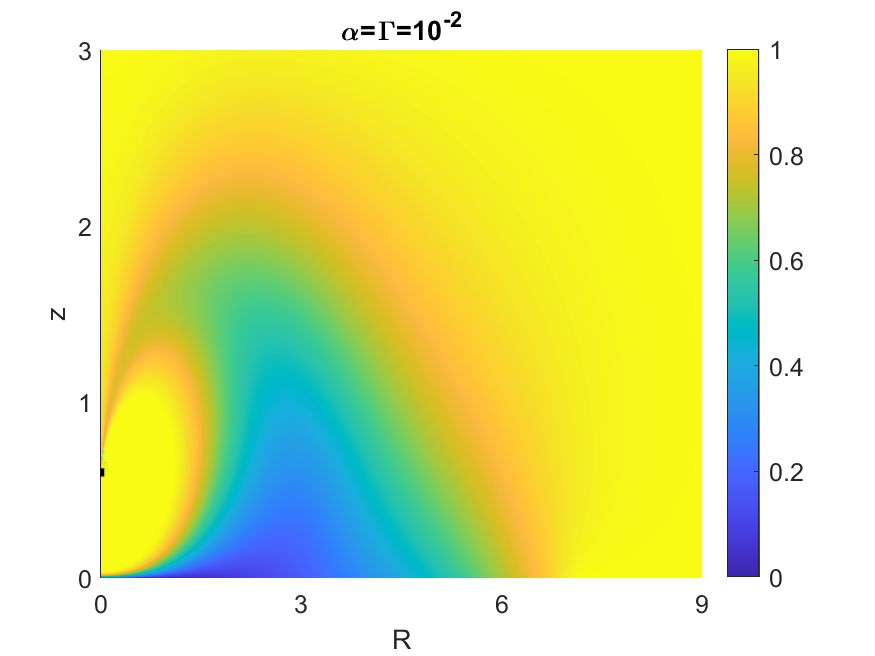}
   \includegraphics[width=0.45\linewidth]{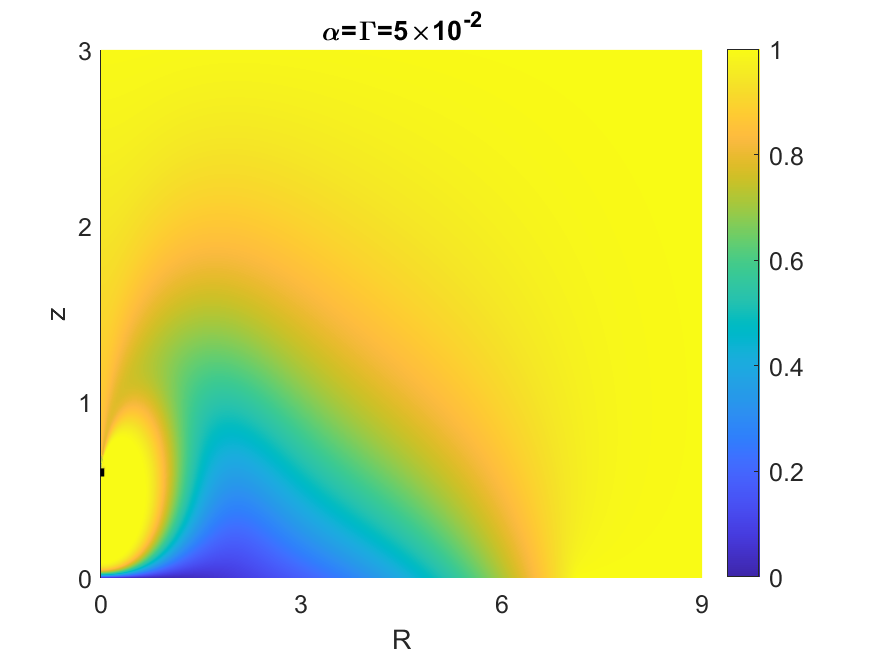}}
   \caption{Voidage field $E(R,z)$, $\alpha=\Gamma=10^{-3}$ (top left), $\alpha=\Gamma=5.0\times10^{-3}$ (top right), $\alpha=\Gamma=10^{-2}$ (bottom left), $\alpha=\Gamma=5.0\times10^{-2}$ (bottom right).}
  \label{fig:inter_alpha}
\end{figure}

To explore this, we begin by fixing all parameter choices as in~(\ref{eqn:martian_param}), with the exception of $\Gamma$, which we decrease from $1.5 \times 10^{-2}$ to $1.0 \times 10^{-3}$, so that $\Gamma/\alpha=1$, as for the examples in~\S\ref{sec:terrestrial}, with parameter choices given by~(\ref{eqn:terrestrial_param}), i.e., here we choose
\begin{equation}
    \Gamma=10^{-3}, \quad \gamma=2\times 10^3, \quad \alpha = 10^{-3}, \quad \Omega=10^0.
    \label{eqn:inter_param}
\end{equation}
The voidage field for this example, for $z_d=0.6$ is shown in the top left plot of figure~\ref{fig:inter_alpha}.
The results in figure~\ref{fig:inter_alpha} were computed by solving (\ref{eqn:h34new}), (\ref{eqn:h35}), (\ref{eqn:h36}), (\ref{eqn:h37new}), together with (\ref{eqn:h38})-(\ref{eqn:h19}) using the numerical method described in \citet[Appendix~B]{NL24}, with discretization parameters $\epsilon=10^{-8}$, $L_R=256$, $L_z=64$, and $N=16$ (as described in \S\ref{sec:model}).  Using these discretization parameters, the number of degrees of freedom is 5,226,112 (for $z_d=0.6$).  As we can see, the ``second swirl'' is still present, leading us to hypothesise that this could be caused by the low value of $\alpha$ rather than the presence of $\Gamma$.  To test this, in the other plots of figure~\ref{fig:inter_alpha}, we repeat the example shown in the top left of figure~\ref{fig:inter_alpha}, with in each case $\Gamma=\alpha$, but now with $\alpha=5.0\times10^{-3}$, $\alpha=1.0\times10^{-2}$, $\alpha=5.0\times10^{-2}$. We see that the swirl is present for $\alpha=\Gamma=5.0\times10^{-3}$, as well as for $\alpha=\Gamma=1.0\times10^{-3}$ but it does not appear to be present for the larger values of $\alpha$.  It seems that the strength of this effect decreases as $\alpha$ increases, when $\alpha=\Gamma$.

\begin{figure}
   \centerline{
   \includegraphics[width=0.45\linewidth]{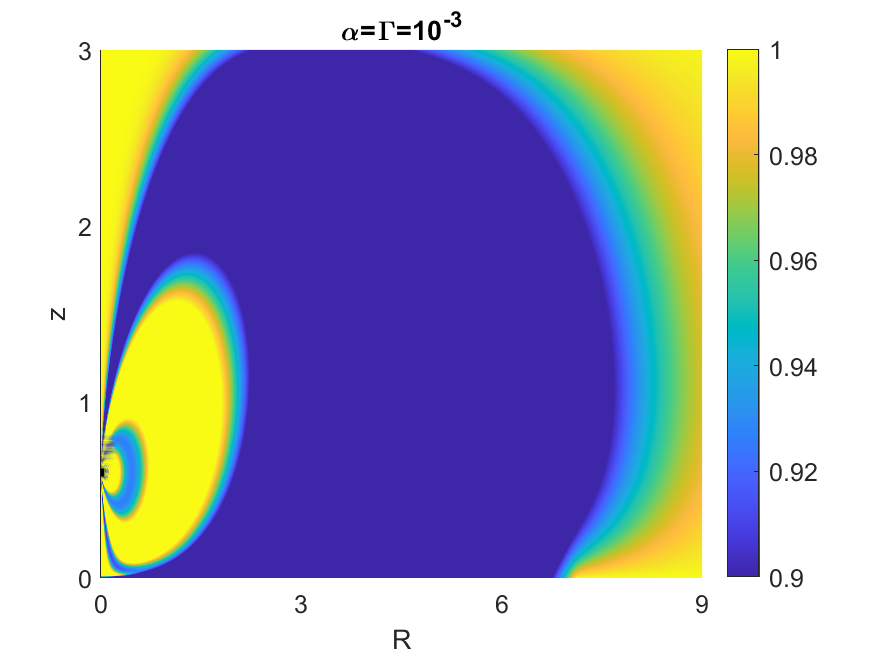}
   \includegraphics[width=0.45\linewidth]{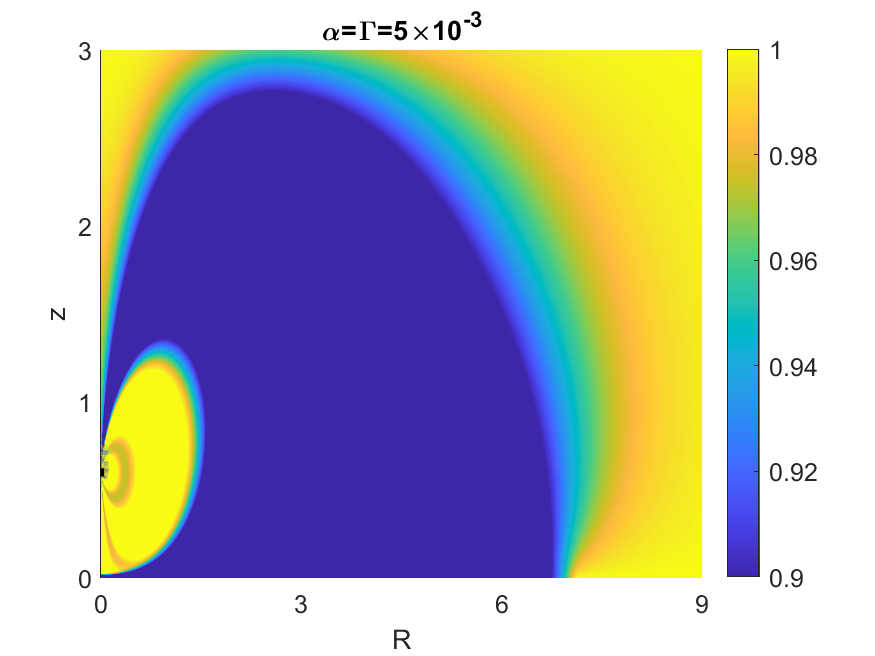}}
   \centerline{
   \includegraphics[width=0.45\linewidth]{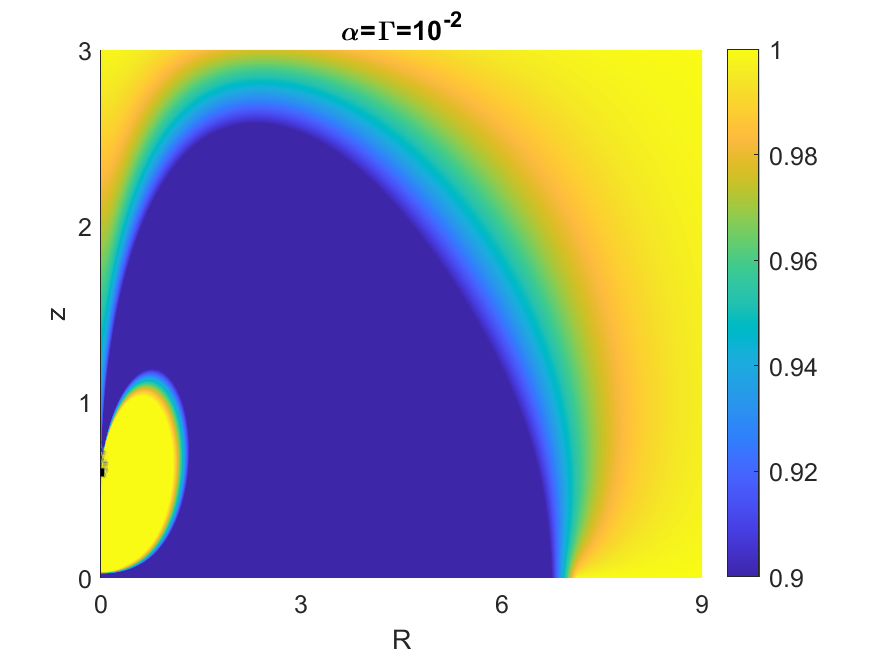}
   \includegraphics[width=0.45\linewidth]{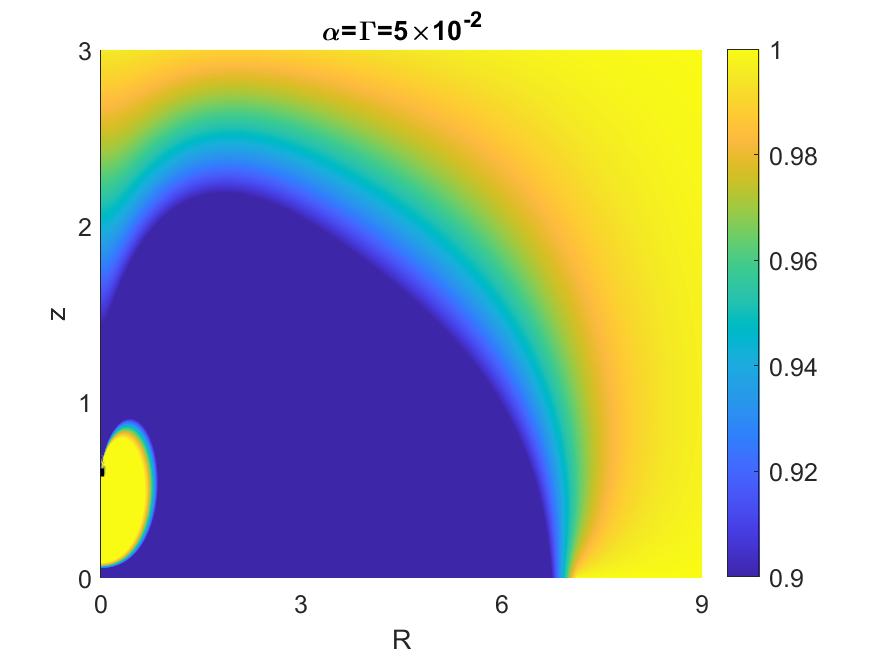}}
   \caption{Voidage field $E(R,z)$, $\alpha=\Gamma=10^{-3}$ (top left), $\alpha=\Gamma=5.0\times10^{-3}$ (top right), $\alpha=\Gamma=10^{-2}$ (bottom left), $\alpha=\Gamma=5.0\times10^{-2}$ (bottom right), reduced colour scale.}
  \label{fig:inter_alpha_clim09}
\end{figure}

This can be seen more clearly in figure~\ref{fig:inter_alpha_clim09}, where we repeat the results in figure~\ref{fig:inter_alpha} but with the colour scale set from 0.9 to 1, rather than from 0 to 1 (as was previously the case), so that all regions where the plot shows blue represent that $E(R,z)\leq 0.9$ in that region.

\begin{figure}
   \centerline{
   \includegraphics[width=0.45\linewidth]{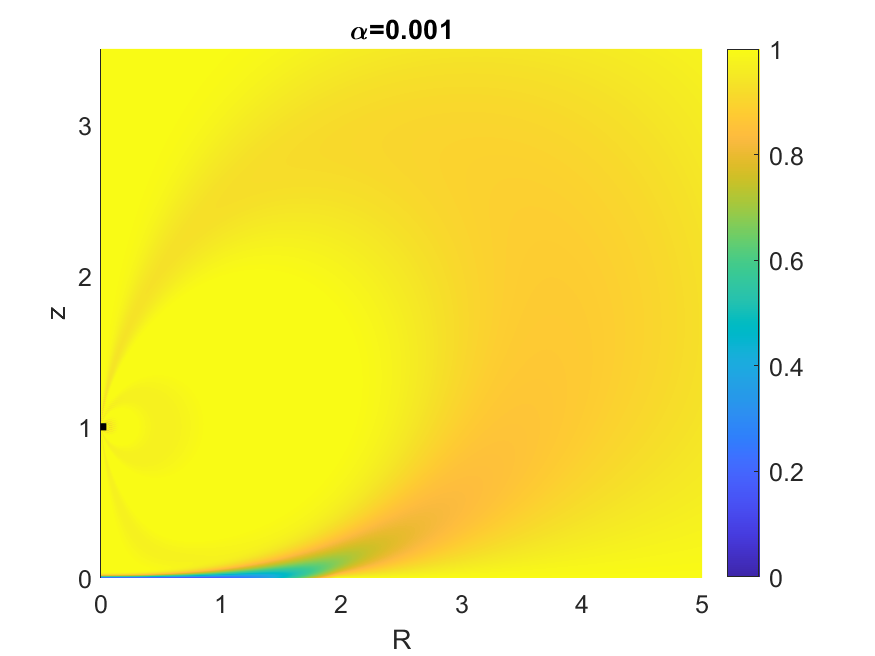}
   \includegraphics[width=0.45\linewidth]{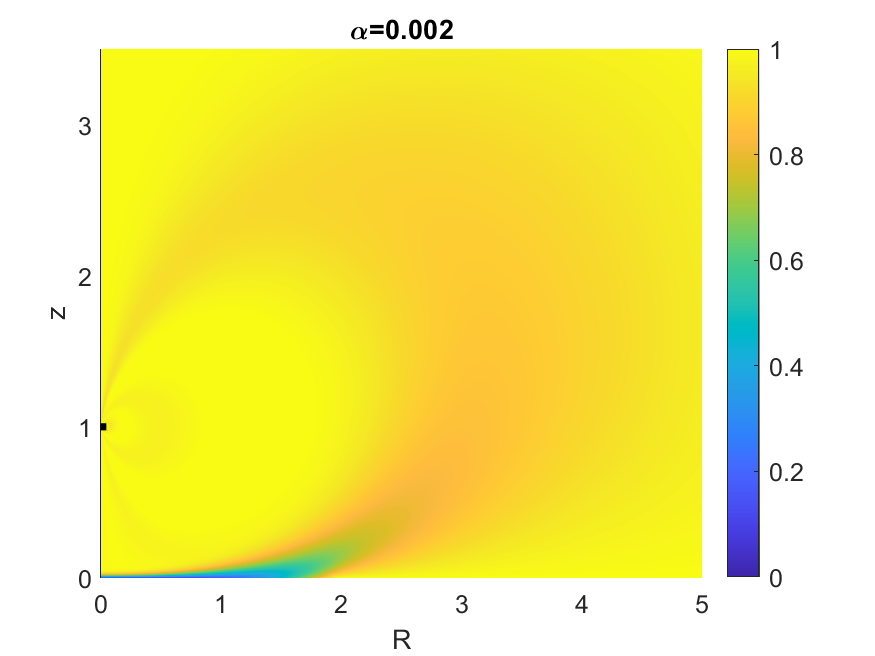}}
   \centerline{
   \includegraphics[width=0.45\linewidth]{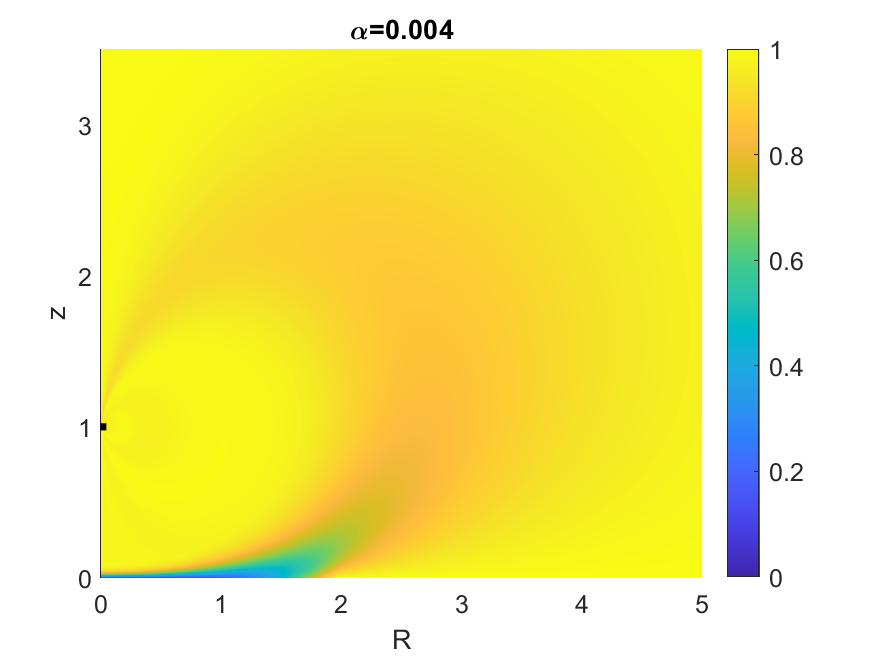}
   \includegraphics[width=0.45\linewidth]{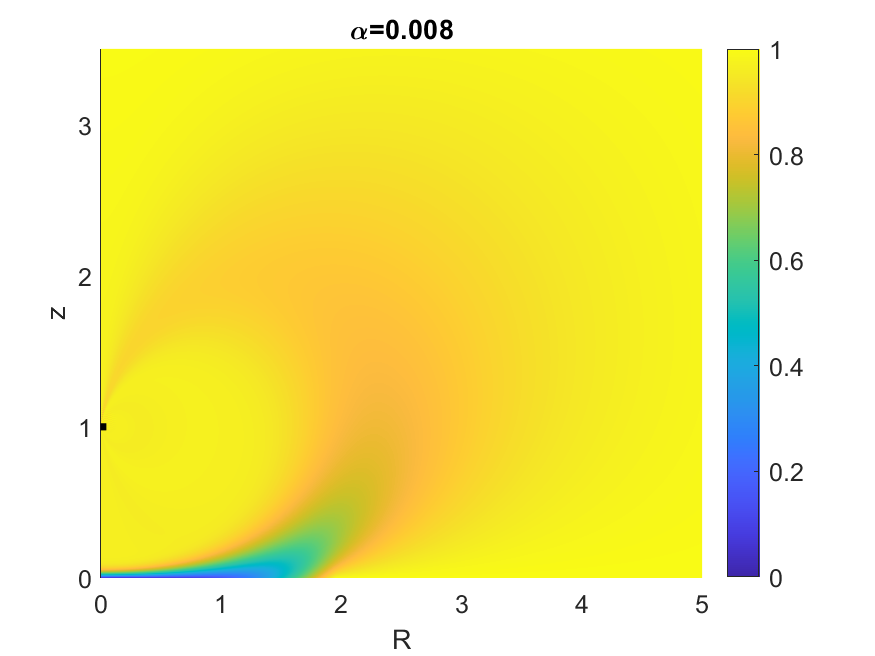}}
   \caption{Voidage field $E(R,z)$, $\alpha=0.001$ (top left), $\alpha=0.002$ (top right), $\alpha=0.004$ (bottom left), $\alpha=0.008$ (bottom right).}
  \label{fig:small_alpha}
\end{figure}

\begin{figure}
   \centerline{
   \includegraphics[width=0.45\linewidth]{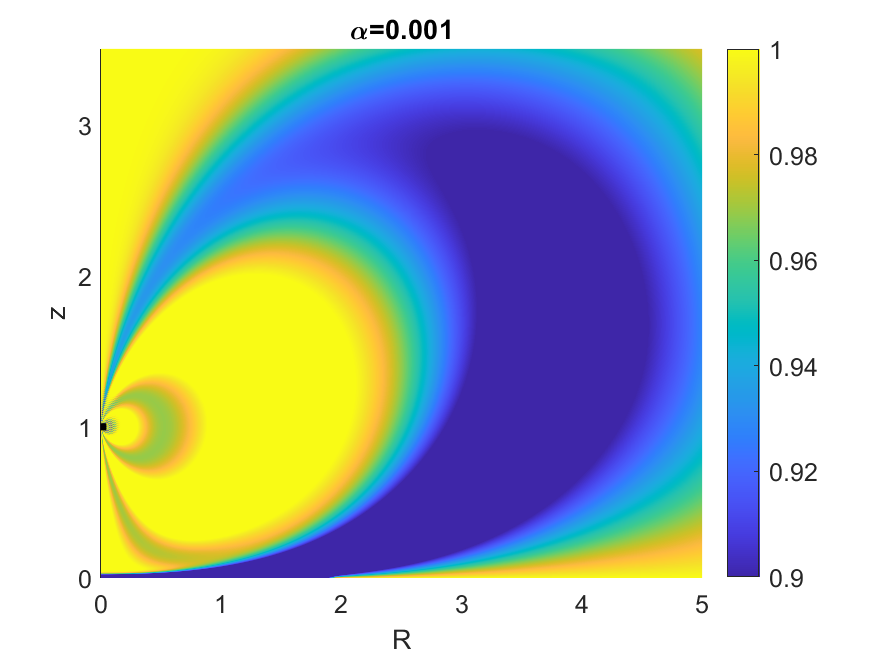}
   \includegraphics[width=0.45\linewidth]{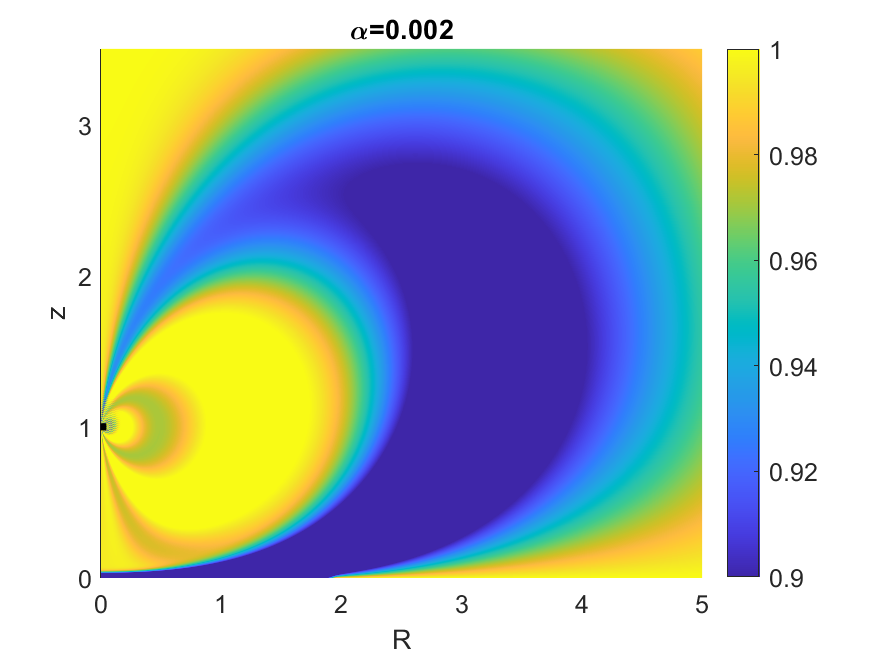}}
   \centerline{
   \includegraphics[width=0.45\linewidth]{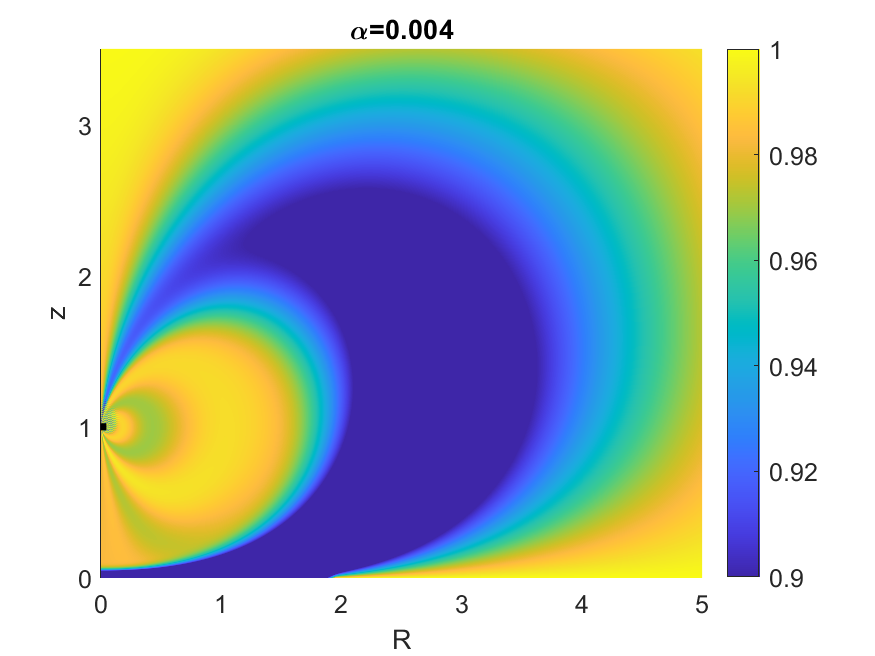}
   \includegraphics[width=0.45\linewidth]{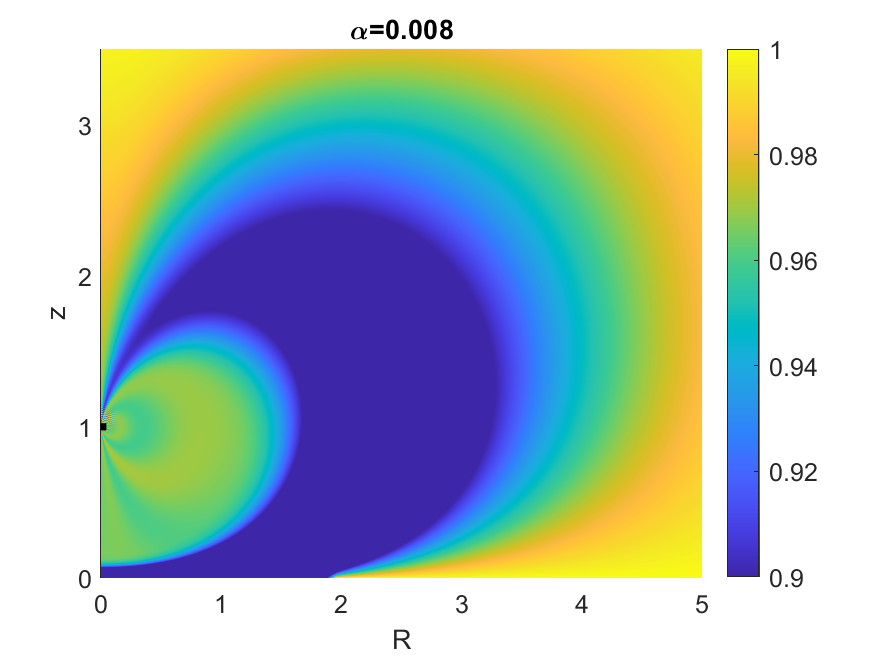}}
   \caption{Voidage field $E(R,z)$, $\alpha=0.001$ (top left), $\alpha=0.002$ (top right), $\alpha=0.004$ (bottom left), $\alpha=0.008$ (bottom right), reduced colour scale.}
  \label{fig:small_alpha_clim09}
\end{figure}

Noting that it appears to be the low value of $\alpha$, rather than the presence of $\Gamma>0$, that causes this phenomenon, we further undertake a series of experiments for the case $\Gamma=0$, replicating the experiments presented in~\citet{NL24} but with much lower values of $\alpha$.  Specifically, we make the following parameter choices:
\begin{equation}
    \Gamma=0, \quad \gamma=500, \quad \Omega=0.5, \quad E_s=0.01, \quad \Delta=0.04, \quad z_d=1.0.
    \label{eqn:small_alpha_param}
\end{equation}
We solve (\ref{eqn:h34})-(\ref{eqn:h19}) using the numerical method described in \citet[Appendix~B]{NL24}, with discretization parameters $\epsilon=10^{-8}$, $L_R=144$, $L_z=72$, and $N=18$ (as described in \S\ref{sec:model}), for which the number of degrees of freedom is 4,577,634 (for $z_d=1.0$).  The voidage field for this set of parameter choices, for a range of choices of $\alpha$, is shown in figure~\ref{fig:small_alpha}.
In figure~\ref{fig:small_alpha_clim09}, we repeat the results in figure~\ref{fig:small_alpha} but with the colour scale set from 0.9 to 1, rather than from 0 to 1, so that all regions where the plot shows blue represent that $E(R,z)\leq 0.9$ in that region.
It is clear from figure~\ref{fig:small_alpha} and figure~\ref{fig:small_alpha_clim09} that this ``second swirl'' occurs when $\alpha$ is sufficiently small even when $\Gamma=0$.

\end{appendix}

\bibliographystyle{jfm}  
\bibliography{helicopter_bib}        

@article{PhBr08,
  title={Eulerian Simulation of the Fluid Dynamics of Helicopter Brownout},
  author={Phillips, C. and Brown, R. E.},
  journal={J. Aircr.},
  fjournal={Journal of Aircraft},
  volume={46},
  number={4},
  pages={1416--1429},
  year={2009},
  doi = {10.2514/1.41999}
}

@article{PhKiBr11,
  title={Helicopter brownout – Can it be modelled?},
  author={Phillips, C. and Kim, H. W. and Brown, R. E.},
  journal={Aeronaut. J.},
  fjournal={The Aeronautical Journal},
  volume={115},
  number={1164},
  pages={123--133},
  year={2011},
  doi = {10.1017/S0001924000005510}
}

@article{PoMiPaPeVe20,
  title = {{HPC} simulations of brownout: A noninteracting particles dynamic model},
  author = {Porc\`u, R. and Miglio, E. and Parolini, N. and Penati, M. and Vergopolan, N.},
  journal = {Int. J. High Perform. Comput. Appl.},
  fjournal = {The International Journal of High Performance Computing Applications},
  volume = {34},
  number = {3},
  pages = {267--281},
  year = {2020},
  doi = {10.1177/1094342020905971}
}

@article{TaGaBaLiZhHu21,
  title = {Novel approach to helicopter brownout based on vortex and discrete element methods},
  author = {Tan, J. F. and Gao, J. and Barakos, G. N. and Lin, C. L. and Zhang, W. G. and Huang, M. Q.},
  journal = {Aerosp. Sci. Technol.},
  fjournal = {Aerospace Science and Technology},
  volume = {116},
  pages = {106839},
  year = {2021},
  doi = {10.1016/j.ast.2021.106839}
}

@article{TaYoHeYuWa22,
  title = {Accelerated method of helicopter brownout with particle-particle collisions},
  author = {Tan, J. F. and Yon, T. and He, L. and Yu, L. J. and Wang, C.},
  journal = {Aerosp. Sci. Technol.},
  fjournal = {Aerospace Science and Technology},
  volume = {124},
  pages = {107511},
  year = {2022},
  doi = {10.1016/j.ast.2022.107511}
}

@article{GLG,
  title = {Particle-clustering algorithms for the prediction of brownout dust clouds},
  author = {Govindarajan, B. and Leishman, J. G. and Gumerov, N. A.},
  journal = {AIAA J.},
  fjournal = {AIAA Journal},
  volume = {51},
  pages = {1080--1094},
  year = {2013},
  number = {5},
  doi = {10.2514/1.J051907}
}

@article{LXJHL,
  title = {Finite particle approach for high-fidelity simulation on helicopter brownout},
  author = {Lin, H. and Xu, C. and Jiang, C. and Hu, S. and Lee, C-H.},
  journal = {AIAA J.},
  fjournal = {AIAA Journal},
  volume = {62},
  pages = {193--208},
  year = {2024},
  number = {1},
  doi = {10.2514/1.J063230}
}

@article{TGZWCW,
  title = {Numerical study on helicopter brownout with crosswind},
  author = {Tan, J. F. and Ge, Y. Y. and Zhang, W. G. and Cui, Z. and Wang, H. W.},
  journal = {Aerosp. Sci. Technol.},
  fjournal = {Aerospace Science and Technology},
  volume = {131},
  pages = {107965},
  year = {2022},
  doi = {10.1016/j.ast.2022.107965}
}

@article{LSXW,
  title = {Research on brownout characteristics of helicopter approach flight based on the unsteady momentum source model coupled with {DEM}},
  author = {Li, Q. and Shi, Y. and Xu, G. and Wu, L.},
  journal = {Int. J. Aeronaut. Space},
  fjournal = {International Journal of Aeronautical and Space Sciences},
  volume = {24},
  pages = {366--380},
  year = {2023},
  doi = {10.1007/s42405-022-00534-8}
}

@article{KPMK,
  title = {The physics of wind-blown sand and dust},
  author = {Kok, J. F. and Parteli, E. J. R. and Michaels, T. I. and Karam, D. B.},
  journal = {Rep. Prog. Phys.},
  fjournal = {Reports on progress in physics. Physical Society (Great Britain)},
  volume = {75},
  pages = {106901},
  year = {2012},
  doi = {10.1088/0034-4885/75/10/106901}
}

@article{NL24,
  title={A mathematical model for wind-generated particle-fluid flow fields with an application to the helicopter cloud problem},
  volume={998},
  DOI={10.1017/jfm.2024.740},
  journal={J. Fluid Mech.},
  fjournal={Journal of Fluid Mechanics},
  author={Needham, D. J. and Langdon, S.},
  year={2024},
  pages={A61}
  }

@article{CaWaJi22,
  title={Study on Influencing Factors of Helicopter Brownout Evolution Based on {CFD-DEM}},
  volume={12},
  DOI={10.3390/app12010126},
  journal={Appl. Sci.},
  fjournal={Applied Sciences},
  author={Cao, Y. and Wang, G. and Jiang, C.},
  year={2022},
  pages={126}
  }

@article{RoBaSt23,
  title = {Comparison between {E}ulerian and {L}agrangian methods to predict brownout clouds},
  author = {Rovere, F. and Barakos, G. N. and Steijl, R.},
  journal = {Aerosp. Sci. Technol.},
  fjournal = {Aerospace Science and Technology},
  volume = {137},
  pages = {108306},
  year = {2023},
  doi = {10.1016/j.ast.2023.108306}
}

@article{LN25,
  title = {Modeling Visibility and Surface Deformation in Helicopter Rotor Generated Particle-Fluid Flow Fields},
  author = {Langdon, S. and Needham, D.},
  journal = {AIAA J.},
  fjournal = {AIAA Journal},
  volume = {63},
  pages = {5461-5475},
  year = {2025},
  number = {12},
  doi = {10.2514/1.J065257}
}

@article{dust_devil_A,
  title = {Behavior of Dust Devils},
  author = {Ives, R. L.},
  journal = {Bull. Am. Meteorol. Soc.},
  fjournal = {Bulletin of the American Meteorological Society},
  volume = {28},
  pages = {168-174},
  year = {1947}
}

@article{dust_devil_B,
  title = {History and Applications of Dust Devil Studies},
  author = {Lorenz, R. D. and Balme, M. R. and Gu, Z. and Kahanp\"{a}\"{a}, H. and Klose, M. and Kurgansky, M. V. and Patel, M. R. and Reiss, D. and Pio Rossi, A. and Spiga, A. and Takemi, T. and Wei, W.},
  journal = {Space Sci. Rev.},
  fjournal = {Space Science Reviews},
  volume = {203},
  pages = {5-37},
  year = {2016},
  doi = {10.1007/s11214-016-0239-2}
}

@article{dust_devil_C,
  title = {The sound of a {M}artian dust devil},
  author = {Murdoch, N. and Stott, A. E. and Gillier, M. and Hueso, R. and Lemmon, M. and Martinez, G. and Ap\'{e}stigue amd Toledo, D. and Lorenz, R. D. and Chide, B. and Munguira, A. and S\'{a}nchez-Lavega, A. and Vicente-Retortillo, A. and Newman, C. E. and Maurice, S. and de la Torre Ju\'{a}rez, M. and Bertrand, T. and Banfield, D. and Navarro, S. and Marin, M. and Torres, J. and Gomez-Elvira, J. and Jacob, X. and Cadu, A. and Sournac, A. and Rodriguez-Manfredi, J. A. and Wiens, R. C. and Mimoun, D.},
  journal = {Nat. Commun.},
  fjournal = {Nature Communications},
  volume = {13},
  pages = {7505},
  year = {2022},
  doi = {10.1038/s41467-022-35100-z}
}

@article{dust_devil_D,
title = {A dynamically consistent model for dust-devil-like flows},
journal = {Aeolian Res.},
fjournal = {Aeolian Research},
volume = {75},
pages = {101030},
year = {2026},
doi = {10.1016/j.aeolia.2025.101030},
author = {Pandey, S. K. and Yadav, K.}
}

@article{dust_devil_E,
  title = {General Characteristics of Dust Devils},
  author = {Sinclair, P. C.},
  journal = {J. Appl. Meteorol.},
  fjournal = {Journal of Applied Meteorology},
  volume = {8},
  pages = {32-45},
  year = {1969}
}

@book{Beeb,
  author = "Beebe, W.",
  title= "Half Mile Down ",
  publisher="Harcout, Brace and Company, New York",
  year="1934"}

@article{Jackson20,
  title = {On the relationship between dust devil radii and heights},
  author = {Jackson, B.},
  journal = {Icarus},
  fjournal = {Icarus},
  volume = {338},
  pages = {113523},
  year = {2020},
  doi = {10.1016/j.icarus.2019.113523}
}

@misc{NASA_ingenuity,
  author = {{National Aeronautics and Space Administration (NASA)}},
  title = {{I}ngenuity {M}ars {H}elicopter - {NASA} {S}cience},
  howpublished = {\url{https://science.nasa.gov/mission/mars-2020-perseverance/ingenuity-mars-helicopter/}},
  year = {2020},
  note = {Accessed: 2026-09-11}
}

@misc{DD_wiki,
  author = {{National Aeronautics and Space Administration (NASA)}},
  title = {A dust devil in {A}rizona, image created 10th {J}une 2005 by {NASA}},
  howpublished = {\url{https://commons.wikimedia.org/wiki/File:Dust\_devil.jpg}},
  year = {2005},
  note = {Accessed: 2026-09-11}
}

@article{Nathan_Green_2009, 
  title={Flow visualisation of the helicopter brown-out phenomenon}, 
  volume={113}, 
  DOI={10.1017/S0001924000003134}, 
  number={1145}, 
  journal={Aeronaut. J.},
  fjournal={The Aeronautical Journal}, 
  author={Nathan, N. D. and Green, R. B.}, 
  year={2009}, 
  pages={467–478}
}

@article{CJC25,
title = {Plume-surface interactions: A review of experimental work},
journal = {Acta Astronaut.},
fjournal = {Acta Astronautica},
volume = {226},
pages = {892-912},
year = {2025},
issn = {0094-5765},
doi = {https://doi.org/10.1016/j.actaastro.2024.09.021},
url = {https://www.sciencedirect.com/science/article/pii/S0094576524005204},
author = {Jimenez Cuesta, C. and Davies, J. and Worrall, K. and Cammarano, A. and Zare-Behtash, H.}
}

@article{JKJKBJL25,
title = {A survey of research on lunar dust dispersal due to rocket plume impingement},
journal = {Acta Astronaut.},
fjournal = {Acta Astronautica},
volume = {227},
pages = {67-95},
year = {2025},
issn = {0094-5765},
doi = {https://doi.org/10.1016/j.actaastro.2024.11.021},
url = {https://www.sciencedirect.com/science/article/pii/S0094576524006672},
author = {Kim, J. and Kim, J. and Lee, B. J.}
}

@article{JC22,
title = {Modeling high-speed gas–particle flows relevant to spacecraft landings},
journal = {Int. J. Multiph. Flow},
fjournal = {International Journal of Multiphase Flow},
volume = {150},
pages = {104008},
year = {2022},
issn = {0301-9322},
doi = {https://doi.org/10.1016/j.ijmultiphaseflow.2022.104008},
url = {https://www.sciencedirect.com/science/article/pii/S0301932222000301},
author = {Capecelatro, J.}
}

@ARTICLE{ZCCW23,
AUTHOR={Cao, Z.  and White, C.  and Agir, M. B.  and Kontis, K. },
TITLE={Lunar plume-surface interactions using {rarefiedMultiphaseFoam}},
JOURNAL={Front. Mech. Eng.},
fJOURNAL={Frontiers in Mechanical Engineering},
VOLUME={9},
YEAR={2023},
URL={https://www.frontiersin.org/journals/mechanical-engineering/articles/10.3389/fmech.2023.1116330},
DOI={10.3389/fmech.2023.1116330},
ISSN={2297-3079}}

@phdthesis{WAC20,
  title = {Experimental and Computational Investigation of Plume Surface Interactions in Vacuum Microgravity},
  author = {Chambers, W. A.},
  school= {University of Central Florida},
  url = {https://stars.library.ucf.edu/etd2020/192},
  year = {2020}
  }

@article{RaLoSlWa21,
title = {Scaling sediment mobilization beneath rotorcraft for {T}itan and {M}ars},
journal = {Aeolian Res.},
fjournal = {Aeolian Research},
volume = {48},
pages = {100653},
year = {2021},
issn = {1875-9637},
doi = {https://doi.org/10.1016/j.aeolia.2020.100653},
url = {https://www.sciencedirect.com/science/article/pii/S1875963720301051},
author = {Rabinovitch, J. and Lorenz, R. and Slimko, E. and Wang, K. C.}
}

@article{GaScMeGi26,
title = {Addressing camera sensors faults in Vision-Based Navigation: Simulation and dataset development},
journal = {Acta Astronaut.},
fjournal = {Acta Astronautica},
volume = {239},
pages = {849-872},
year = {2026},
issn = {0094-5765},
doi = {https://doi.org/10.1016/j.actaastro.2025.10.071},
url = {https://www.sciencedirect.com/science/article/pii/S0094576525007520},
author = {Gallon, R. and Schiemenz, F. and Menicucci, A. and Gill, E.}
}

@article{CaBaDiYo25,
author = {Caprace, D. and Baker, E. and Diaz, P. V. and Yoon, S.},
title = {Planetary Helicopter Brownout Simulation of Ingenuity Using a {E}ulerian Dust Transport Model},
journal = {AIAA J.},
fjournal = {AIAA Journal},
volume = {63},
number = {10},
pages = {4393-4408},
year = {2025},
doi = {10.2514/1.J065017},
URL = {https://doi.org/10.2514/1.J065017}
}

@inbook{BaCoBrDe19,
author = {Bayard, D. S. and Conway, D. T. and Brockers, R. and Delaune, J. H. and Matthies, L. H. and Grip, H. F. and Merewether, G. B. and Brown, T. L. and San Martin, A. M.},
title = {Vision-Based Navigation for the NASA Mars Helicopter},
booktitle = {AIAA Scitech 2019 Forum},
chapter = {},
pages = {},
publisher = {AIAA Scitech 2019 Forum},
year = {2019},
doi = {10.2514/6.2019-1411},
URL = {https://arc.aiaa.org/doi/abs/10.2514/6.2019-1411}
}

@article{TaYaXuZhWaFa26,
author = {Tan, J. F. and Yan, Y. Z. and Xu, S. G. and Zhang, W. G. and Wang, C. and Fan, F.},
title = {Numerical Investigation of Blade Erosion in Helicopter Brownout},
journal={J. Aircr.},
fjournal={Journal of Aircraft},
volume = {63},
number = {4},
pages = {1418-1435},
year = {2026},
doi = {10.2514/1.C038327},
URL = {https://doi.org/10.2514/1.C038327}
}

@Article{LiDoSoYa25,
AUTHOR = {Li, Q. and Dong, L. and Song, C. and Yang, W.},
TITLE = {Sand Ingestion Behavior of Helicopter Engines During Hover in Ground Effect},
JOURNAL = {Aerospace},
FJOURNAL = {Aerospace},
VOLUME = {12},
YEAR = {2025},
NUMBER = {10},
ARTICLE-NUMBER = {927},
URL = {https://www.mdpi.com/2226-4310/12/10/927},
ISSN = {2226-4310},
DOI = {10.3390/aerospace12100927}
}

@article{HuCa26,
author = {Huang, Z. and Cao, Y.},
title ={Numerical study of the effect of trailing edge flap on helicopter brownout based on the CFD-DEM coupling method},
journal = {Proc. Inst. Mech. Eng. G},
fjournal = {Proceedings of the Institution of Mechanical Engineers, Part G: Journal of Aerospace Engineering},
volume = {240},
number = {1},
pages = {177-187},
year = {2026},
doi = {10.1177/09544100251367331},
URL = {https://doi.org/10.1177/09544100251367331}
}

@article{WuRe26,
title = {Concentration field characteristics of sediments entrained by rotor downwash flow},
journal = {Chin. J. Aeronaut.},
fjournal = {Chinese Journal of Aeronautics},
pages = {104121},
year = {2026},
issn = {1000-9361},
doi = {https://doi.org/10.1016/j.cja.2026.104121},
url = {https://www.sciencedirect.com/science/article/pii/S1000936126000580},
author = {Wu, Z. and Ren, J. and Niu, J. and Li, W. and Tan, H. and Zhang, T. and Xie, M.}
}

\end{document}